\documentclass{LMCS}

\def\Rlap#1{\rlap{\mathsurround=0 pt$#1$}}

\usepackage{ifthen,calc,proof}
\usepackage[arrow,matrix,curve,frame,rotate]{xy}  
\usepackage{fancybox,multicol}
\usepackage{caption}
\usepackage{array}
\usepackage{pifont}
\usepackage{pstricks,pst-node,pst-tree}
\usepackage{graphicx,picins}
\usepackage{amsmath,amssymb,stmaryrd}
\usepackage{ulsy}
\usepackage[greek,english]{babel}
\usepackage{prooftree}
\usepackage{enumerate,hyperref}

\newcommand{\dfny}[2][\all]{D^{#1}_{#2}}
\newenvironment{eqn}{\begin{equation}}{\end{equation}}
\newenvironment{eqn*}{\begin{equation*}}{\end{equation*}}

\newcommand{\nzcirc}{\text{\raisebox{-.25ex}{$^{_{_\circ}}$}}}
\newcommand{\NZvbar}{\mid}
\newcommand{\NZvdash}{\mathrel|\joinrel\relbar}
\newcommand{\nzvddash}{\models}
\newcommand{\oted}[1]{(#1)^{\circ}}

\newcounter{quad_i}
\newcounter{newl_i}

\newcommand{\HY}{\hspace{2pt}---\hspace{2pt}}
\newcommand{\boldemph}[1]{\textbf{\emph{#1}}}
\newcommand{\sublist}{\preceq}

\newcommand{\MODEL}[2][\all\in{\NAA[\#]}]{(#2)_{#1}}

\newcommand{\all}{\ensuremath{{\bar{a}}}}

\newcommand{\allal}{\ensuremath{{\all\al}}}

\newcommand{\pln}{\setminus}
\newcommand{\plin}[1]{{^{{\scriptstyle\smallsetminus}#1}}}
\newcommand{\listplin}{\smallsetminus}
\newcommand{\TY}{{\rm TY}}
\newcommand{\QQ}[1][]{Q^2\!#1}
\newcommand{\QQQ}{Q^3}
\newcommand{\TT}[1][]{T^2#1}

\newcommand{\TTT}{T^3}
\newcommand{\bbfamily}{\fontencoding{U}\fontfamily{bbold}\selectfont}
\newcommand{\abs}[1][\al]{\text{\bbfamily{\char 94}}\,#1\,\text{\bbfamily{\char 95}}\,} 
\newcommand{\trn}[2][]{{\llbracket#2\rrbracket_{#1}}}
\newcommand{\trnn}[1]{{|#1|}}

\newcommand{\ifo}[3]{\mathtt{if0}\  #1\ \mathtt{then}\ #2\ \mathtt{else}\ #3}

\newcommand{\na}[1][]{\mathtt{a}\ifthenelse{\equal{#1}{}}{}{^{#1}}}
\newcommand{\nb}[1][]{\mathtt{b}\ifthenelse{\equal{#1}{}}{}{^{#1}}}
\newcommand{\ena}{\mathbbold{1}}
\renewcommand{\1}{\mathop{;}}
\newcommand{\nm}[1]{\tilde{#1}}
\newcommand{\n}{\nm{n}}

\newcommand{\proj}[1]{\mathtt{proj}_{#1}}
\newcommand{\incl}[1]{\mathtt{incl}_{#1}}
\newcommand{\sskip}{\mathtt{skip}}
\newcommand{\sstop}{\mathtt{stop}}
\newcommand{\fst}{\mathtt{fst}\,}
\newcommand{\snd}{\mathtt{snd}\,}
\newcommand{\pred}{\mathtt{pred}\,}
\renewcommand{\succ}{\mathtt{succ}\,}

\newcommand{\iifo}[1][]{\mathtt{if0}_{#1}}
\newcommand{\eq}[1][]{\mathtt{eq}_{#1}}
\newcommand{\drf}[1][]{\mathtt{drf}_{#1}}

\newcommand{\upd}[1][]{\mathtt{upd}_{#1}}

\newcommand{\upp}[1][]{\mathtt{up}_{#1}}
\newcommand{\puu}[1][]{\mathtt{pu}_{#1}}
\newcommand{\dnn}[1][]{\mathtt{dn}_{#1}}
\newcommand{\stt}[1][]{\mathtt{st}_{#1}}
\newcommand{\isom}{\mathord{\cong}}

\newcommand{\la}{\lambda}
\newcommand{\La}{\Lambda}
\newcommand{\LaT}[1][]{\mathord{\Lambda_{#1}^{T}}}

\newcommand{\LaQT}[1][]{\mathord{\Lambda_{#1}^{Q,T}}}
\newcommand{\LaQQT}[1]{\mathord{\Lambda^{#1,T}}}
\newcommand{\Laa}[1][\all]{\mathord{\Lambda^{#1}}}

\newcommand{\ev}{\mathsf{ev}}
\newcommand{\evT}[1][]{\ev^{T}_{#1}}
\newcommand{\new}[1][]{\nu\ifthenelse{\equal{#1}{}}{}{#1.}}
\newcommand{\nw}{\mathtt{nu}}
\newcommand{\frs}{\mathtt{new}}

\newcommand{\tims}{\!\times\!}
\newcommand{\arnt}{\raisebox{.3ex}{$\scriptscriptstyle-$}}
\newcommand{\comp}{\mathop{\|}}
\newcommand{\cmp}{\mathop{;}}
\newcommand{\mix}{\mathop{\bullet}}

\newcommand{\hrp}{\upharpoonright}
\newcommand{\sig}{\sigma}
\newcommand{\st}{\sigma\cmp\tau}
\newcommand{\iseq}{\mathsf{ISeq}}
\newcommand{\nlist}{\mathsf{nlist}}
\newcommand{\play}[1]{P_{#1}}
\newcommand{\iplay}[1]{P_{#1}^{\mathtt{i}}}

\newcommand{\tenn}[1][]{\mathord{\otimes}}
\newcommand{\ten}[1][]{\mspace{1.5mu}#1\mathord{\otimes}\mspace{1.5mu}#1}
\newcommand{\tenf}{(\!\_\otimes\!\_\,)}
\newcommand{\bigtenf}{(\bigotimes\!\_\,)}
\newcommand{\Ten}[2][i]{\bigotimes\nolimits_{#1}\!#2_#1}
\newcommand{\Tenn}[2][i]{\bigotimes\nolimits_{#1}\!#2}
\newcommand{\pls}[1][]{+}
\newcommand{\Lub}[1][i]{\bigsqcup\nolimits_{#1}}
\newcommand{\himf}{(\!\_\him\!\_\,)}
\newcommand{\implf}{(\!\_\impl\!\_\,)}
\newcommand{\botf}{(\!\_\,)_\bot}
\newcommand{\bS}{\bar{S}}
\newcommand{\BI}[1]{{\bar{I}_{#1}{}}}
\newcommand{\BJ}[1]{{\bar{J}_{#1}{}}}

\newcommand{\ypob}{\mathrel{\lessapprox}}
\newcommand{\Rypob}{\mathrel{\gtrapprox}}
\newcommand{\Eypob}{\mathrel{\approxeq}}
\newcommand{\ypoc}[1][\all]{\mathrel{\lesssim^{#1}}}

\newcommand{\Eypoc}{\mathrel{\backsimeq}}

\renewcommand{\obs}[1][\all]{O^{#1}}
\newcommand{\rred}[1][]{\mathrel{\overset{#1}{\longrightarrow\hspace{-3mm}\rightarrow}}}
\newcommand{\viewf}{\mathtt{viewf}}
\newcommand{\strat}{\mathtt{strat}}
\newcommand{\Ma}[1][]{\ensuremath{\MM^{#1}}}
\newcommand{\T}{\mathcal{T}}

\newcommand{\Vt}[1][]{\mathcal{V}_{\mathtt{t}}^{#1}}
\newcommand{\Vtt}[1][]{\mathcal{V}_{\mathtt{tt}}^{#1}}

\newcommand{\Vtts}[1][]{\mathcal{V}_{\mathtt{tt}*}^{#1}}

\newcommand{\Vts}[1][]{\mathcal{V}_{\mathtt{t}*}^{#1}}
\newcommand{\Vnr}[1][]{\mathcal{V}_{\nurho}}
\newcommand{\Lift}{(\uscore)_\bot}
\renewcommand{\o}{\!:\!}

\newcommand{\uscore}{\,\!\_\!\_\,}
\newcommand{\lnza}{\rightarrow}
\newcommand{\paei}{\lnza}

\newcommand{\modl}{\nzvddash}
\newcommand{\seq}[3]{#1\NZvbar#2\NZvdash#3}
\newcommand{\Gseq}[3][]{\seq{#2}{\Ga\ifthenelse{\equal{#1}{}}{}{,#1}}{#3}}
\newcommand{\Aseq}[3][]{\seq{\all\ifthenelse{\equal{#1}{}}{}{#1}}{#2}{#3}}
\newcommand{\AGseq}[2][]{\seq{\all\ifthenelse{\equal{#1}{}}{}{#1}}{\Ga}{#2}}
\newcommand{\GAseq}[2][]{\seq{\all}{\Ga\ifthenelse{\equal{#1}{}}{}{,#1}}{#2}}
\newcommand{\GAseqq}[2][]{\seq{\all'}{\Ga'\ifthenelse{\equal{#1}{}}{}{,#1}}{#2}}
\newcommand{\str}[2][2cm]{\save[]-<#1,0cm>*{#2}\restore}
\newcommand{\jst}[2][]{\ar@{-}@/_#2/#1}
\def\ColorA{gray}
\def\ColorB{black}

\newcommand{\cc}[3][5pt]{\ncbar[angle=-90,nodesep=3pt,linewidth=0.5pt,linecolor=darkgray,armA=#1,armB=#1]{#2}{#3}}
\newcommand{\ccolour}[4][5pt]{\ncbar[angle=-90,nodesep=3pt,linewidth=0.5pt,linecolor=#4,armA=#1,armB=#1]{#2}{#3}}
\newcommand{\ccA}[3][5pt]{\ccolour[#1]{#2}{#3}{\ColorA}}
\newcommand{\ccB}[3][5pt]{\ccolour[#1]{#2}{#3}{\ColorB}}

\newcommand{\nccdu}[3][]{\nccurve[nodesep=1.5pt,linewidth=0.4pt,angleA=90,angleB=270,linecolor=darkgray#1]{#2}{#3}}
\newcommand{\dunode}[3]{\rnode{#1}{#2}\nccdu{#1}{#3}}

\newcommand{\biglinehere}[2]{\psline[linewidth=1mm,linestyle=dotted,linecolor=gray](#1)(#2)}
\newcommand{\cnd}[1][]{\mathtt{cnd}_{#1}}

\newcommand{\per}{\pi}
\newcommand{\zet}{\zeta}

\newcommand{\zzet}{\bar\zeta{}}
\newcommand{\eps}{\varepsilon}

\newcommand{\de}{\delta}
\newcommand{\prn}[1][\all]{Q^{#1}}

\newcommand{\laNR}{\ensuremath{\nurho}}
\newcommand{\nurho}{\ensuremath{\nu\mspace{-2mu}\rho\mspace{.5mu}}}
\newcommand{\ner}{\ensuremath{\nu\hspace{-0.4mm}\varepsilon\hspace{-0.4mm}\rho}}

\newcommand{\ixi}{\circledast}
\newcommand{\indx}[1]{_{\scriptscriptstyle(#1)}}
\newcommand{\stin}[1]{{}^{\,#1}}
\newcommand{\eistin}[2]{#1\stin{#2}}

\newcommand{\trunc}{\mathtt{trunc}}
\newcommand{\trnat}[1]{_{\leq#1}}
\newcommand{\aaaa}[2]{*++[Fo]{\txt{#1\\#2}}}
\newcommand{\td}[1][q]{\tilde{#1}}

\newcommand{\bfi} {\begin{figure}}
\newcommand{\efi} {\end{figure}}
\newcommand{\bfa} {\begin{array}}
\newcommand{\efa} {\end{array}}
\newcommand{\bs} {\begin{split}}
\newcommand{\es} {\end{split}}
\newcommand{\boe} {\begin{equation}}
\newcommand{\eoe} {\end{equation}}
\newcommand{\ba} {\begin{eqnarray}}
\newcommand{\ea} {\end{eqnarray}}
\newcommand{\bdm} {\begin{displaymath}}
\newcommand{\edm} {\end{displaymath}}

\newcommand{\ul} {\underline}
\newcommand{\ara}{\;\therefore\;}
\newcommand{\contra}{\blitza}
\newcommand{\newlines}[1]{\setcounter{newl_i}{0}\whiledo{\value{newl_i}<#1}{\nada\\\stepcounter{newl_i}}}
\newcommand{\newlinesminusone}[1]{\setcounter{newl_i}{1}\whiledo{\value{newl_i}<#1}{\nada\\\stepcounter{newl_i}}}

\theoremstyle{plain}
\newtheorem{propdefn}[thm]{Proposition and Definition}
\newtheorem{lemma}[thm]{Lemma}
\theoremstyle{definition}
\newtheorem{remark}[thm]{Remark}
\newtheorem{excercise}[thm]{Excercise}
\newtheorem{defin}[thm]{Definition}
\newtheorem{notation}[thm]{Notation}
\newtheorem{example}[thm]{Example}
\newcommand{\bethm}[1][]{%
    \ifthenelse{\equal{#1}{}}{\begin{thm}}{\begin{thm}[\textbf{#1}]}}
\newcommand{\enthm}{\end{thm}\noindent}
\newcommand{\beprop}[1][]{%
    \ifthenelse{\equal{#1}{}}{\begin{prop}}{\begin{prop}[\textbf{#1}]}}
\newcommand{\enprop}{\end{prop}\noindent}
\newcommand{\bepropdefn}[1][]{%
    \ifthenelse{\equal{#1}{}}{\begin{propdefn}}{\begin{propdefn}[\textbf{#1}]}}
\newcommand{\enpropdefn}{\end{propdefn}\noindent}
\newcommand{\befact}[1][]{%
    \ifthenelse{\equal{#1}{}}{\begin{fact}}{\begin{fact}[\textbf{#1}]}}
\newcommand{\enfact}{\end{fact}}
\newcommand{\bdefn}[1][]{%
    \ifthenelse{\equal{#1}{}}{\begin{defin}}{\begin{defin}[\textbf{#1}]}\text{}}
\newcommand{\edefn}[1][]{\ifthenelse{\equal{#1}{}}{\deq}{}\end{defin}\noindent}
\newcommand{\berem}[1][]{%
    \ifthenelse{\equal{#1}{}}{\begin{remark}}{\begin{remark}[\textbf{#1}]}\text{}}
\newcommand{\enrem}{\end{remark}\noindent}
\newcommand{\benotn}[1][]{%
    \ifthenelse{\equal{#1}{}}{\begin{notation}}{\begin{notation}[\textbf{#1}]}\text{}}
\newcommand{\ennotn}{\end{notation}\noindent}
\newcommand{\beexc}[1][]{%
    \ifthenelse{\equal{#1}{}}{\begin{excercise}}{\begin{excercise}[\textbf{#1}]}\text{}}
\newcommand{\enexc}{\end{excercise}}
\newcommand{\becor}[1][]{%
    \ifthenelse{\equal{#1}{}}{\begin{cor}}{\begin{cor}[\textbf{#1}]}}
\newcommand{\encor}{\end{cor}}
\newcommand{\beexam}[1][]{%
    \ifthenelse{\equal{#1}{}}{\begin{example}}{\begin{example}[\textbf{#1}]}\text{}}
\newcommand{\enexam}{\end{example}\noindent}
\newcommand{\belem}[1][]{%
    \ifthenelse{\equal{#1}{}}{\begin{lemma}}{\begin{lemma}[\textbf{#1}]}} 
\newcommand{\enlem}{\end{lemma}\noindent}

\newcommand{\qd}[1][]{\ifthenelse{\equal{#1}{}}{\quad}{\setcounter{quad_i}{0}\whiledo{\value{quad_i}<#1}{\quad\stepcounter{quad_i}}}}

\newcommand{\nada}{$\text{}$}

\newcommand{\N}{\mathbb N}

\newcommand{\A}{\mathbb A}

\DeclareMathAlphabet{\mathbbold}{U}{bbold}{m}{n}
\newcommand{\GA}[1][\all]{\mathbbold{A}^{#1}}
\newcommand{\GAA}[1]{\mathbbold{A}_{#1}}
\newcommand{\GN}{\mathbbold{N}}

\newcommand{\ypo}{\subseteq}
\newcommand{\ypoo}{\sqsubseteq}
\newcommand{\ypop}[1][]{\trianglelefteq_{#1}}
\newcommand{\Land}{\,\land\,}

\newcommand{\plhn}{\setminus}

\newcommand{\keno}{\varnothing}
\newcommand{\impl}{\mathbin{\Rightarrow}}

\newcommand{\defn}{\,\triangleq\,}

\newcommand{\LL}{\mathcal{L}}
\newcommand{\PP}{\mathcal{P}}
\newcommand{\MM}{\mathcal{M}}
\newcommand{\CC}{\mathcal{C}}
\newcommand{\OO}{\mathcal{O}}
\newcommand{\GG}{\mathcal{G}}
\newcommand{\RR}{\mathcal{R}}
\newcommand{\VV}{\mathcal{V}}
\newcommand{\al}{a}
\newcommand{\bee}{{\bar{b}}}
\newcommand{\gaa}{{\bar{c}}}
\newcommand{\dee}{{\bar{d}}}
\newcommand{\be}{b}
\newcommand{\ga}{c}
\newcommand{\ee}{\epsilon}
\newcommand{\om}{\omega}
\newcommand{\Ga}{\Gamma}
\newcommand{\De}{\Delta}

\newcommand{\zz}{\bar{z}}

\newcommand{\dom}{\mathtt{dom}}

\newcommand{\id}{\mathtt{id}}

\newcommand{\bang}{\mathop{!}}
\newcommand{\GOQ}{{\scriptstyle\color{gray}OQ}}
\newcommand{\GOA}{{\scriptstyle\color{gray}OA}}
\newcommand{\GPA}{{\scriptstyle\color{gray}PA}}
\newcommand{\GPQ}{{\scriptstyle\color{gray}PQ}}

\newcommand{\NA}[1]{\A_{#1}}
\newcommand{\NAA}[1][\bar\al]{\A^{#1}}
\newcommand{\pit}[2]{\tfrac{#1}{#2}}
\newcommand{\pitt}[2]{\frac{#1}{#2}}
\newcommand{\nom}{\mathbf{Nom}}
\newcommand{\ang}[1]{\langle#1\rangle}

\newcommand{\supp}{\mathtt{S}}
\newcommand{\perm}{\normalfont\text{\small PERM}}

\newcommand{\PPfs}{\PP\!\mathsf{_{fs}}}
\newcommand{\PPfn}{\PP\!\mathsf{_{fin}}}
\newcommand{\fin}{{\mathsf{fin}}}

\newcommand{\actn}{\mathop{\nzcirc}}
\newcommand{\actnsz}{\mathop{\nzcirc}}
\newcommand{\pact}[1][]{\pi#1\actn}
\newcommand{\pactsz}[1][]{\pi#1\actnsz}
\newcommand{\sw}[2]{(#1\ #2)}
\newcommand{\swn}[2]{(#1\ #2)\actn}

\newcommand{\swab}{(\al\ \be)\actn}

\newcommand{\frale}[2]{\forall #1\!\in\!#2.\,}

\newcommand{\flabel}[2]{{\text{\scriptsize\txt{$#1$}}}\ar@{.}#2}

\renewcommand{\o}{\!:\!}
\newcommand{\4}{\mathbin{\#}}

\newcommand{\pv}[1]{\mathop{\ulcorner\!#1\!\urcorner}}
\newcommand{\ov}[1]{\oview{#1}}

\newcommand{\ctx}{\mathrm{C}}

\newcommand{\ctxE}{\mathrm{E}}

\newcommand{\nzaa}[1]{\xrightarrow{#1}}
\newcommand{\Rnza}[1]{\xleftarrow{#1}}
\newcommand{\Lnza}[1]{\xrightarrow{#1}}
\newcommand{\lred}[2][]{\xrightarrow[#1]{#2}}

\newcommand{\him}{\mathbin{\text{\raisebox{.3ex}{\scalebox{.5}{$\boldsymbol\relbar\!\!\boldsymbol\relbar\joinrel\mathrel{\boldsymbol\otimes}$}}}}}
\newcommand{\tote}{\rightarrow}

\newcommand{\ltote}[1]{\overset{#1}{\longrightarrow}}
\newcommand{\lrred}[1]{\overset{#1}{\longrightarrow\hspace{-3mm}\rightarrow}}
\renewcommand{\proof}[1][]{\noindent\textsl{Proof#1:} }
\newcommand{\deqsymbol}{$\blacktriangle$}
\newcommand{\myqed}[1][-1]{\ifthenelse{#1<0}{\nada\\[-2\baselineskip]\nada\hfill\makebox[0mm][r]{\rm\qedsymbol}\newlinesminusone{-#1}}%
                                                  {\nada\hfill\makebox[0mm][r]{\rm\qedsymbol}\newlines{#1}}}
\newcommand{\deq}[1][0]{\ifthenelse{#1<0}{\nada\\[-1.5\baselineskip]\nada\hfill\makebox[0mm][r]{\rm\deqsymbol}\newlinesminusone{-#1}}%
                                                  {\nada\hfill\makebox[0mm][r]{\rm\deqsymbol}\newlines{#1}}}
\newcommand{\mindeq}[1][0]{\nada\\[-2.25\baselineskip]\nada\hfill\makebox[0mm][r]{\rm\deqsymbol}\newlines{#1}}

\newcommand{\markedeq}[3][b]{%
  \begin{equation*}
  \makebox[\columnwidth]{%
    \parbox[#1]{.98\columnwidth}{%
      \setlength{\abovedisplayskip}{0pt}%
      \setlength{\belowdisplayskip}{0pt}%
      \noindent
      \begin{equation*}#2\end{equation*}%
    }\hfill
    \makebox[0pt][r]{#3}%
  }%
  \end{equation*}
}

\newcommand{\diagr}{\xymatrix@+5mm}

\newcommand{\myproof}[5][dummy]{%
    \prooftree 
    \ifthenelse{\equal{#2}{}}{#3}{#2\qquad #3}
    \justifies #4
    \thickness=0.05em
    \using \text{\ifthenelse{\equal{#1}{dummy}}{\scriptsize $#5$}{$^{#5}_{#1}$}}
    \endprooftree}
\newcommand{\myproofdoesitwork}[5][dummy]{%
    \prooftree
    {#2\qquad #3}
    \justifies #4
    \thickness=0.05em
    \using \text{\ifthenelse{\equal{#1}{dummy}}{\scriptsize $#5$}{$^{#5}_{#1}$}}
    \endprooftree}
\newcommand{\lproof}[6][]{%
    {\begin{Bflushleft}[b]
    \ifthenelse{\equal{#3}{}}{\textsuperscript{#6}}{\scriptsize #6}
    \end{Bflushleft}}\hspace{4pt}
    \prooftree \vspace{-1mm}
    \ifthenelse{\equal{#2}{}}{#3}{#2\quad #3}
    \justifies #4
    \thickness=0.05em
    \using \raisebox{-1mm}{$^{#5}_{#1}$}
    \endprooftree}
\newcommand{\Myproof}[4]{%
    \prooftree
    \ifthenelse{\equal{#1}{}}{#2}{#1\quad #2}
    \Justifies #3
    \thickness=0.05em
    \using #4
    \endprooftree}
\newcommand{\myprof}[4]{%
    \prooftree
    \ifthenelse{\equal{#1}{}}{#2}{#1\quad #2} \vspace{-1mm}
    \justifies \vspace{-2mm} #3
    \thickness=0.05em
    \using #4
    \endprooftree}

\newenvironment{mylefteqn*}[1][0pt]{\begin{equation*}\hspace{5.8ex}\hspace{#1}}{\hspace{\linewidth minus\linewidth}\end{equation*}}
\renewenvironment{lefteqn}{\begin{equation}}{\end{equation}}
\newenvironment{lefteqn*}{\begin{equation*}}{\end{equation*}}
\newenvironment{leftalign}{\begin{lefteqn*}\begin{aligned}}{\end{aligned}\end{lefteqn*}}
\newenvironment{myaligned}[1][2.5mm]{\setlength{\jot}{#1}\begin{aligned}}{\end{aligned}}
\newenvironment{myalign}[1][2.5mm]{\begin{equation*}\setlength{\jot}{#1}\begin{aligned}[#1]}{\end{aligned}\end{equation*}}

\makeatletter
\def\loverlay#1#2{\mathpalette\@overlay{{#1}{#2}{}{\hfil}}}
\def\overlay#1#2{\mathpalette\@overlay{{#1}{#2}{\hfil}{\hfil}}}
\def\roverlay#1#2{\mathpalette\@overlay{{#1}{#2}{\hfil}{}}}
\def\@overlay#1#2{\@@overlay#1#2}
\def\@@overlay#1#2#3#4#5{{%
    \def\overlaystyle{#1}%
    \setbox0=\hbox{\m@th$\overlaystyle#2$}%
    \setbox1=\hbox{\m@th$\overlaystyle#3$}%
    \ifdim \wd0<\wd1 \setbox2=\box1 \setbox1=\box0 \setbox0=\box2\fi
    \rlap{\hbox to\wd0{#4\box1\relax#5}}\box0}}
\makeatother

\newenvironment{RmEnumerate}[1][xxx]{\begin{list}{}%
    {\setcounter{enumi}{1}\renewcommand\makelabel[1]{\bf\Roman{enumi}.\hfill\stepcounter{enumi}}%
     \setlength\itemsep{2mm} \settowidth\labelwidth{#1} \setlength\leftmargin{\labelwidth+\labelsep}}}{\end{list}}%
\newenvironment{cEnumerate}[2][\bf\arabic{enumi}.]{\begin{list}{}%
    {\setcounter{enumi}{1}\renewcommand\makelabel[1]{#1\stepcounter{enumi}}%
     \settowidth\labelwidth{#2} \setlength\leftmargin{\labelwidth+\labelsep} }}{\end{list}}%
\newenvironment{ccEnumerate}[2][\bf\arabic{enumii}.]{\begin{list}{}%
    {\setcounter{enumii}{1}\renewcommand\makelabel[1]{#1\stepcounter{enumii}}%
     \settowidth\labelwidth{#2} \setlength\leftmargin{\labelwidth+\labelsep} }}{\end{list}}%
\newenvironment{DFNitemize}
    {\begin{list}{}{\renewcommand\makelabel[1]{\;\;$\bullet$##1}%
                    \setlength\labelwidth{3.666ex}%
                    \setlength\leftmargin{\labelwidth+\labelsep}}}%
    {\end{list}}

\newenvironment{Itemize}[1][$\bullet$]
    {\begin{list}{}{\renewcommand\makelabel[1]{#1}%
                    \setlength\labelwidth{5pt}%
                    \setlength\leftmargin{\labelwidth+\labelsep}}}%
    {\end{list}}
\newenvironment{Description}[2][]
    {\begin{list}{}{\ifthenelse{\equal{#1}{}}{\renewcommand\makelabel[1]{\rm\bf##1 \hfill}}{\renewcommand\makelabel[1]{#1 \hfill}}%
                    \setlength\parsep{0pt}%
                    \settowidth\labelwidth{\makelabel{#2}}%
                    \setlength\leftmargin{\labelwidth+\labelsep}}}%
    {\end{list}}
\newenvironment{aDescription}[2][]
    {\begin{list}{}{\ifthenelse{\equal{#1}{}}{\renewcommand\makelabel[1]{\rm##1 \hfill}}{\renewcommand\makelabel[1]{#1 \hfill}}%
                    \settowidth\labelwidth{\makelabel{#2}}%
                    \setlength\leftmargin{\labelwidth+\labelsep}}}%
    {\end{list}}
    {\end{list}}
    {\end{list}}
    {\end{list}}
    {\end{list}}

\newlength{\extraht}%
\newlength{\viewdrop}%
\newcommand{\oview}[1]{
    \settodepth{\viewdrop}{\makebox{$#1$}}
    \addtolength{\viewdrop}{-0.3\extraht}%
    \raisebox{-\viewdrop}{
      \makebox{$\llcorner$}}
      \!#1\!
    \settodepth{\viewdrop}{\makebox{$#1$}}
    \addtolength{\viewdrop}{-0.3\extraht}%
    \raisebox{-\viewdrop}{
      \makebox{$\lrcorner$}}
}
\newcommand{\xinode}[2][\ixi]{\pnode(-0.5pt,-5pt){#2*}\rnode{#2}{#1}\pnode(0.5pt,-5pt){#2**}}
\newcommand{\xicc}[3][5pt]{\ifdim#1>0pt%
    \ncline[linewidth=0.5pt,linecolor=darkgray,nodesepA=1pt]{#2}{#2*}
    \ncline[linewidth=0.5pt,linecolor=darkgray,nodesepA=1pt]{#3}{#3*}
    \ncbar[armA=#1,armB=#1,angle=-90,nodesep=0pt,linewidth=0.5pt,linecolor=darkgray]{#2*}{#3*}%
   \else%
    \ncline[linewidth=0.5pt,linecolor=gray,nodesepA=1pt]{#2}{#2*}
    \ncline[linewidth=0.5pt,linecolor=gray,nodesepA=1pt]{#3}{#3*}
    \ncbar[armA=-#1,armB=-#1,angle=-90,nodesep=0pt,linewidth=0.5pt,linecolor=gray]{#2*}{#3*}\fi}
\newcommand{\xiccl}[3][5pt]{\ifdim#1>0pt%
    \ncline[linewidth=0.5pt,linecolor=darkgray,nodesepA=1pt]{#2}{#2*}
    \ncbar[armA=#1,armB=#1,angle=-90,nodesepA=0pt,nodesepB=3pt,linewidth=0.5pt,linecolor=darkgray]{#2*}{#3}%
   \else%
    \ncline[linewidth=0.5pt,linecolor=gray,nodesepA=1pt]{#2}{#2*}
    \ncbar[armA=-#1,armB=-#1,angle=-90,nodesepA=0pt,nodesepB=3pt,linewidth=0.5pt,linecolor=gray]{#2*}{#3}\fi}
\newcommand{\xiccr}[3][5pt]{\ifdim#1>0pt%
    \ncline[linewidth=0.5pt,linecolor=darkgray,nodesepA=1pt]{#3}{#3*}
    \ncbar[armA=#1,armB=#1,angle=-90,nodesepB=0pt,nodesepA=3pt,linewidth=0.5pt,linecolor=darkgray]{#2}{#3*}%
   \else%
    \ncline[linewidth=0.5pt,linecolor=gray,nodesepA=1pt]{#3}{#3*}
    \ncbar[armA=-#1,armB=-#1,angle=-90,nodesepB=0pt,nodesepA=3pt,linewidth=0.5pt,linecolor=gray]{#2}{#3*}\fi}
\newcommand{\xxicc}[3][5pt]{\ifdim#1>0pt%
    \ncline[linewidth=0.5pt,linecolor=darkgray,nodesepA=1pt]{#2}{#2**}\ncline[linewidth=0.5pt,linecolor=darkgray,nodesepA=1pt]{#3}{#3**}
    \ncbar[armA=#1,armB=#1,angle=-90,nodesep=0pt,linewidth=0.5pt,linecolor=darkgray]{#2**}{#3**}%
   \else%
    \ncline[linewidth=0.5pt,linecolor=gray,nodesepA=1pt]{#2}{#2**}\ncline[linewidth=0.5pt,linecolor=gray,nodesepA=1pt]{#3}{#3**}
    \ncbar[armA=-#1,armB=-#1,angle=-90,nodesep=0pt,linewidth=0.5pt,linecolor=gray]{#2**}{#3**}\fi}
\newcommand{\xxiccl}[3][5pt]{\ifdim#1>0pt%
    \ncline[linewidth=0.5pt,linecolor=darkgray,nodesepA=1pt]{#2}{#2**}
    \ncbar[armA=#1,armB=#1,angle=-90,nodesep=0pt,linewidth=0.5pt,linecolor=darkgray]{#2**}{#3}%
   \else%
    \ncline[linewidth=0.5pt,linecolor=gray,nodesepA=1pt]{#2}{#2**}
    \ncbar[armA=-#1,armB=-#1,angle=-90,nodesep=0pt,linewidth=0.5pt,linecolor=gray]{#2**}{#3}\fi}

\def\doi{5 (3:8) 2009}
\lmcsheading%
{\doi}
{1--69}
{}
{}
{Jan.~\phantom{0}2, 2008}
{Sep.~11, 2009}
{} 

\begin{document}
\title[Full abstraction for nominal general references]%
  {{\ \phantom{x}}\\{\ \phantom{x}}\\ \vskip-6 pt
   Full abstraction for nominal general references}
\author[N.~Tzevelekos]{Nikos Tzevelekos}
\address{Oxford University Computing Laboratory}
\email{nikt@comlab.ox.ac.uk}
\thanks{Research financially supported by the Engineering and Physical Sciences Research Council, the Eugenides Foundation, the A.~G.~Leventis Foundation and Brasenose College.}

%
\begin{abstract}
Game semantics has been used with considerable success in formulating fully abstract semantics for languages with higher-order procedures and a wide range of computational effects. Recently, nominal games have been proposed for modelling functional languages with names. These are ordinary, stateful games cast in the theory of nominal sets developed by Pitts and Gabbay. Here we take nominal games one step further, by developing a fully abstract semantics for a language with nominal general references.
\end{abstract}

\maketitle

\begin{figure}[h]\fbox{\parbox{.95\linewidth}{\small\renewcommand\multicolsep{8pt}\renewcommand\columnseprule{.2pt}
\begin{multicols}{2}
\tableofcontents
\listoffigures
\end{multicols}}\qd}
\end{figure}
\vfill\eject
\section{Introduction}%
\noindent
Functional languages constitute a programming paradigm built around the intuitive notion of a \emph{computational function}, that is, an effectively specified entity assigning values from a codomain to elements of a domain in a \emph{pure manner}: a pure function is not allowed to carry any notion of state or side-effect. This simple notion reveals great computational power if the domains considered are \emph{higher-order}, i.e~sets of functions: with the addition of recursive constructs, higher-order functional computation becomes \emph{Turing complete} (PCF~\cite{Scott:LCF,Plotkin:77:PCF}). In practice, though, functional programming languages usually contain \emph{impure} features that make programming simpler (\emph{computational effects}), like references, exceptions, etc. While not adding necessarily to its computational power, these effects affect the \emph{expressivity} of a language: two functions which seem to accomplish the same task may have different inner-workings which can be detected by use of effects (e.g.~exceptions can distinguish constant functions that do or do not evaluate their inputs). The study of denotational models for effects allows us to better understand their expressive power and to categorise languages with respect to their expressivity.

A computational effect present in most functional programming languages is that of general references.
General references are references which can store not only values of ground type (integers, booleans, etc.) but also of higher-order type (procedures,
higher-order functions) or references themselves. They constitute a very powerful and useful programming construct, allowing us not only the encoding of recursion (see example~\ref{exam:Y}) but also the simulation of a
wide range of computational effects and programming paradigms (e.g.~object-oriented programming~\cite[section~2.3]{Abramsky+:GamesReferences} or
aspect-oriented programming~\cite{Sanjabi_Ong:07}).  The denotational modelling of general references is quite demanding since, on top of phenomena of dynamic update and interference, one has to cope with the inherent cyclicity of higher-order storage. In this paper we provide a fully abstract semantics for a language with general references called the \emph{$\nurho$-calculus}.

The $\nurho$-calculus is a functional language with dynamically allocated general references, reference-equality tests and ``good variables", which faithfully reflects the practice of real programming languages such as ML~\cite{SML}. In particular, it extends the basic nominal language of Pitts and Stark~\cite{Pitts_Stark}, the $\nu$-calculus, by using \boldemph{names} for general references. That is, names in $\nurho$ are atomic entities which can be (cf.~\cite{Pitts_Stark}):
\begin{quote}
        \it created with local scope, updated and dereferenced, tested for equality and passed around via function application, but that is all.
\end{quote}
The fully abstract model of $\nurho$ is the first such for a language with general references and good variables.\footnote{In fact, the $\nurho$-calculus and its fully abstract model were first presented in~\cite{Tze:lics07}, of which the present paper is an extended and updated version.}

Fully abstract models for general references were given via game semantics in~\cite{Abramsky+:GamesReferences} and via abstract categorical semantics (and games) in~\cite{Laird:Cat_Semantics_Store}. Neither approach used names. The model of~\cite{Abramsky+:GamesReferences} is based on the idea of relaxing strategy conditions in order to model computational effects. In particular, it models references as variables of a read/write product type and it uses strategies which violate visibility in order to use values assigned to references previously in a play. The synchronisation of references is managed by \emph{cell strategies} which model fresh-reference creation.
Because references are modelled by products, and in order to produce a fully abstract semantics, the examined language needs to include
\emph{bad variables}, which in turn yield unwanted behaviours affecting severely the expressivity of the language and prohibit the use of equality tests for
references.\footnote{By ``bad variables" we mean read/write constructs of reference type which are not references. They are necessary for obtaining definability and full-abstraction in~\cite{Abramsky+:GamesReferences} since read/write-product semantical objects may not necessarily denote references.}
On the other hand, the approach in~\cite{Laird:Cat_Semantics_Store} bypasses the bad-variables problem by not including types for references (variables and references of the same type coincide). This contributes new intuitions on sequential categorical behaviour (\emph{sequoidal category}), but we think that it is somehow distanced from the common notion of reference in functional programming.

The full-abstraction problem has also been tackled via trace semantics in~\cite{Laird:Icalp07}. The language examined is a version of that
in~\cite{Abramsky+:GamesReferences} without bad variables. The latter are not needed since the modelling of references is achieved by names pointing to a store (which is analogous to our approach).
Of relevance is also the fully abstract trace model for a language with nominal threads and nominal objects presented
in~\cite{JeffreyRathke:lics02:ConcObj}. An important difference between trace models and game models is that the former are defined operationally (i.e.~traces are computed by using the operational semantics), whereas game models are defined in a purely compositional manner. Nonetheless, trace models and game models have many similarities, deriving mainly from their sequential-interactive representation of computation, and in particular there are connections between~\cite{Laird:Icalp07} and the work herein that should be further examined.

\paragraph{\it The approach} We model nominal computation in \boldemph{nominal games}.
These were introduced independently in~\cite{AGMOS,Laird:fossacs04} for producing fully abstract models of the $\nu$-calculus and its extension with pointers respectively. Here we follow the formulation of~\cite{AGMOS} with rectifications pertaining to the issue of \emph{unordered state} (see remark~\ref{rem:SS}).\footnote{The nominal games of~\cite{AGMOS} use moves attached with finite sets of names. It turns out, however, that this yields discrepancies, as unordered name-creation is incompatible with the deterministic behaviour of strategies and, in fact, nominal games in~\cite{AGMOS} do not form a category. Here (and also in~\cite{Tze:lics07}), we
recast nominal games using moves attached with name-lists instead of name-sets. This
allows us to restrict our attention to \emph{strong nominal sets} (v.~definition~\ref{d:Ssupp}), a restriction necessary for overcoming the complications with determinacy.}
Thus, our nominal games constitute a stateful (cf.~Ong~\cite{Ong:lics02}) version of Honda-Yoshida call-by-value games~\cite{Honda:CBV} built inside the universe of nominal sets of Gabbay and Pitts~\cite{Gabbay_Pitts:FM02,Pitts_Nominal_Logic}. 

A particularly elegant approach to the modelling of names is by use of
\boldemph{nominal sets}~\cite{Gabbay_Pitts:FM02,Pitts_Nominal_Logic}. These are sets whose elements involve a finite number of \boldemph{atoms}, and which can be acted upon by finite atom-permutations. The expressivity thus obtained is remarkable: in the realm (the category) of nominal sets, notions like \emph{atom-permutation}, \emph{atom-freshness} and \emph{atom-abstraction} are built inside the underlying structure. We therefore use nominal sets, with atoms playing the role of names, as a \textbf{general foundation for reasoning about names}.

The essential feature of nominal games is the appearance of names \emph{explicitly} in plays as constants (i.e.~as atoms), which allows us to directly model names and express name-related notions (name-equality, name-privacy, scope-extrusion, etc.) in the games setting. Thus nominal games can capture the essential features of nominal computation and, in particular, they model the $\nu$-calculus. From that model we can move to a model of $\nurho$ by an appropriate \emph{effect-encapsulation} procedure, that is, by use of a \emph{store-monad}. A fully abstract model is then achieved by enforcing appropriate store-discipline conditions on the games.

The paper is structured as follows. In section~2 we briefly present nominal sets and some of their basic properties. We finally introduce \emph{strong nominal sets}, that is, nominal sets with ``ordered involvement" of names, and prove the \emph{strong support lemma}. In section~3 we introduce the $\nurho$-calculus and its operational semantics. We then introduce the notion of a \emph{$\laNR$-model}, which provides abstract categorical conditions for modelling $\nurho$ in a setting involving \emph{local-state comonads} and a \emph{store-monad}. We finally show definability and, by use of a quotienting procedure, full-abstraction in a special class of $\laNR$-models. In section~4 we introduce nominal games and show a series of results with the aim of constructing a category $\Vt$ of \emph{total, innocent} nominal strategies. In the end of the section we attempt a comparison with the nominal games presented by Laird in~\cite{Laird:fossacs04,Laird:Names_Pointers}. In section~5 we proceed to construct a specific fully abstract $\laNR$-model in the category $\Vt$. The basic ingredients for such a construction have already been obtained in the previous section, except for the construction of the store-monad, which involves solving a recursive domain equation in $\Vt$. Once this has been achieved and the $\laNR$-model has been obtained, we further restrict legal strategies to \emph{tidy} ones, i.e.~to those that obey a specific store-related discipline; for these strategies we show definability and full-abstraction. We conclude in section~6 with some further directions.

The contributions of this paper are: a) the identification of strong nominal sets as the adequate setting for nominal language semantics;
b) the abstract categorical presentation in a monadic-comonadic setting of models of a language with nominal general references; c) the rectification of nominal games of~\cite{AGMOS} and their use in constructing a specific such model; d) the introduction of a game-discipline (\emph{tidiness}) to capture computation with names-as-references, leading to a definable and hence fully abstract game model.

\section{Theory of nominal sets}
\noindent
We give a short overview of nominal sets, which form the basis of all constructions presented in this paper; our presentation generally follows~\cite{Pitts_Nominal_Logic}. Nominal sets are an inspiring paradigm of
the universality (and reusability) of good mathematics: invented in the 1920's and 1930's by Fraenkel and Mostowski as a model of set theory with atoms (ZFA) for showing its independence from the Axiom of Choice, they were reused in the late 1990's by Gabbay and Pitts~\cite{Gabbay_Pitts:FM02} as the foundation of a general theory of syntax with binding constructs. The central notion of nominal sets is that of \emph{atoms}, which are to be seen as basic `particles' present in elements of nominal sets, and of atom-permutations which can act upon those elements. Moreover, there is an infinite supply of atoms, yet each element of a nominal set `involves' finitely many of them, that is, it has \emph{finite support} with regard to atom-permutations.


We will be expressing the intuitive notion of names by use of atoms, both in the abstract syntax of the language and in its denotational semantics.
Perhaps it is not clear to the reader why nominal sets should be used\HY couldn't we simply model names by natural numbers? Indeed, numerals could
be used for such semantical purposes (see e.g.~\cite{Laird:Names_Pointers}), but they would constitute an overspecification: numerals carry a linear
order and a bottom element, which would need to be carefully nullified in the semantical definitions. Nominal sets factor out this burden by providing
the minimal solution to specifying names; in this sense, nominal sets are \emph{the intended model} for names.

\subsection{Nominal sets}

Let us fix a countably infinite family
$(\A_i)_{i\in\om}$
of pairwise disjoint, countably infinite sets of \boldemph{atoms}, and let us denote by $\perm(\A_i)$ the group of finite permutations of $\A_i$. Atoms are denoted by $\al,\be,\ga$ and variants; permutations are denoted by $\pi$ and variants; $\id$ is the identity permutation and $(\al\ \be)$ is the permutation swapping $\al$ and $\be$ (and fixing all other atoms). We write $\A$ for the union of all the $\A_i$'s. We take
\begin{eqn}
    \perm(\A)\defn\bigoplus_{i\in I}\perm(\A_i)
\end{eqn}%
to be the direct sum of the groups $\perm(\A_i)$, so $\perm(\A)$ is a group of finite permutations of $\A$ which act separately on each constituent $\A_i$. In particular, each $\pi\in\perm(\A)$ is an $\omega$-indexed list of permutations, $\pi\in\prod_{i\in\om}\perm(\A_i)$, such that $(\pi)_i\neq\id_{\A_i}$ holds for finitely many indices $i$. In fact, we will write (non-uniquely) each permutation $\pi$ as a finite composition
\[ \pi = \pi_1\circ\cdots\circ\pi_n \]
such that each $\pi_i$ belongs to some $\perm(\A_{j_i})$\HY note that $j_i$'s need not be distinct.



\bdefn\label{d:2:NomSet} A \boldemph{nominal set} $X$ is a set $|X|$ (usually denoted $X$) equipped with an action of $\perm(\A)$, that is, a
function\, $\uscore\actn\uscore\ :\perm(\A)\tims X\tote X$\, such that, for any $\pi,\pi'\in\perm(\A)$ and $x\in X$,
\[ \pi\actn(\pi'\actn x)=(\pi\circ\pi')\actn x\,, \qd \id\actn x = x\,. \]
Moreover, for any $x\in X$ there exists a finite set $S$ such that, for all permutations $\pi$,
\[ (\frale{\al}{S}\pi(\al)=\al)\implies \pact x=x\,. \]
\mindeq[-1]\edefn[1]%
For example, $\A$ with the action of permutations being simply permutation-application is a nominal set. Moreover, any set can be trivially rendered into a nominal set of elements with empty support.

Finite support is closed under intersection and hence there is a least finite support for each element $x$ of a nominal set; this we call
\boldemph{the support of $x$} and denote by $\supp(x)$.

\bepropdefn[\cite{Gabbay_Pitts:FM02}] Let $X$ be a nominal set and
$x\in X$. For any finite $S\ypo\A$, $S$ supports $x$ iff \
$\frale{\al,\al'}{(\A\plhn S)}\swn{\al}{\al'}x=x$\,.

Moreover, if finite $S,S'\ypo\A$ support $x$ then $S\cap S'$ also supports $x$. Hence, we can define
\begin{align*}
    \supp(x) &\defn\bigcap\{\,S\ypo_\fin\A\,|\,S\text{ supports }x \,\}\,,
\intertext{which can be expressed also as:}
   \supp(x)&=\{\,\al\in\A\,|\,\text{for infinitely many $\be$. }\swab x\neq x\,\}\,.
\end{align*}%
\myqed\enpropdefn%
For example, for each $\al\in\A$, $\supp(\al)=\{\al\}$.
We say that $\al$ \emph{\bfseries is fresh for} $x$, written $\al\4x$, if $\al\notin\supp(x)$.
$x$ is called \boldemph{equivariant} if it has empty support. It follows from the definition that
\begin{equation}
    \al\4x \iff \text{for cofinitely many $\be$. }\swab x=x\,.
\end{equation}
There are several ways to obtain new nominal sets from given nominal sets $X$ and $Y$:
\begin{DFNitemize}
    \item The disjoint union $X\uplus Y$ with permutation-action inherited from $X$ and $Y$ is a nominal set. This extends to infinite disjoint unions.
    \item The cartesian product $X\tims Y$ with permutations acting componentwise is a nominal set; if $(x,y)\in X\tims Y$ then
    $\supp(x,y)=\supp(x)\cup\supp(y)$.
    \item The fs-powerset $\PPfs(X)$, that is, the set of subsets of $X$ which have finite support, with permutations acting on subsets of $X$ elementwise. In particular, $X'\ypo X$ is a \boldemph{nominal subset} of $X$ if it has empty support, i.e.~if for all $x\in X'$ and permutation $\pi$, $\pact x\in X'$.
\end{DFNitemize}
Apart from $\A$, some standard nominal sets are the following.
\begin{DFNitemize}
    \item Using products and infinite unions we obtain the nominal set
        \begin{eqn}
            \NAA[\#]\defn\bigcup_n\{\,a_1\dots a_n\,|\,\forall i,j\in1..n.\ \ a_i\in\NAA[]\Land (j\neq i\implies a_j\neq a_i)\,\}\,,
        \end{eqn}%
        that is, the set of \boldemph{finite lists of distinct atoms}. Such lists we denote by $\all,\,\bee,\,\gaa$ and variants.
    \item The fs-powerset $\PPfs(\A)$ is the set of finite and cofinite sets of atoms, and has $\PPfn(\A)$ as a
          nominal subset (the set of finite sets of atoms).
\end{DFNitemize}
For $X$ and $Y$ nominal sets, a relation $\RR\ypo X\tims Y$ is a \boldemph{nominal relation} if it is a nominal subset of $X\tims Y$. Concretely, $\RR$ is a nominal relation iff, for any permutation $\pi$ and $(x,y)\in X\tims Y$,
\[ x \RR y\iff (\pact x)\RR(\pact y)\,. \]
For example, it is easy to show that $\4\ypo\A\tims X$ is a nominal relation. Extending this reasoning to functions we obtain the notion of \boldemph{nominal functions}.
\bdefn[The category $\nom$] We let $\nom$ be the category of nominal sets and nominal functions, where a function $f:X\paei Y$ between nominal sets is nominal
if $f(\pact x)=\pact f(x)$ for any $\pi\in\perm(\A)$ and $x\in X$. \edefn For example, the support function, $\supp(\_):X\paei \PPfn(\A)$\,, is a
nominal function since
\[ \supp(\pact x)=\pact\supp(x)\,. \]
$\nom$ inherits rich structure from $\textbf{Set}$ and is in particular a topos. More importantly, it contains atom-abstraction mechanisms; we will concentrate on the following.

\bdefn[Nominal abstraction]\label{NS:d:NomAbstr} Let $X$ be a nominal set and $x\in X$. For any finite $S\ypo\A$, we can \emph{abstract $x$ to $S$}, by forming
\[    [x]_S \defn\{\,y\in X\,|\,\exists\pi.\ (\forall\al\in S\cap\supp(x).\ \pi(a)=a)\land y=\pi\actn x\,\}\,. \]
\deq[-1] \edefn[q]%
The abstraction restricts the support of $x$ to $S\cap\supp(x)$ by appropriate orbiting of $x$ (note that $[x]_S\in\PPfs(X)$). In particular, we can show the following.%
\belem[\cite{Tze_PhD}] For any $x\in X$, $S\ypo_\fin\A$ and $\pi\in\perm(\A)$,
\[ \pact[][x]_S=[\pact x]_{\pactsz S}\ \land \ \supp([x]_S)=\supp(x) \cap S\,. \]
\myqed\enlem%
Two particular subcases of nominal abstraction are of interest. Firstly, in case $S\ypo\supp(x)$ the abstraction becomes
\begin{lefteqn*}{}\tag{$*$}
    [x]_S = \{\,y\in X\,|\,\exists\pi.\ (\forall a\in S.\ \pi(a)=a)\land y=\pact x\,\}\,.
\end{lefteqn*}%
This is the mechanism used in~\cite{Tze:lics07}. Note that if $S\nsubseteq\supp(x)\land\supp(x)\nsubseteq S$ then ($*$) does not yield
$\supp([x]_S)=S\cap\supp(x)$.
The other case is the simplest possible, that is, of $S$ being empty; it turns out that this last constructor is all we need from nominal abstractions in this paper. We define:
\begin{lefteqn}
    [x] \defn \{\,y\in X\,|\,\exists\pi.\, y=\pact x\,\}\,.
\end{lefteqn}%

\subsection{Strong support}

Modelling local state in sets of atoms yields a notion of \emph{unordered state}, which is inadequate for our intended semantics. Nominal game semantics is defined by means of nominal strategies for games that model computation. These strategies, however, are deterministic up to choice of fresh names, a feature which is in direct conflict to unordered state. For example, in unordered state the consecutive creation of two atoms $\al,\be$ is modelled by
adding the set $\{\al,\be\}$ to the local state; on the other hand, by allowing strategies to play such moves we lose determinism in
strategies.\footnote{The problematic behaviour of nominal games in weak support is discussed again in remark~\ref{rem:SS}.}

Ordered state is therefore more appropriate for our semantical purposes and so we restrict our attention to nominal sets with \emph{ordered presence} of atoms in their elements. This notion is described as \emph{strong support}.\footnote{An even stricter notion of support is \emph{linear support}, introduced in~\cite{Tze_Murawski:RML}: a nominal set $X$ is called \emph{linear} if for each $x\in X$ there is a linear order $<_x$ of $\supp(x)$ such that $\al<_x\be \implies \pi(\al)<_{\pact{x}}\pi(\be)$.}
\bdefn\label{d:Ssupp} For any nominal set $X$, any $x\in X$ and any $S\ypo\A$, $S$ \boldemph{strongly supports} $x$ if, for any permutation $\pi$,
\[    (\frale{\al}{S}\pi(\al)=\al)\iff\pact x=x\,. \]
We say that $X$ is a \boldemph{strong nominal set} if it is a nominal set with all its elements having strong support.
\edefn%
Compare the last assertion above with that of definition~\ref{d:2:NomSet}, which employs only the left-to-right implication. In fact, strong support coincides with weak support when the former exists.
\beprop%
If $X$ is a nominal set and $x\in X$ has strong support $S$ then $S=\supp(x)$.%
\enprop%
\proof By definition, $S$ supports $x$, so $\supp(x)\ypo
S$. Now suppose there exists $\al\in S\plhn\supp(x)$. For any fresh $\be$, $\sw{\al}{\be}$ fixes $\supp(x)$ but not $S$, so it doesn't fix $x$,
\contra.\qed\noindent%
Thus, for example, the set
$\{\al,\be\}\ypo\A_i$ of the previous paragraph does not have strong support, since the permutation $(a\ b)$ does not fix the atoms in its support (the set $\{a,b\}$) but still $\swab\{a,b\}=\{a,b\}$. On the other hand, $\{a,b\}$ strongly supports the list $ab$. In fact, all lists of (distinct) atoms have strong support and therefore $\NAA[\#]$ is a strong nominal set (but $\PPfn(\A)$ is not).

The main reason for introducing strong nominal sets is the following result, which is a specialised version of the Strong Support Lemma of~\cite{Tze_PhD} (with $S=\keno$).

\belem[Strong Support Lemma]\label{l:strongsupp}%
Let $X$ be a strong nominal set and let\linebreak $x_1,x_2,y_1,y_2,z_1,z_2\in X$. Suppose also that $\supp(y_i)\cap\supp(z_i)\ypo\supp(x_i)$\,, for $i=1,2$, and that there exist $\pi_y,\pi_z$ such that
\[
\pi_y\actn x_1=\pi_z\actn x_1=x_2\,,\quad \pi_y\actn y_1=y_2\,,\quad \pi_z\actn z_1=z_2\,.
\]
Then, there exists some $\pi$ such that $\pi\actn x_1=x_2$\,, $\pi\actn y_1=y_2$ and $\pi\actn z_1=z_2$.
\enlem
\proof
Let
$\Delta_i\defn\supp(z_i)\plhn\supp(x_i)$\,, $i=1,2$\,, so $\Delta_2=\pact[_z]\Delta_1$, and let $\pi'\defn\pi_y^{-1}\circ\pi_z$. By assumption,
$\pact[']x_1=x_1$, and therefore by strong support $\pi'(\al)=\al$ for all $\al\in\supp(x_1)$. Take any $\be\in\Delta_1$. Then $\pi'(\be)\4\pact[']x_1=x_1$ and
$\pi_z(\be)\in\pact[_z]\Delta_1=\Delta_2$, $\ara\pi_z(\be)\4y_2$, $\ara\pi'(\be)\4\pact[_y^{-1}]y_2=y_1$. Hence,
\[ b\in\De_1\implies \be,\pi'(\be)\4x_1,y_1\,. \]
Now assume $\Delta_1=\{\be_1,...,\be_N\}$ and define $\pi_1,...,\pi_N$ by recursion:
\[ \pi_0\defn\id \,,\quad \pi_{i+1}\defn\sw{\be_{i+1}}{\pact[_i]\pact[']\be_{i+1}}\circ\pi_i\,. \]
We claim that, for each $0\leq i\leq N$ and $1\leq j\leq i$, we have
\[ \pact[_i]\pact[']\be_j=\be_j \,,\quad \pact[_i]x_1=x_1 \,,\quad \pact[_i]y_1=y_1\,. \]
We do induction on $i$; the case of $i=0$ is trivial. For the inductive step, if $\pact[_i]\pact[']\be_{i+1}=\be_{i+1}$ then $\pi_{i+1}=\pi_i$, and
$\pact[_{i+1}]\pact[']\be_{i+1}=\pact[_{i}]\pact[']\be_{i+1}=\be_{i+1}$. Moreover, by IH, $\pact[_{i+1}]\pact[']\be_{j}=\be_j$ for all $1\leq j\leq i$,
and $\pact[_{i+1}]x_1=x_1$ and $\pact[_{i+1}]y_1=y_1$. If $\pact[_i]\pact[']\be_{i+1}=\be'_{i+1}\neq\be_{i+1}$ then, by construction,
$\pact[_{i+1}]\pact[']\be_{i+1}=\be_{i+1}$. Moreover, for each $1\leq j\leq i$, by IH,
$\pact[_{i+1}]\pact[']\be_{j}=\swn{\be_{i+1}}{\be'_{i+1}}\be_{j}$, and the latter equals $\be_j$ since $\be_{i+1}\neq\be_j$ implies
$\be'_{i+1}\neq\pact[_i]\pact[']\be_{j}=\be_j$. Finally, for any $\al\in\supp(x_1)\cup\supp(y_1)$,
$\pact[_{i+1}]\al=\swn{\be_{i+1}}{\be'_{i+1}}\pact[_i]\al=\swn{\be_{i+1}}{\be'_{i+1}}\al$, by IH, with $\al\neq\be_{i+1}$. But the latter equals $\al$
since $\pi'(\be_{i+1})\neq\al$ implies that $\be'_{i+1}\neq\pact[_i]\al=\al$, as required.

Hence, for each $1\leq j\leq N$,
\[ \pact[_N]\pact[']\be_j=\be_j \,,\quad \pact[_N]x_1=x_1 \,,\quad \pact[_N]y_1=y_1\,. \]
Moreover, $\pact[_N]\pact[']z_1=z_1$, as we also have
\[ \be\in\supp(z_1)\cap\supp(x_1)\implies \pact[_N]\pact[']\be=\pact[_N]\be=\be \]
(again by strong support). Thus, considering $\pi\defn\pi_y\circ\pi_N^{-1}$ we have:
\begin{gather*}
\pact[_y]\pact[_N^{-1}]x_1=\pact[_y]x_1=x_2 \,,\quad \pact[_y]\pact[_N^{-1}]y_1=\pact[_y]y_1=y_2\,, \\
\pact[_y]\pact[_N^{-1}]z_1=\pact[_y]\pact[_N^{-1}]\pact[_N]\pact[']z_1=\pact[_y]\pact[']z_1=\pact[_y]\pact[_y^{-1}]\pact[_z]z_1=z_2\,,
\end{gather*}
as required.
\qed\noindent%
A more enlightening formulation of the lemma can be given in terms of abstractions, as in the following table.
In the context of nominal games later on, the strong support lemma
will guarantee us that composition of abstractions of plays can be reduced to composition of plays.

\[\fbox{\parbox{.85\linewidth}{\it%
{\rm\bf Strong Support Lemma.}\\
Let $X$ be a strong nominal set and $x_1,x_2,y_1,y_2,z_1,z_2\in X$. Suppose also that $\supp(y_i)\cap\supp(z_i)\ypo\supp(x_i)$\,, for
$i=1,2$, and moreover that
\[ [x_1,y_1]=[x_2,y_2] \,,\quad [x_1,z_1]=[x_2,z_2]\,.\qd[6] \]
Then, $[x_1,y_1,z_1]=[x_2,y_2,z_2]$.}}
\]

\section{The language}
\noindent
The language we examine, the $\nurho$-calculus, is a call-by-value $\lambda$-calculus with nominal general references. It constitutes an extension of the $\nu$-calculus~\cite{Pitts_Stark} and Reduced ML~\cite[chapter 5]{Stark:PhD} in which names are used for general references. It is essentially the same calculus of~\cite{Laird:Icalp07}, that is, the $\mathsf{mkvar}$-free fragment of the language of~\cite{Abramsky+:GamesReferences} extended with reference-equality tests and names.

\subsection{Definitions}

The syntax of the language is built inside $\nom$. In particular, we assume there is a set of names (atoms) $\A_A\in(\A_i)_{i\in\omega}$ for each type $A$ in the language. Types include types for commands, naturals and references, product types and arrow types.

\bdefn
The $\nurho$-calculus is a typed functional language of nominal references. Its types, terms and values are given as follows.
\begin{leftalign}
    \text{TY}\ni A,B ::= &\ \ena\ |\ \N\ |\ [A]\ |\ A\tote B\ |\ A\times B \\
\\[-4mm]
\text{TE}\ni M,N ::= &\  \ x\ |\  \la x.M\ |\  M\,N\  \ang{M,N}\ |\  \fst M\ |\ \snd N     && \text{$\lambda$-calculus}\\
& |\ n\ |\  \pred M\ |\ \succ N   && \text{arithmetic}\\
& |\ \sskip\ |\  \ifo{M}{N_1}{N_2}  && \text{return / if\_then\_else}\\
& |\ \al  && \text{reference to type $A$ ($\al\in\NA{A}$)}\\
& |\  [M=N] && \text{name-equality test}\\
& |\  \new[\al]M  && \text{$\nu$-abstraction}\\
& |\  M:= N  && \text{update}\\
& |\  \bang M   && \text{dereferencing} \\
\\[-4mm]
 \text{VA}\ni V,W ::= &\ \ n\ |\ \sskip\ |\ \al\ |\ x\ |\ \la x.M\ |\ \ang{V,W}
\end{leftalign}%
The typing system involves terms in environments $\all\NZvbar\Ga$, where $\all$ a list of (distinct) names and $\Ga$ a finite set of variable-type pairs. Typing rules are given in figure~\ref{f:Typing}.
\edefn%
\begin{figure}
\fbox{\small\parbox{.95\linewidth}{\[\setlength{\extrarowheight}{8mm}\begin{array}{c c}
\\[-20mm]
\myproof{}{}{\AGseq{n:\N}}{}\qquad \myproof{}{}{\GAseq[x\o A]{x:A}}{} &%
\myproof{}{}{\AGseq{\sskip:\ena}}{}
\\
\myproof{}{\GAseq{M:A\times B}}{\GAseq{\fst{M}:A}}{}\qquad \myproof{}{\GAseq{M:A\times B}}{\GAseq{\snd{M}:B}}{} &
\myproof{\GAseq{M:A}}{\GAseq{N:B}}{\GAseq{\ang{M,N}:A\times B}}{}
\\
\myproof{}{\GAseq{M:\N}}{\GAseq{\pred{M}:\N}}{}\qquad \myproof{}{\GAseq{M:\N}}{\GAseq{\succ{M}:\N}}{} &%
\myproof{\GAseq{M:\N}}{\GAseq{N_i:A}\qd{\scriptstyle(i=1,2)}}{\GAseq{\ifo{M}{N_1}{N_2}:A}}{}
\\
\myproof{}{\GAseq[x\o A]{M:B}}{\GAseq{\la x.M:A\tote B}}{} & \myproof{\GAseq{M:A\tote B}}{\GAseq{N:A}}{\GAseq{M\,N:B}}{}
\\
\myproof[\land\al\,\in\,\all]{}{}{\AGseq{\al:[A]}}{\al\,\in\,\NA{A}}\qquad \myproof{}{\AGseq[\al]{M:B}}{\GAseq{\new[\al]M:B}}{} &
\myproof{\GAseq{M:[A]}}{\GAseq{N:[A]}}{\GAseq{[M=N]:\N}}{}
\\
\myproof{\GAseq{M:[A]}}{\GAseq{N:A}}{\GAseq{M:=N:\ena}}{} & \myproof{}{\GAseq{M:[A]}}{\GAseq{\bang M:A}}{}
\end{array}
\]}}\caption{Typing rules.}\label{f:Typing}
\end{figure}%
The $\nu$-constructor is a \boldemph{name-binder}: an occurrence of a name $\al$ inside a term $M$ is \emph{bound} if it is in the scope of some $\nu\al\,$. We follow the standard convention of equating terms up to $\alpha$-equivalence, the latter defined with respect to both variable- and name-binding.

Note that TE and VA are strong nominal sets: each name $\al$ of type $A$ is taken from $\NA{A}$ and all terms contain
finitely many atoms\HY be they free or bound\HY which form their support. Note also the notion of \emph{ordered state} that is imposed by use of name-lists (instead of name-sets) in type-environments.
In fact, we could have used unordered state at the level of syntax (and operational semantics) of $\nurho$, and ordered state at the level of denotational semantics. This already happens with contexts: a context $\Gamma$ is a set of premises, but $\trn{\Gamma}$ is an (ordered) product of type-translations. Nevertheless, we think that ordered state does not add much complication while it saves us from some informality.

The operational semantics of the calculus involves computation in some \emph{store environment} where created names have their values stored. Formally, we define store environments $S$ to be lists of the form:
\begin{equation}
    S::= \ee\ |\ \al,S\ |\ \al::V,S\,.
\end{equation}%
Observe that the store may include names that have been created but remain as yet unassigned a value. For each store environment $S$ we define its
domain to be the name-list given by:
\begin{equation}
    \dom(\ee)\defn\ee\,,\qd \dom(\al,S)\defn\al,\dom(S)\,,\qd \dom(\al::V,S)\defn\al,\dom(S)\,.
\end{equation}
We only consider environments whose domains are lists of distinct names. We write \hbox{$S\modl_{\Ga,A}M$}, or simply $S\modl M$, only if $\Gseq{\dom(S)}{M:A}$ is valid (i.e.,~derivable).
\bdefn%
The operational semantics is given in terms of a small-step reduction, the rules of which are given in figure~\ref{f:Reduction}.
Evaluation contexts $\ctxE[\uscore]$ are of the form:
\begin{gather*}
[\uscore=N]\ ,\ [\al=\uscore]\ ,\ !\uscore\ ,\ \uscore:=N\ ,\ \al:=\uscore\ ,\ \ifo{\uscore}{N_1}{N_2}\ ,\\
(\la x.N)\,\uscore\ ,\ \uscore\, N\ ,\ \fst\uscore\ ,\ \snd\uscore\ ,\ \pred\uscore\ ,\ \succ\uscore\ ,\ \ang{\uscore,N}\ ,\ \ang{V,\uscore}
\end{gather*}
\mindeq[-1]\edefn[1]%
\begin{figure}[b]
\fbox{\small\parbox{.95\linewidth}{\[
\begin{array}{c@{\qquad} c}
\lproof{}{}{S\modl\new[\al]M\longrightarrow S,\al\modl M}{\al\4S}{NEW} & \lproof{}{}{S\modl\succ n \longrightarrow S\modl n\mathord{+}1}{}{SUC}
\\\\
\lproof[n=1\text{ if }\al\neq\be]{}{}{S\modl [\al=\be] \longrightarrow S\modl n}{n=0\text{ if }\al=\be}{EQ} &
\lproof{}{}{S\modl\pred(n\mathord{+}1) \longrightarrow S\modl n}{}{PRD}
\\\\
\lproof[j=2\text{ if }n>0]{}{}{S\modl\ifo{n}{N_1}{N_2} \longrightarrow S\modl N_j}{j=1\text{ if }n=0}{IF0} &
\lproof{}{}{S\modl\pred0\longrightarrow S\modl0}{}{PRD}
\\\\
\lproof{}{}{S,\al(::W),S'\modl \al:=V \longrightarrow S,\al::V,S'\modl \sskip}{}{UPD}
& \lproof{}{}{S\modl\fst\ang{V,W} \longrightarrow S\modl V}{}{FST}
\\\\
\lproof{}{}{S,\al::V,S'\modl \bang \al \longrightarrow S,\al::V,S'\modl V}{}{DRF} & \lproof{}{}{S\modl\snd\ang{V,W} \longrightarrow S\modl W}{}{SND}
\\\\
\lproof{}{}{S\modl(\la x.M)\,V \longrightarrow S\modl M\{V/x\}}{}{LAM} & \lproof{}{S\modl M\longrightarrow S'\modl M'}{S\modl \ctxE[M]\longrightarrow S'\modl \ctxE[M']}{}{CTX}
\end{array}
\]}}\caption{Reduction rules.}\label{f:Reduction}
\end{figure}%
We can see that $\nurho$ is not strongly normalising with the following example. Recall the standard CBV encoding of sequencing:
\begin{equation}
  M\1N \defn (\la z.N)M
\end{equation}
with $z$ not free in $N$.

\beexam\label{exam:stop}%
For each type $A$, take
\[ \sstop_A\defn\new[\be](\be:=\la x.(\bang\be)\sskip)\1(\bang\be)\sskip\]
with $\be\in\NA{\ena\tote A}$. We can see that $\sstop_A$ diverges, since:
\begin{align*}
  \modl\sstop_A &\lrred{} \be::\la x.(\bang\be)\sskip\modl(\bang\be)\sskip
    \ltote{} \be::\la x.(\bang\be)\sskip\modl(\la x.(\bang\be)\sskip)\sskip \\
    &\ltote{} \be::\la x.(\bang\be)\sskip\modl(\bang\be)\sskip\,.
\end{align*}\myqed[-1]
\enexam%
%
%
%
%
The great expressive power of general references is seen in the fact that we can encode the \textbf{Y} combinator. The following example is adapted from~\cite{Abramsky+:GamesReferences}.

\beexam\label{exam:Y} Taking $\al\in\NA{A\tote A}$, define:
\[
\textbf{Y}\!_A\defn\la f.\new[\al](\al:=\la x.f(\bang\al)x)\1\bang\al\,.
\]
$\textbf{Y}\!_A$ has type $((A\tote A)\tote A\tote A)\tote A\tote A$ and, for any relevant term $M$ and value $V$, we
have
\begin{align*}
\modl (\textbf{Y}\!_A(\la y.M))V &\lrred{}\ \al::\la x.(\la y.M)(\bang\al)x\modl (\bang\al)V \\
    &\ltote{}\ \al::\la x.(\la y.M)(\bang\al)x\modl (\la x.(\la y.M)(\bang\al)x)V \\
    &\ltote{}\ \al::\la x.(\la y.M)(\bang\al)x\modl (\la y.M)(\bang\al)V\,,
\end{align*}
and also \ $\modl (\la y.M)(\textbf{Y}\!_A(\la y.M))V \lrred{}\ \al::\la x.(\la y.M)(\bang\al)x\modl (\la y.M)(\bang\al)V$\,.
\\
For example, setting
\begin{align*}
  \mathsf{addrec}_\mathtt{x} &\defn \la x.\ \ifo{\snd x}{x}{\mathtt{x}\ang{\succ\fst x,\pred\snd x}}\,, \\
  \mathsf{add} &\defn \textbf{Y}(\la h.\mathsf{addrec}_h)\,, \\
  S &\defn \al:: \la x. (\la h.\mathsf{addrec}_h)(\bang\al)x\,,
\end{align*}
where $\mathtt{x}$ is a metavariable of relevant type, we have that, for any $n,m\in\N$,
\begin{align*}
  \modl\mathsf{add}\ang{n,m} &\lrred{}\ S\modl (\la h.\mathsf{addrec}_h)(\bang\al)\ang{n,m} \lrred{}\ S\modl\mathsf{addrec}_{S(\al)}\ang{n,m}\\
    &\lrred{}\ S\modl\ifo{m}{\ang{n,m}}{S(\al)\ang{\succ\fst\ang{n,m},\pred\snd\ang{n,m}}} \\
    &\lrred{}\ S\modl S(a)\ang{n\mathord{+}1,m\mathord{-}1} \ltote{}\ S\modl (\la h.\mathsf{addrec}_h)(\bang\al)\ang{n\mathord{+}1,m\mathord{-}1} \\
\cdots & \lrred{}\ S\modl (\la h.\mathsf{addrec}_h)(\bang\al)\ang{n\mathord{+}m,0}\lrred{}\ S\modl\ang{n\mathord{+}m,0}\,.
\end{align*}
\myqed\enexam%
The notions of \boldemph{observational approximation} and \boldemph{observational equivalence} are built around the \emph{observable type} $\N$.
Two terms are equivalent if, whenever they are put inside a variable- and name-closing context of resulting type $\N$, called a \boldemph{program context}, they reduce to the same natural number.
The formal definition follows; note that we usually omit $\all$ and $\Ga$ and write simply $M\ypob N$.
\bdefn
For typed terms $\GAseq{M:A}$ and $\GAseq{N:A}$\,, define
\begin{align*}
\GAseq{M\ypob N} &\iff \forall\ctx.\,(\exists S'.\,\modl \ctx[M]\rred S'\modl0)\implies(\exists S''.\,\modl \ctx[N]\rred S''\modl0)
\end{align*}
where $\ctx$ is a program context. Moreover,\, $\Eypob\ \defn\ \ypob\cap\Rypob$\,.%
\edefn

\subsection{Categorical semantics}\label{s:CatSem}
We now examine sufficient conditions for a fully abstract semantics of $\nurho$ in an abstract categorical setting. Our aim is to construct fully abstract models in an appropriate categorical setting, pinpointing the parts of structure needed for such a task. In section~\ref{S:model} we will apply this knowledge in constructing a concrete such model in nominal games.

Translating each term $M$ into a semantical entity $\trn{M}$ and assuming a preorder ``$\ypoc[]$" in the semantics, full-abstraction amounts to the assertion:
\[ M\ypob N\iff \trn{M}\ypoc[]\trn{N} \tag{FA}\label{e:FA}\]
Note that this formulation is weaker than equational full abstraction, which is given by:
\[ \tag{EFA}\label{e:EFA}
    M\Eypob N\iff \trn{M}=\trn{N}\,.
\]
Nevertheless, once we achieve~\eqref{e:FA} we can construct an \emph{extensional model}, via a quotienting construction, for which~\ref{e:EFA}
holds. Being a quotiented structure, the extensional model does not have an explicit, simple description, and for this reason we prefer working with
the intensional model (i.e.,~the unquotiented one). Of course, an intensional model satisfying~\eqref{e:EFA} would be preferred but this
cannot be achieved in our nominal games. Therefore, our categorical models will be guided by the~\eqref{e:FA} formulation.

\subsubsection{Monads and comonads}
The abstract categorical semantics we put forward is based on the notions of monads and comonads.
These are standard categorical notions (v.~\cite{MacLane:Categories}, and~\cite[\emph{Triples}]{Barr_Wells:Categories}) which have been
used extensively in denotational semantics of programming languages. We present here some basic definitions and properties.

\paragraph{\it Monads} Monads were introduced in denotational semantics through the work of Moggi~\cite{Moggi:89_computational,Moggi:91:Notions}
as a generic tool for encapsulating computational effects. Wadler~\cite{Wadler:92:essence} popularised monads in programming as a means of simulating effects in functional programs, and nowadays monads form part and parcel of the Haskell programming language~\cite{Haskell}.

\bdefn\label{d:monad}
A \boldemph{strong monad} over a category $\CC$ with finite products is a quadruple $(T,\eta,\mu,\tau)$, where $T$ is an endofunctor in $\CC$ and
$\eta:Id_{\CC}\paei T$, $\mu:\TT\paei T$ and $\tau:\uscore\times T\uscore\paei T(\uscore\times\uscore)$ are natural transformations
such that the following diagrams commute.
  \[
  \xymatrix@C=12mm{\TTT A\ar[r]^-{\mu_{TA}}\ar[d]_{T\mu_{A}} & \TT A\ar[d]^{\mu_A}\\ \TT A\ar[r]_-{\mu_A} & TA}\
  \xymatrix@C=12mm{TA\ar[r]^-{\eta_{TA}}\ar[dr]_{\id_{TA}}\ar[d]_{T\eta_A} & \TT A\ar[d]^{\mu_A}\\
                \TT A\ar[r]_{\mu_A} & TA}\ 
  \xymatrix@C=12mm{A\times B\ar[d]_{\id_A\times\eta_B}\ar[dr]^(.63){\eta_{A\times B}} &\!\!\!1\times TA\ar[r]^-{\tau_{1,A}}\ar[dr]_(.37){\isom}
                & T(1\times A)\ar[d]^{T\isom}\\
                A\times TB\ar[r]_-{\tau_{A,B}} & T(A\times B)\!\!\! & TA}
  \]
  \[
  \xymatrix@C=7mm{(A\times B)\times TC\ar[r]^{\tau_{A\times B,C}}\ar[d]_{\isom} & T((A\times B)\times C)\ar[dr]^(.63){T\isom}
        &\!\!\! A\times\TT B\ar[dr]_(.37){\id_A\times\mu_B\ }\ar[r]^-{\tau_{A,TB}} & T(A\times TB)\ar[r]^-{T\tau_{A,B}}& \TT(A\times B)\ar[d]_{\mu_{A\times B}} \\
        A\times (B\times TC)\ar[r]_{\id_A\times\tau_{B,C}} & A\times T(B\times C)\ar[r]_-{\tau_{A,B\times C}} & T(A\times(B\times C))\!\!\!
        & A\times TB\ar[r]_-{\tau_{A,B}} & T(A\times B)}
  \]
We say that $\CC$ has \boldemph{$T$-exponentials} if, for every pair $B,C$ of objects, there exists an object $TC\stin{B}$
such that for any object $A$ there exists a bijection
\[ \LaT[A,B,C]: \CC(A\times B,TC)\lred{\isom}\CC(A,TC\stin{B}) \]
natural in $A$.
\edefn
Given a strong monad $(T,\eta,\mu,\tau)$, we can define the following transformations.
\begin{equation}\begin{aligned}
  \tau'_{A,B} &\defn TA\times B\lred{\isom}B\times TA\lred{\tau_{A,B}}T(B\times A)\lred{\isom}T(A\times B)\,,\\
  \psi_{A,B}  &\defn TA\times TB\lred{\tau'_{A,TB}}T(A\times TB)\lred{T\tau_{A,B}}\TT(A\times B)\lred{\mu_{A\times B}}T(A\times B)\,, \\
  \psi'_{A,B} &\defn TA\times TB\lred{\tau_{TA,B}}T(TA\times B)\lred{T\tau'_{A,B}}\TT(A\times B)\lred{\mu_{A\times B}}T(A\times B)\,.
\end{aligned}\end{equation}
Moreover, $T$-exponentials supply us with \boldemph{$T$-evaluation arrows}, that is,
\begin{equation}
    \evT[B,C]: TC\stin{B}\times B\paei TC \defn \LaT^{-1}(\id_{TC^B})
\end{equation}%
so that, for each $f:A\times B\paei TC$,
\[ f=\LaT(f)\times_B\1\evT[B,C]\,. \]
In fact, $T$-exponentiation upgrades to a functor $(T\uscore)^{-}:\CC\stin{op}\times\CC\paei\CC$ which takes each $f:A'\paei A$ and $g:B'\paei B$ to
\begin{equation}
    Tg\stin{f} : TB'\stin{A}\paei TB\stin{A'}\defn\LaT(TB'\stin{A}\times A'\nzaa{\id\times f}TB'\stin{A}\times A\nzaa{\evT}TB'\nzaa{Tg}TB)\,.
\end{equation}
Naturality of $\LaT[A,B,C]$ in $A$ implies its naturality in $B,C$ too, by use of the above construct.

\paragraph{\it Comonads} Comonads are the dual notion of monads. They were first used in denotational semantics by Brookes and Geva~\cite{BrookesGeva:91:ComputationalComonads}
for modelling programs \emph{intensionally}, that is, as mechanisms which receive external computation data and decide on an output. Monadic-comonadic approaches were examined by Brookes and van Stone~\cite{BrookesVanStone:93:MonadsComonads}.
\bdefn A \boldemph{comonad} on a category $\CC$ is a triple $(Q,\eps,\de)$, where $Q$ is an endofunctor in $\CC$ and $\eps:Q\tote Id_{\CC}$,
$\de:Q\tote\QQ$ are natural transformations such that the following diagrams commute.
\[
  \xymatrix@C=15mm{\QQQ A & \QQ A\ar[l]_-{\de_{QA}}\\ \QQ A\ar[u]^{Q\de_{A}} & QA\ar[l]^-{\de_A}\ar[u]_{\de_A}}\qd[3]
  \xymatrix@C=15mm{QA & \QQ A\ar[l]_-{\eps_{QA}}\ar[r]^-{Q\eps_A} & QA\\ & QA\ar[u]^{\de_A}\ar[ul]^{\id_{QA}}\ar[ur]_{\id_{QA}}}
\]
Now assume $\CC$ has binary products. We define a transformation $\zzet:Q(\uscore\times\uscore)\paei\uscore\times Q(\uscore)$,
\[
    \zzet_{A,B}  \defn Q(A\times B)\lred{\ang{Q\pi_1,Q\pi_2}}QA\times QB\lred{\eps_A\times\id_{QB}}A\times QB\,.
\]
$Q$ is called a \boldemph{product comonad} if $\zzet$ is a natural isomorphism, and is written $(Q,\eps,\de,\zet)$ where $\zet$ is the inverse of $\zzet$.
\edefn%
It is easy to see that the transformation $\zzet$ makes the relevant (dualised) diagrams of definition~\ref{d:monad} commute, even without stipulating the existence of the inverse $\zet$. Note that we write $\zet',\zzet'$ for the  symmetric counterparts of $\zet,\zzet$.

Product comonads are a stronger version of ``strong comonads" of~\cite{BrookesVanStone:93:MonadsComonads}.
A product comonad $Q$ can be written as:
\[ Q\uscore \cong Q1\times\uscore \]
hence the name.\footnote{Note this is an isomorphism between comonads, not merely between functors.} We say that $Q1$ is the \boldemph{basis of the
comonad}.
\paragraph{\it Monadic-comonadic setting}
In the presence of both a strong monad $(T,\eta,\mu,\tau)$ and a product comonad $(Q,\eps,\de,\zet)$ in a cartesian category $\CC$, one may want to
solely consider arrows from some initial computation data (i.e.,~some \emph{initial state}) of type $A$ to some computation of type $B$, that is, arrows of type:
\[ QA \paei TB \]
This amounts to applying the \emph{biKleisli} construction on $\CC$, that is, defining the category $\CC_Q^T$ with the same objects as $\CC$, and arrows
\[ \CC_Q^T(A,B)\defn\CC(QA,TB)\,. \]
For arrow composition to work in the biKleisli category, we need a distributive law between $Q$ and $T$, that is, a natural transformation $\ell:QT\paei
TQ$ making the following diagrams commute.
\[ \xymatrix@C=15mm{
        QA\ar[r]^{Q\eta_A}\ar[dr]_{\eta_{QA}} & QT A\ar[d]_{\ell_A}\ar[r]^{\eps_{TA}} & TA \\
        & TQ A\ar[ur]_{T\eps_A}}\qd[2]
    \xymatrix@C=15mm{
        Q\TT A\ar[r]^{Q\mu_A}\ar[d]_{\ell_{TA}\1 T\ell_A} & QTA\ar[d]_{\ell_A}\ar[r]^{\de_{TA}} & Q^2T A\ar[d]^{Q\ell_A\1\ell_{QA}} \\
        \TT QA\ar[r]_{\mu_{QA}} & TQA\ar[r]_{T\de_A} & TQ^2A}
\]
In this case, composition of $f:QA\paei TB$ and $g:QB\paei TC$ is performed as:
\[ QA\lred{\de}\QQ A\lred{Qf}QTB\lred{\ell_B}TQB\lred{Tg}\TT C\lred{\mu_C}TC \]
Since we are examining a monadic-comonadic setting for strong monad $T$ and product comonad $Q$, a distributive law amounts to a natural transformation
\[ \ell:Q1\times T\uscore\paei T(Q1\times \uscore)\,, \]
which is therefore given for free: take \,$\ell\defn\tau_{Q1,\uscore}$\,. The distributivity equations follow straightforwardly from the monadic
equations.

\paragraph{\it Exponentials and the intrinsic preorder} The notion of $T$-exponentials can be generalised to the monadic-comonadic setting as follows.%
\bdefn Let $\CC$ be a category with finite products and let $(T,\eta,\mu,\tau)$, $(Q,\eps,\de)$ be a strong monad and comonad, respectively, on $\CC$.
We say that $\CC$ has \boldemph{$(Q,T)$-exponentials} if, for each pair $B,C$ of in $\CC$ there exists an object $(Q,T)C\stin{B}$ such that, for each
object $A$, there exists a bijection
\[ \phi_{A,B,C}:\CC(Q(A\times B),TC)\lred{\isom}\CC(QA,(Q,T)C\stin{B}) \]
natural in $A$. \edefn %
Assume now we are in a monadic-comonadic setting $(\CC,Q,T)$ with $T$ a strong monad with $T$-exponentials and $Q$ a product comonad. $(Q,T)$-exponentials then come for free. %
\beprop In the setting of the previous definition, if $T$ is a strong monad with exponentials
and $Q$ is a product comonad then $\CC$ has $(Q,T)$-exponentials defined by:
\begin{equation*}\begin{aligned}
  (Q,T)C\stin{B} &\defn TC\stin{B},\\
  \phi(f)&\defn \LaT(QA\times B\lred{\zet'}Q(A\times B)\lred{f} TC)\,. 
\end{aligned}\end{equation*}%
$\phi$ is a bijection with its inverse sending each $g:QA\paei TC\stin{B}$ to the arrow:
\[ Q(A\times B)\lred{\zzet'}QA\times B\lred{g\times\id}TC\stin{B}\times B\lred{\evT}TC\,. \]
\myqed \enprop%
In the same setting, we can define a notion of \boldemph{intrinsic preorder}. Assuming an object $O$ of \emph{observables} and a collection $\OO\ypo\CC(1,TO)$ of observable arrows, we can have the following.%
\bdefn\label{MonComon:d:QTintrinsic} %
Let $\CC,Q,T,O,\OO$ be as above. We define $\ypoc[]$ to be the union, over all objects $A,B$, of relations $\ypoc[]_{A,B}\ypo\CC(QA,TB)^2$ defined by:
\[
    f\ypoc[]_{A,B}g \iff \forall\rho\in\CC(Q(TB\stin{A}),TO).\ \ \LaQT(f);\rho\in\OO\implies\LaQT(g);\rho\in\OO\,, \]
where \ $\LaQT(f) \defn Q1\lred{\de}\QQ1\lred{Q\LaT(\zet'\1f)}Q(TB\stin{A})$\,.
\edefn%
We have the following enrichment properties. %
\beprop\label{p:MonComon_enrich}%
Let $\CC,Q,T,O,\OO$ and $\ypoc[]$ be as above. Then, for any $f,g:QA\paei TB$ and any arrow $h$, if $f\ypoc[]g$ then:
\begin{DFNitemize}
  \item if $h:QB\paei TB'$ then \ $\de\1Qf\1\ell\1Th\1\mu\ypoc[]\de\1Qg\1\ell\1Th\1\mu$\,,\\[-3mm]
  \item if $h:QA'\paei TA$ then \ $\de\1Qh\1\ell\1Tf\1\mu\ypoc[]\de\1Qh\1\ell\1Tg\1\mu$\,,\\[-3mm]
  \item if $h:QA\paei TC$ then \ $\ang{f,h}\1\psi\ypoc[]\ang{g,h}\1\psi \text{\qd and\qd }\ang{h,f}\1\psi\ypoc[]\ang{h,g}\1\psi$\,,\\[-3mm]
  \item if $A=A_1\times A_2$ then \ $\LaT[QA_1,A_2,B](\zet'\1f)\1\eta\ypoc[]\LaT[QA_1,A_2,B](\zet'\1g)\1\eta$\,.\qed
\end{DFNitemize}
\enprop
\subsubsection{Soundness}
We proceed to present categorical models of the $\nurho$-calculus. The approach we take
is a {monadic} and {comonadic} one, over a computational monad $T$ and a family of local-state comonads $\prn[]=\MODEL{\prn}$, so that the morphism related to each $\GAseq{M:A}$ be of the form $\trn{M}:\prn\trn{\Ga}\paei T\trn{A}$. Computation in $\nurho$ is store-update and fresh-name creation, so $T$ is a store monad, while initial state is given by product comonads.
\bdefn\label{d:laM}\label{d:laNR} %
A \boldemph{$\laNR$-model} $\MM$ is a structure $(\MM,T,\prn[])$ such that:
\begin{cEnumerate}[\bf\Roman{enumi}.]{123}
\item $\MM$ is a category with finite products, with 1 being the terminal object and $A\times  B$ the product of $A$ and $B$.
\item $T$ is a strong monad $(T,\eta,\mu,\tau)$ with exponentials.
\item $\MM$ contains an appropriate natural numbers object $\GN$ equipped with successor and predecessor arrows and $\nm{n}:1\!\paei\GN$, each $n\in\N$. Moreover, for each object $A$, there is an arrow $\cnd[A]:\N\times TA\times TA\paei TA$ for zero-equality tests.
\item $\prn[]$ is a family of product comonads $(\prn,\eps,\de,\zet)_{\all\in\NA{}^\#}$ on $\MM$ such that:
    \begin{ccEnumerate}[\normalfont(\alph{enumii})]{123}
      \item the basis of $\prn[\ee]$ is $1$, and $\prn[\all]=\prn[\all']$ whenever $[\all]=[\all']$ (i.e.,~whenever $\pact\all=\all'$),
      \item if $\supp(\all')\ypo\supp(\all)$ then there exists a comonad morphism $\pit{\all}{\all'}:\prn\paei\prn[\all']$ such that
            $\pit{\all}{\ee}=\eps$, $\pit{\all}{\all}=\id$  and, whenever $\supp(\all')\ypo\supp(\all'')\ypo\supp(\all)$,
            \[ \pitt{\all}{\all''}\1\pitt{\all''}{\all'}=\pitt{\all}{\all'} \]
      \item for each $\all\al\in\NAA[\#]$ there exists a natural transformation $\nw^{\all\al}:\prn\paei T\prn[\all\al]$ such that,
      for each $A,B\in Ob(\MM)$ and $\all\al,\all'\!\al$ with $\supp(\all\al)\ypo\supp(\all'\!\al)$, the following diagrams commute.
      \[\label{e:new_nrMod} \tag{N2}\hspace{-2mm}
        \xymatrix@C=11mm{ \prn[\all']A\ar[r]^{\pit{\all'}{\all}}\ar[d]_{\nw^{\all'\!\!\al}}
                    & \prn A\ar[d]_{\nw^{\all\al}}\ar[r]^-{\ang{\id,\nw^{\all\al}}} & \prn A\times T\prn[\all\al]A\ar[d]^{\tau} \\
                    T\prn[\all'\!\al]A\ar[r]_{T\pit{\all'\!\al}{\all\al}} & T\prn[\all\al]A\ar[r]_-{T\ang{\pit{\all\al}{\all},\id}}
                    & T(\prn A\times\prn[\all\al]A) }\
        \xymatrix@C=10mm{ A\times\prn B\ar[d]_{\id\times\nw_B}\ar[r]^-{\zet} & \prn (A\times B)\ar[d]^{\nw_{A\times B}} \\
                    A\times T\prn[\all\al]B\ar[r]_-{\tau\1T\zet} & T\prn[\all\al](A\times B) } \]
    \end{ccEnumerate}
\item Setting $\GAA{A}\defn\prn[\al]1$, for each $\al\in\NA{A}$, there is a name-equality arrow $\eq[A]:\GAA{A}\times\GAA{A}\paei\GN$ such that, for any distinct $\al,\be\in\NA{A}$, the following diagram commutes.
        \[ \tag{N1}
        \xymatrix@C=2cm{ \prn[\al]1 \ar[r]^-{\De}\ar[d]_{!} & \GAA{A}\times\GAA{A} \ar[d]^{\eq[A]} &
                                \prn[\al\be]1 \ar[l]_-{\ang{\pit{\al\be}{\al},\pit{\al\be}{\be}}}\ar[d]_{!} \\
                                1 \ar[r]^{\nm{0}} & \GN & 1 \ar[l]_{\nm{1}} }
        \]
\item Setting $\trn{\ena}\defn1$, $\trn{\N}\defn\GN$, $\trn{[A]}\defn\GAA{A}$, $\trn{A\tote B}\defn T\trn{B}\stin{\trn{A}}$,
    $\trn{A\times B}\defn\trn{A}\times\trn{B}$,
    $\MM$ contains, for each $A\in\TY$, arrows
    \[ \drf[A]:\GAA{A}\paei T\trn{A}\quad\text{and}\quad \upd[A]:\GAA{A}\times\trn{A}\paei T1 \]
    such that the following diagrams commute,
    \[\tag{NR}
    \begin{gathered}
    \xymatrix@C=25mm{
        \GAA{A}\times\trn{A}\ar[r]^-{\ang{\id,\upd[A]}\1\tau\1\isom}
        & T(\GAA{A}\times\trn{A})\ar@/^/[r]^-{T(\pi_1\1\drf[A])\1\mu}\ar@/_/[r]_-{T\pi_2} & T\trn{A} }
    \\
    \xymatrix@C=17mm{
        \GAA{A}\times\trn{A}\times\trn{A}\ar[rr]^-{\ang{\id\times\pi_1;\upd[A],\id\times\pi_2;\upd[A]}} && T1\times T1
        \ar@/^/[r]^-{\psi\1\isom}\ar@/_/[r]_-{\pi_2} & T1}
    \\
    \xymatrix@C=17mm{
        \prn[\al\be]1\times\trn{A}\times\trn{B}\ar[rr]^-{\ang{\pit{\al\be}{\al}\times\pi_1;\upd[A],\pit{\al\be}{\be}\times\pi_2;\upd[B]}}
        && T1\times T1\ar@/^/[r]^-{\psi\1\isom}\ar@/_/[r]_-{\psi'\1\isom} & T1}
    \end{gathered}\]
    and, moreover,
    \[\tag{SNR} (\nw^{\all\al}_A\times\upd[B])\1\psi = (\nw^{\allal}_A\times\upd[B])\1\psi'\,, \]
    i.e.,~updates and fresh names are independent effects. \deq
\end{cEnumerate}
\edefn[1]%
The second subcondition of~\eqref{e:new_nrMod} above essentially states that, for each object $A$, $\nw_A$ can be expressed as:
\[ \prn A\lred{\isom}\prn1\times A\lred{\nw_1\times\id}T\prn[\all\al]1\times A\lred{\tau'}T(\prn[\allal]1\times A)\lred{\isom}T\prn[\allal]A \]
It is evident that the role reserved for $\nw$ in our semantics is that of fresh
name creation. Accordingly, $\nw$ gives rise to a categorical name-abstraction operation: for any arrow $f:\prn[\allal]A\paei TB$ in $\MM$, we define
\begin{equation}
\abs f\defn\prn A\lred{\nw_A}T\prn[\allal]A\lred{Tf}\TT B\lred{\mu}TB\,.
\end{equation}%
The (NR) diagrams give the basic equations for dereferencings and updates (cf.~\cite[definition 1]{Plotkin_Power:Notions}
and~\cite[section 5.8]{Stark:PhD}). The first diagram stipulates that by dereferencing an updated reference we get  the value of the update. The second diagram ensures that the value of a reference is that of the last update: doing two consecutive updates to the same reference is the same as doing only the last one. The last diagram states that updates of distinct references are independent effects.

Let us now proceed with the semantics of $\nurho$ in $\laNR$-models.
\bdefn %
Let $(\Ma,T,\prn[])$ be a $\laNR$-model. Recall the type-translation:
\[ \trn{\ena}\defn1\,,\quad\trn{\N}\defn\GN\,,\quad\trn{[A]}\defn\GAA{A}\,,\quad\trn{A\tote B}\defn\eistin{T\trn{B}}{\trn{A}}\,,
    \quad\trn{A\times B}\defn\trn{A}\times \trn{B}\,. \]
A typing judgement $\GAseq{M:A}$ is translated to an arrow $\trn[\all\scriptstyle\NZvbar\Ga]{M}:\prn\trn{\Ga}\paei T\trn{A}$\, in $\MM$, which we write simply as $\trn{M}:\prn \Ga\paei TA$, as in figure~\ref{f:SemTrans}. \edefn%
\begin{table}\small%
\fbox{$\begin{array}{@{}c|c@{}}%
\begin{array}{c}
\begin{aligned}
&\trn{\n}\defn\prn{\Ga}\lred{\prn!}\prn1\lred{\pit{\all}{\ee}}1\lred{\n}\GN\lred{\eta}T\GN
\\
&\trn{x}\defn\prn{\Ga}\lred{\prn\pi}\prn{A}\lred{\pit{\all}{\ee}}A\lred{\eta}TA
\\
&\trn{\al}\defn\prn{\Ga}\lred{\prn!}\prn1\lred{\pit{\all}{\al}}\GAA{A}\lred{\eta}T\GAA{A}
\\
&\trn{\sskip}\defn\prn{\Ga}\lred{\prn!}\prn{1}\lred{\pit{\all}{\ee}}1\lred{\eta}T1
\end{aligned}
\\[5mm]\hline\\[-1mm]
\myproof{}{\trn{M}:\prn[\all\al]\Ga\lred{}TA}{\trn{\new[\al]M}:\prn\Ga\lred{\abs\trn{M}}TA}{}
\\[5mm]\hline\\
\myproof{}{\trn{M}:\prn(\Ga\times{A})\lred{}T{B}}
    {\xymatrix@C=25mm@R=7mm{\prn{\Ga}\ar@{-->}[dr]_{\trn{\la x.M}}\ar[r]^{\LaT{}(\zet'\1\trn{M})} & \eistin{TB}{A}\ar[d]^{\eta}\\ &T(\eistin{TB}{A})}}{}
\\\hline\\
\myproof{}{\begin{aligned} \trn{M}&:\prn\Ga\lred{}T(\eistin{TB}{A}) \\ \trn{N}&:\prn\Ga\lred{}TA \end{aligned}}%
    {\xymatrix@C=25mm@R=7mm{\prn\Ga\ar@{-->}[ddr]_{\trn{M\,N}}\ar[r]^-{\ang{\trn{M},\trn{N}}}&T(\eistin{TB}{A})\times TA\ar[d]^{\psi}\\
    &T(\eistin{TB}{A}\times A)
        \ar[d]^{T\evT\1\mu}\\ &TB}}{}
\\\hline\\
\myproof{}{\trn{M}:\prn\Ga\lred{}T(A\times B)}{
 \xymatrix@C=25mm@R=7mm{\prn\Ga\ar@{-->}[dr]_{\trn{\fst M}}\ar[r]^{\trn{M}}&T(A\times B)\ar[d]^{T\pi_1}\\&TA}}{}
\\\hline\\
\myproof{}{%
\begin{aligned}
\trn{M}&:\prn\Ga\lred{}TA\\
\trn{N}&:\prn\Ga\lred{}TB
\end{aligned}}%
{\xymatrix@C=25mm@R=7mm{\prn\Ga\ar@{-->}[dr]_{\trn{\ang{M,N}}}\ar[r]^-{\ang{\trn{M},\trn{N}}}& TA\times TB\ar[d]^{\psi}\\&T(A\times B)}}{}
\end{array}
&
\begin{array}{c}
\myproof{}{\trn{M}:\prn\Ga\lred{}T\GN}{\xymatrix@C=25mm@R=7mm{\prn\Ga\ar@{-->}[dr]_{\trn{\succ M}}\ar[r]^{\trn{M}}&T\GN\ar[d]^{T\succ}\\&T\GN}}{}
\\\hline\\
\myproof{}{\begin{aligned}
\trn{M}&:\prn\Ga\lred{}T\GAA{A}\\
\trn{N}&:\prn\Ga\lred{}T\GAA{A}
\end{aligned}}
    {\xymatrix@C=25mm@R=7mm{\prn\Ga\ar@{-->}[ddr]_{\trn{[M=N]}}\ar[r]^-{\ang{\trn{M},\trn{N}}} & T\GAA{A}\times T\GAA{A}\ar[d]^{\psi}\\
        &T(\GAA{A}\times\GAA{A})\ar[d]^{T\eq}\\&T\GN}}{}
\\\hline\\
\myproof{}{\begin{aligned}
\trn{M}&:\prn\Ga\lred{}T\GAA{A}\\
\trn{N}&:\prn\Ga\lred{}TA
\end{aligned}}%
    {\xymatrix@C=25mm@R=7mm{\prn\Ga\ar@{-->}[ddr]_{\trn{M:=N}}\ar[r]^{\ang{\trn{M},\trn{N}}} & T\GAA{A}\times TA\ar[d]^{\psi}\\
        &T(\GAA{A}\times A)\ar[d]^{T\upd[A]\1\mu}\\ &T1}}{}
\\\hline\\
\myproof{}{\trn{M}:\prn\Ga\lred{}T\GAA{A}}{\xymatrix@C=25mm@R=7mm{\prn\Ga\ar@{-->}[dr]_{\trn{!M}}\ar[r]^{\trn{M}} &T\GAA{A}\ar[d]^{T\drf[A]\1\mu}\\
    &TA}}{}
\\\hline\\
\myproof{}{\begin{aligned}\trn{M}&:\prn\Ga\lred{}T\GN\\ \trn{N_i}&:\prn\Ga\lred{}TA\end{aligned}}%
    {\xymatrix@C=25mm@R=7mm{\prn\Ga\ar@{-->}[ddr]_{\!\!\!\!\!\!\trn{\ifo{M}{N_1}{N_2}}\quad}\ar[r]^-{\ang{\trn{M},\trn{N_1},\trn{N_2}}}&T\GN\times TA^2\ar[d]^{\tau'}\\
        &T(\GN\times TA^2)\ar[d]^{\cnd[A]\1\mu}\\&TA}}{}
\end{array}
\end{array}$}

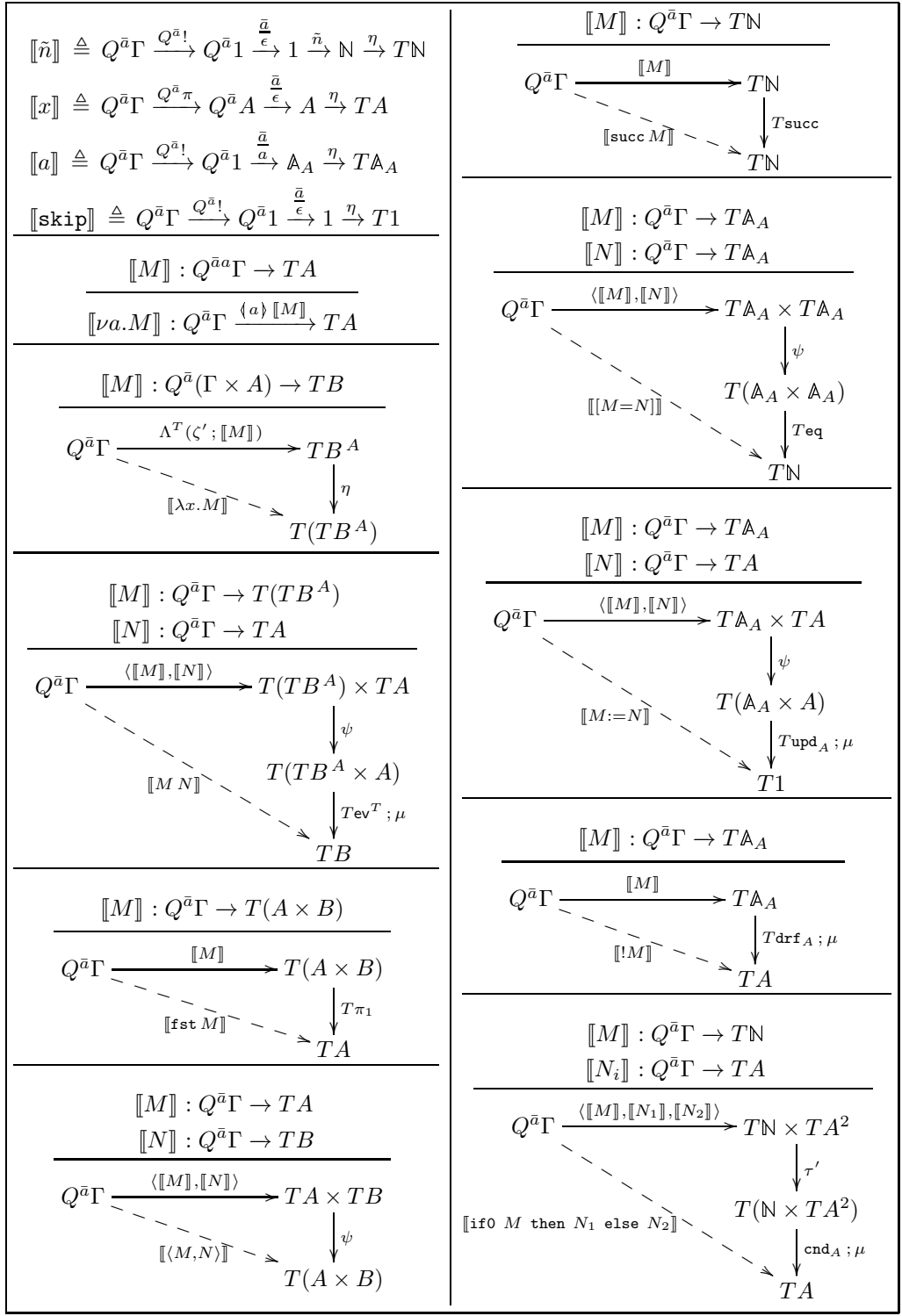
\captionof{figure}{The semantic translation.}\label{f:SemTrans}
\end{table}%
We note that the translation of values follows a common pattern: for any $\GAseq{V:B}$, we have
$\trn{V}=\trnn{V}\1\eta$\,, where
\begin{equation}
\begin{aligned}
\trnn{x}&\defn \prn\pi\1\pit{\all}{\ee}
    & \trnn{\n}&\defn \prn!\1\pit{\all}{\ee}\1\n
& \trnn{\la x.M}&\defn \LaT{}(\zet'\1\trn{M}) \\
\trnn{\al}&\defn \prn!\1\pit{\all}{\al}
& \trnn{\sskip}&\defn \prn!\1\pit{\all}{\ee}
& \trnn{\ang{V,W}}&\defn \ang{\trnn{V},\trnn{W}}\,.
\end{aligned}
\end{equation}
We can show the following lemmas, which will be used in the proof of Correctness.%
\belem\label{l:abs_terms}%
For any $\AGseq{M:A}$ and $\supp(\all)\ypo\supp(\all')$, \
$\trn[\all'|\Ga]{M}= \pit{\all'}{\all}\1\trn[\all|\Ga]{M}$\,. \\
Moreover, if $\Ga=x_1\o B_1,...,x_n\o B_n$\,, and $\GAseq{M:A}$\, and $\GAseq{V_i:B_i}$ are derivable,
\[ \trn{M\{\vec{V}/\vec{x}\}}=\prn\Ga\lred{\ang{\id,\trnn{V_1},...,\trnn{V_n}}}\prn\Ga\times\Ga\lred{\zet'\1\prn\pi_2}\prn\Ga\lred{\trn{M}}TA\,. \]
\myqed[-1]\enlem
%
\belem\label{l:abs_arrows} For any relevant $f,g$,
\begin{align*}
  &\abs\Big(\prn[\all\al]A\lred{\ang{f,\pit{\all\al}{\all}\1g}}TB\times TC\lred{\psi}T(B\times C)\Big) =
    \prn A\lred{\ang{\abs f,g}}TB\times TC\lred{\psi}T(B\times C)\,, \\
  &\abs\left(\prn[\all\al]A\lred{f}TB\lred{Tg}\TT C\lred{\mu}TC\right) = \prn A\lred{\abs f}TB\lred{Tg}\TT C\lred{\mu}TC\,.
\end{align*}
\myqed\enlem%
\belem\label{l:EvalCtx}%
Let $\GAseq{M:A}$ and $\GAseq{\ctxE[M]:B}$ be derivable, with $\ctxE[\uscore]$ being an evaluation context. Then $\trn{\ctxE[M]}$ is equal to:
\[\prn\Ga\lred{\ang{\id,\trn{M}}}\prn\Ga\times TA\lred{\tau}T(\prn\Ga\times A)\lred{T\zet'}T\prn(\Ga\times A)\lred{T\trn{\ctxE[x]}}\TT B\lred{\mu}TB\,.\]
\myqed\enlem%
We write $S\modl M\lred{\text{r}}S'\modl M'$ with $\text{r}\in\{$NEW,SUC,EQ,...,LAM$'\}$ if the last non-CTX rule in the related
derivation is r. Also, to any store $S$, we relate the term $\bar{S}$ of type $\ena$ as:
\begin{equation}
\bar{\ee}\defn\sskip\,,\quad \overline{\al,S}\defn\bar{S}\,,\quad \overline{\al::V,S}\defn(\al:=V\1\bar{S})
\end{equation}%
%
\beprop[Correctness]\label{p:Correctness} For any typed term $\GAseq{M:A}$, and $S$ with $\dom(S)=\all$, and \emph{r} as above,
\begin{ccEnumerate}{12}
\item if $\mathrm{r}\notin\{\mathrm{NEW,UPD,DRF}\}$ then\;  $S\modl M\lred{\mathrm{r}}S\modl M'\;\implies\; \trn{M}=\trn{M'}$\,,
\item if $\mathrm{r}\in\{\mathrm{UPD,DRF}\}$ then\;  $S\modl M\lred{\mathrm{r}}S'\modl M'\;\implies\; \trn{\bar S\1M}=\trn{\bar S'\1M'}$\,,
\item $S\modl M\lred{\mathrm{NEW}}S,\al\modl M'\;\implies\; \trn{\bar S\1M}=\abs\trn{\bar S\1M'}$\,.
\end{ccEnumerate}%
Therefore,\; $S\modl M\lred{}S'\modl M'\;\implies\; \trn{\bar S\1M}=\abs[\all']\trn{\bar S'\1M'}$\,, with $\dom(S')=\all\all'$. %
\enprop %
\proof The last assertion follows easily from 1-3. For 1-3 we do induction on the size of the reduction's derivation. The base case follows from the
specifications of definition~\ref{d:laM} and lemma~\ref{l:abs_terms}. For the inductive step we have that, for any $S,M,\ctxE$, the following diagram
commutes.
{\small\[\xymatrix@C=1.5cm{ \prn\Ga \ar[drr]_{\ang{\id,\trn{\bar
S\1M}}}\ar@/_1cm/[ddrrr]_(.4){\ang{\LaT(\zet'\1\trn{\ctxE[x]})\1\eta,\trn{\bar S\1M}}\1\psi'\qd[4]\text{}}
    \ar[r]^-{\ang{\id,\trn{\bar S}}} & \prn\Ga\times T1\ar[r]^-{\tau\1T\zet'}& T\prn\Ga\ar[r]^-{T\ang{\id,\trn{M}}\1T\tau}
    & \TT(\prn\Ga\times A)\ar[r]^-{\TT(\zet'\1\trn{\ctxE[x]})}\ar[d]^\mu& \TTT B\ar[d]^{T\mu}\ar@/_/[d]_\mu \\
    && \prn\Ga\times TA\ar[r]^\tau\ar[dr]_(.4){\LaT(\zet'\1\trn{\ctxE[x]})\times\id\1\tau\qd\text{ }}
    & T(\prn\Ga\times A)\ar[d]^{T(\LaT(\zet'\1\trn{\ctxE[x]})\times\id)}\ar[r]^-{T(\zet'\1\trn{\ctxE[x]})}& \TT B\ar[d]^\mu \\
    &&& T((A\him TB)\times A)\ar[r]_-{T\evT\1\mu} & TB} \]}%
By the previous lemma, the upper path is equal to $\ang{\id,\trn{\bar S}}\1\tau\1T\zet'\1T\trn{\ctxE[M]}\1\mu$ and therefore to $\trn{\bar
S\1\ctxE[M]}$. Hence, we can immediately show the inductive steps of 1-2. For 3, assuming $S\modl \ctxE[M]\lred{\mathrm{NEW}}S,\al\modl \ctxE[M']$ and
$\trn{\bar S\1M}=\abs\trn{\bar S\1M'}$\,, we have, using also lemmas~\ref{l:abs_terms} and~\ref{l:abs_arrows}, %
\markedeq{ \begin{aligned}
\abs\trn{\bS\1\ctxE[M']}&=\abs(\ang{\LaT(\zet'\1\trn{\ctxE[x]})\1\eta,\trn{\bar S\1M'}}\1\psi'\1T\evT\1 \mu) \\
    &=\abs(\ang{\LaT(\zet'\1\trn{\ctxE[x]})\1\eta,\trn{\bar S\1M'}}\1\psi')\1T\evT\1\mu \\
    &=\ang{\LaT(\zet'\1\trn{\ctxE[x]})\1\eta,\abs\trn{\bar S\1M'}}\1\psi'\1T\evT\1\mu \\
    &=\ang{\LaT(\zet'\1\trn{\ctxE[x]})\1\eta,\trn{\bar S\1M}}\1\psi'\1T\evT\1\mu = \trn{\bS\1\ctxE[M]}\,.
\end{aligned}}{\qedsymbol}
%
%
In order for the model to be sound, we need computational adequacy. This is added explicitly as a specification.
\bdefn\label{d:adeq} %
Let $\MM$ be a $\laNR$-model and $\trn{\uscore}$ the respective translation of~$\nurho$. $\MM$ is \boldemph{adequate} if
\[ \exists S,\bee.\ \trn{M}=\abs[\bee]\trn{\bS\1\nm{0}} \implies \exists S'.\ \all\modl M\rred S'\modl\nm{0}\,, \]
for any typed term $\Aseq{\keno}{M\o\N}$.
\edefn
\beprop[Equational Soundness]
If $\MM$ is an adequate \laNR-model, %
\[ \trn{M}=\trn{N}\implies M\ypob N\,. \]
\myqed[-1]\enprop%
%
%
\subsubsection{Completeness}\label{s:Compl} We equip the semantics with a preorder to match the observational preorder of the syntax as in~\eqref{e:FA}.
The chosen preorder is the intrinsic preorder with regard to a collection of observable arrows in the biKleisli monadic-comonadic setting
(cf.~definition~\ref{MonComon:d:QTintrinsic}). In particular, since we have a collection of monad-comonad pairs, we also need a collection of sets of
observable arrows.
\bdefn%
An adequate $\laNR$-model $\MM=(\MM,T,\prn[])$ is \boldemph{observational} if, for all $\all$:
\begin{DFNitemize}
\item There exists $\obs\ypo\MM(\prn1,T\GN)$ such that, for all $\Aseq{\keno}{M\o\N}$,\,
    \[ \trn{M}\!\in\! \obs \iff \exists S,\bee.\,\trn{M}\!=\!\abs[\bee]\trn{\bS\1\nm{0}}\,. \]
\item The induced intrinsic preorder on arrows in $\MM(\prn A,TB)$ defined by
  \[ f\ypoc g\iff \forall\rho:\prn(\eistin{TB}{A})\paei T\GN.\,(\Laa(f)\1 \rho\in\obs\!\implies\!\Laa(g)\1\rho\in\obs) \]
    with $\Laa(f)\defn\LaQQT{\prn}(f)$, satisfies, for all relevant $\al,\all',f,f'$,
    \[ f\ypoc[\all\al]f' \implies \abs f\ypoc\abs f' \qd\land\qd
        f\ypoc f'\implies \pit{\all'}{\all}\1f\ypoc[\al']\pit{\all'}{\all}\1f'\,. \]
\end{DFNitemize}
We write $\MM$ as $(\MM,T,\prn[],\obs[])$. \edefn
Recurring to $\LaQQT{\prn}$ of definition~\ref{MonComon:d:QTintrinsic}, we have that $\Laa(f)$ is the arrow:
\begin{equation}
\prn1\lred{\de}\prn\prn1\lred{\prn{\LaT(\zet'\1f)}}\prn(\eistin{TB}{A})\,.
\end{equation}
Hence, $O^\all$ contains those arrows that have a specific \emph{observable behaviour} in the model, and the semantic preorder is built over this notion. In particular, terms that yield a number have observable behaviour.

In order to make good use of the semantic preorder we need it to be a \emph{congruence} with regard to the semantic translation. Congruences for
$\nurho$, along with typed contexts, are defined properly in~\cite{Tze_PhD}. For now, we state the following.
\belem\label{l:congr}
Let $(\MM,T,\prn[],\obs[])$ be an observational $\laNR$-model. Then, for any pair $\GAseq{M,N:A}$ of typed terms and any context $\ctx$ such that
$\GAseqq{\ctx[M],\ctx[N]:B}$ are valid,
\[ \trn{M}\ypoc\trn{N} \implies \trn{\ctx[M]}\ypoc[\all']\trn{\ctx[N]}\,. \]
\myqed \enlem
Assuming that we translate $\nurho$ into an observational $\laNR$-model, we can now show one direction of (\ref{e:FA}).
\beprop[Inequational Soundness] For typed terms $\Aseq{\Ga}{M,N:A}$,
\[ \trn{M}\ypoc[]\trn{N} \implies M\ypob N\,. \]
\enprop %
\proof Assume $\trn{M}\ypoc\trn{N}$ and $\modl\ctx[M]\rred S'\modl\nm{0}$\,, so $\trn{\ctx[M]}=\abs[\all']\trn{\bS'\1\nm{0}}$ with $\all'=\dom(S')$.
$\trn{M}\ypoc[\all]\trn{N}$ implies $\trn{\ctx[M]}\ypoc[]\trn{\ctx[N]}$\,, and hence $\trn{\ctx[N]}\in \obs[\ee]$. Thus, by adequacy,
there exists $S''$ such that $\modl\ctx[N]\rred S''\modl\nm{0}$\,. %
\qed\noindent%
In order to achieve completeness, and hence full-abstraction, we need our semantic translation to satisfy some definability requirement with regard to the intrinsic preorder.
\bdefn%
Let $(\MM,T,\prn[],\obs[])$ be an observational $\laNR$-model and let $\trn{\uscore}$ be the semantic translation of $\nurho$ to $\MM$. $\MM$
satisfies \boldemph{ip-definability} if, for any $\all,A,B$, there exists $\dfny{A,B}\ypo\MM(\prn\trn{A},T\trn{B})$ such that:
\begin{DFNitemize}
\item For each $f\in \dfny{A,B}$ there exists a term $M$ such that $\trn{M}=f$.
\item For each $f,g\in\MM(\prn A,TB)$,
 \[ f\ypoc[\all] g\iff\forall\rho\in \dfny{A\tote B,\N}\,.\,(\Laa(f)\1\rho\in\obs\implies\Laa(g)\1\rho\in\obs)\,.\]
\end{DFNitemize}
We write $\MM$ as $(\MM,T,\prn[],\obs[],\dfny[]{})$.
\edefn
%
For such a model $\MM$ we achieve full abstraction.
\bethm[FA] For typed terms $\Aseq{\Ga}{M,N:A}$,
\[  \trn{M}\ypoc[]\trn{N} \iff M\ypob N\,. \]
\enthm \proof
Soundness is by previous proposition. For completeness (``$\Longleftarrow$"), we do induction on the size of $\Ga$.\\
For the base case suppose $\Aseq{\keno}{M\ypob N}$ and take any \ $\rho\in D\,_{\ena\tote A,\N}$ \ such that $\Laa(\trn{M})\1\rho\in\obs$. Let \
$\rho=\trn{\Aseq{y:\ena\tote A}{L:\N}}$\,, some term $L$, so \ $\Laa(\trn{M})\1\rho$ \ is
\[ \Laa(\trn{M})\1\trn{L}=\de\1\prn\trnn{\la z.M}\1 \trn{L}=\trn{(\la y.L)(\la z.M)} \]
for some $z\o\ena$. The latter being in $\obs$ implies that it equals $\abs[\bee]\trn{\bS\1\nm{0}}$, some $S$. Now, $M\ypob N$ implies $(\la y.L)(\la
z.M)\ypob (\la y.L)(\la z.N)$\,, hence $\new[\bee](\bS\1\nm{0})\ypob (\la y.L)(\la z.N)$\,, by soundness. But this implies that $\all\modl (\la
y.L)(\la z.N)\rred S'\modl\nm{0}$\,, so $\trn{(\la y.L)(\la z.N)}\in \obs$, by correctness. Hence, $\Laa(\trn{N})\1\rho\in \obs$,
so $\trn{M}\ypoc\trn{N}$, by ip-definability.

For the inductive step, if $\Ga=x\o B,\Ga'$ then
\begin{align*}
\Aseq{\Ga}{M\ypob N} &\implies \Aseq{\Ga'}{\la x.M \ypob \la x.N}\overset{IH}{\implies} \trn{\la x.M}\ypoc\trn{\la x.N} \\
    &\implies \trn{M}=\trn{(\la x.M)x} \ypoc \trn{(\la x.N)x}=\trn{N}
\end{align*}
where the last approximation follows from lemma~\ref{l:congr}. \qed\noindent%

\section{Nominal games}
\noindent
In this section we introduce nominal games and strategies, and construct the basic structure from which a fully abstract model of $\nurho$ will be
obtained in the next section. We first introduce nominal arenas and strategies, which form the category $\GG$. We afterwards refine $\GG$ by
restricting to \emph{innocent, total} strategies, obtaining thus the category $\Vt$.

$\Vt$ is essentially a semantical basis for call-by-value nominal computation in general. In fact, from it we can obtain not only fully abstract models of
$\nurho$, but also of the $\nu$-calculus~\cite{AGMOS}, the $\ner$-calculus~\cite{Tze:apal08} ($\nurho+$exceptions), etc.
\subsection{The basic category \texorpdfstring{$\GG$}{GG}}
The basis for all constructions to follow is the category $\nom$ of nominal sets. We proceed to arenas.
\bdefn%
A \boldemph{nominal arena} $A\defn(M_A,I_A,\vdash_A,\lambda_A)$ is given by:
\begin{DFNitemize}
  \item a strong nominal set $M_A$ of \boldemph{moves},
  \item a nominal subset $I_A\ypo M_A$ of \boldemph{initial moves},
  \item a nominal \boldemph{justification relation} $\vdash_A\ypo M_A\times (M_A\setminus I_A)$,
  \item a nominal \boldemph{labelling function} $\la_A:M_A\tote \{O,P\}\times\{A,Q\}$,
  which labels moves as \emph{Opponent} or \emph{Player moves}, and as \emph{Answers} or \emph{Questions}.
\end{DFNitemize}
 An arena $A$ is subject to the following conditions.
\begin{aDescription}{aai}
  \item[\,(f)\,] For each $m\in M_A$, there exists unique $k\geq0$ such that $I_A\ni m_1\vdash_A\cdots\vdash_A m_k\vdash_A m$\,, for some $m_l$'s in $M_A$.
  $k$ is called the \emph{level} of $m$, so initial moves have level 0.
  \item[\,(l1)\,] Initial moves are P-Answers.
  \item[\,(l2)\,] If $m_1,m_2\in M_A$ are at consecutive levels then they have complementary OP-labels.
  \item[\,(l3)\,] Answers may only justify Questions. \deq
\end{aDescription}
\edefn[1]%
We let level-1 moves form the set $J_A$\,; since $\vdash_A$ is a nominal relation, $J_A$ is a nominal subset of $M_A$. Moves in $M_A$ are denoted by $m_A$ and variants, initial moves by $i_A$ and variants, and level-1 moves by
$j_A$ and variants. By $\bar{I}_A$ we denote $M_A\plhn I_A$, and by $\bar{J}_A$ the set $M_A\plhn J_A$.

Note that, although the nominal arenas of~\cite{AGMOS} are defined by use of a set of weaker conditions than those above, the actual arenas used there fall within the above definition. We move on to {prearenas}, which are the `boards' on which nominal games are played.

\bdefn A \boldemph{prearena} is defined exactly as an arena, with the only exception of condition (l1): in a prearena initial moves are O-Questions.

Given arenas $A$ and $B$, construct the prearena $A\tote B$ as:
\begin{leftalign}
M_{A\tote B}&\defn M_A + M_B\\
I_{A\tote B}&\defn I_A\\
\la_{A\tote B}&\defn [\,(i_A\mapsto OQ \;,\; m_A\mapsto \bar{\la}_A(m_A)) \;,\; \la_B] \\
\vdash_{A\tote B}&\defn \{(i_A,i_B)\}\cup\{\,(m,n)\,|\,m\vdash_{A,B}n\,\}
\end{leftalign}%
where $\bar{\la}_A$ is the $OP$-complement of $\la_A$. \edefn
It is useful to think of the (pre)arena $A$ as a vertex-labelled directed graph with vertex-set $M_A$ and edge-set $\vdash_A$ such that the
labels on vertices are given by $\la_A$ (and satisfying (l1-3)). It follows from (f) that the graph so defined is \emph{levelled}: its vertices
can be partitioned into disjoint sets L0, L1, L2,\dots\ such that the edges may only travel from level $i$ to level $i+1$ and only level-0 vertices
have no incoming edges (and therefore (pre)arenas are directed acyclic). Accordingly, we will be depicting arenas by levelled graphs or triangles.

The simplest arena is $0\defn(\keno,\keno,\keno,\keno)$. Other (flat) arenas are $1$ (\emph{unit arena}), $\GN$ (\emph{arena of naturals}) and $\GA$ (\emph{arena of $\all$-names}), for any $\all\in\A^\#$, which we define as
\begin{equation}
    M_{1}=I_{1}\defn\{*\}\,, \qd[2]
    M_{\GN}=I_{\GN}\defn\N\,, \qd[2]
        M_{\GA}=I_{\GA}\defn\A^\all\,,
\end{equation}
where $\A^\all\defn\{\,\pact\all\,|\,\pi\in\perm(\A)\,\}$.
Note that for $\all$ empty we get $\GA[\ee]=1$, and that we write $\GA[]_i$ for $\GA[\al]$ with $\al$ being of type $i$.

More involved are the following constructions. For arenas $A,B$, define the arenas $A\ten B$, $A_\bot$, $A\him B$ and $A\impl B$ as follows.
\\
\parbox[c]{.83\linewidth}{\qd$\begin{aligned}
M_{A\ten B}&\defn I_A\tims I_B + \BI{A} + \BI{B}\\
I_{A\ten B}&\defn I_A\tims I_B \\
\la_{A\ten B}&\defn [\,((i_A,i_B)\mapsto PA)\,, \la_A\hrp\BI{A}\,, \la_B\hrp\BI{B}] \\
\vdash_{A\ten B} &\defn \{\,((i_A,i_B),m)\,|\,i_A\vdash_A m\lor i_B\vdash_B m\,\}\cup(\vdash_A\hrp\BI{A}^2)\cup(\vdash_B\hrp\BI{B}^2)
\end{aligned}$}
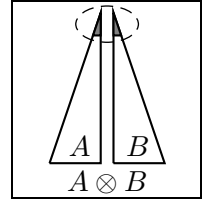
\fbox{\parbox[c][24mm][c]{.15\linewidth}{\qd\rput(0,-0.15){%
\psline(0,-0.7)(0.68,-0.7)(0.68,1.3)(0,-0.7) \pspolygon[fillstyle=solid,fillcolor=gray](0.68,1)(0.578,1)(0.68,1.3)
\psline(0.85,-0.7)(1.53,-0.7)(0.85,1.3)(0.85,-0.7) \pspolygon[fillstyle=solid,fillcolor=gray](0.85,1)(0.952,1)(0.85,1.3)
\psellipse[linestyle=dashed,linewidth=0.5pt](0.765,1.15)(0.425,0.25) \rput[bl]{0}(0.255,-0.6){$A$} \rput[bl](1.02,-0.6){$B$}
\rput[b](0.765,-1.1){$A\otimes B$}}}}
\\[2mm]
\parbox[c]{.83\linewidth}{\qd$\begin{aligned}
M_{A\him B}&\defn I_B+I_A\tims J_B + \BI{A} + \BI{B}\cap \BJ{B} \\
I_{A\him B}&\defn I_B \\
\la_{A\him B}&\defn [\,(i_B\mapsto PA)\,,((i_A,j_B)\mapsto OQ)\,,\bar{\la}_A\hrp\BI{A}\,,\la_B\hrp(\BI{B}\cap\BJ{B})\,] \\
\vdash_{A\him B} &\defn \{\,(i_B,(i_A,j_B))\,|\,i_B\vdash_B j_B\,\}\cup\{\,((i_A,j_B),m)\,|\,i_A\vdash_A m\,\} \\
    &\qd\;\,\cup\{\,((i_A,j_B),m)\,|\,j_B\vdash_B m\,\}\cup(\vdash_A\hrp\BI{A}^2) \cup (\vdash_B\hrp(\BI{B}\cap\BJ{B})^2)
\end{aligned}$}
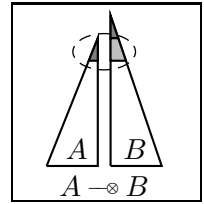
\fbox{\parbox[c][24mm][c]{.15\linewidth}{\qd\rput(-3.1,-0.15){%
\psline(3.06,-0.7)(3.74,-0.7)(3.74,1)(3.06,-0.7) \pspolygon[fillstyle=solid,fillcolor=gray](3.74,1)(3.62,0.7)(3.74,0.7)
\psline(3.91,-0.7)(4.59,-0.7)(3.91,1.3)(3.91,-0.7) \pspolygon[fillstyle=solid,fillcolor=gray](3.91,1)(4.012,1)(3.91,1.3)
\pspolygon[fillstyle=solid,fillcolor=lightgray](3.91,0.7)(3.91,1)(4.012,1)(4.114,0.7)
\psellipse[linestyle=dashed,linewidth=0.5pt](3.825,0.82)(0.425,0.25) \rput[bl]{0}(3.315,-0.6){$A$} \rput[bl](4.08,-0.6){$B$}
\rput[b](3.825,-1.1){$A\him B$}}}}
\\[2mm]
\parbox[c]{.67\linewidth}{\qd$\begin{aligned}
M_{A_\bot}&\defn\{*_1\}+\{*_2\}+M_A \\
I_{A_\bot}&\defn\{*_1\} \\
\la_{A_\bot}&\defn [\,(*_1\mapsto PA)\,, (*_2\mapsto OQ)\,, \la_A] \\
\vdash_{A_\bot} &\defn \{(*_1,*_2),(*_2,i_A)\}\cup(\vdash_A\hrp {M_A}^2) \\
A\impl B &\defn A\him B_\bot
\end{aligned}$}%
\fbox{\parbox[c][24mm][c]{.15\linewidth}{\qd\rput(-1.5,-0.15){%
\psline(1.955,-0.7)(2.635,-0.7)(2.295,0.7)(1.955,-0.7) \psline[linewidth=0.5pt](2.295,0.7)(2.295,1.3) \rput[b]{0}(2.295,-0.6){$A$}
\rput(2.295,1){$*$}\rput(2.295,1.3){$*$} \rput[b](2.295,-1.1){$A_\bot$}}}}%
\fbox{\parbox[c][24mm][c]{.15\linewidth}{\qd\rput(-5,-0.15){%
\psline(5.015,-0.7)(5.695,-0.7)(5.78,1)(5.015,-0.7) \pspolygon[fillstyle=solid,fillcolor=gray](5.78,1)(5.7647,0.7)(5.66015,0.7)
\psline(5.865,-0.7)(6.545,-0.7)(5.78,0.7)(5.865,-0.7) \rput[bl]{0}(5.27,-0.6){$A$} \rput[bl](6.035,-0.6){$B$} \rput[b](5.78,-1.1){$A\impl B$}
\rput(5.78,1.3){$*$} \psline[linewidth=0.5pt](5.78,1.3)(5.78,1)}}}
In the constructions above it is assumed that all moves which are not hereditarily justified by initial moves are discarded. Hence, for example, for
any $A,B$
\[
J_B=\keno\implies A\him B=B
\]
Moreover, we usually identify arenas with graph-isomorphic structures; for example,
\[  1\him A=A\,,\qd 0\impl A = A_\bot\,.  \]
Using the latter convention, the construction of $A\impl B$ in the previous definition is equivalent to $A\impl B$ of \cite{Honda:CBV,AGMOS}\,;
concretely, it is given by:
\begin{lefteqn}
\begin{aligned}[t]
M_{A\impl B}&\defn\{*\}+I_A + \BI{A} + M_B \\
I_{A\impl B}&\defn\{*\} \\
\la_{A\impl B}&\defn[\,(*\mapsto PA)\,,(i_A\mapsto OQ)\,, \bar{\la}_A\,, \la_B]\\
\vdash_{A\impl B}&\defn \{(*,i_A)\}\cup\{\,(i_A,m)\,|\,i_A\vdash_A m\lor m\in I_B\,\}\cup(\vdash_A\hrp\BI{A}{}^2)\cup(\vdash_B\hrp{M_B}^2) \hspace{-20pt}
\end{aligned}
\end{lefteqn}%
Of the previous constructors all look familiar apart from $\him$ (which in~\cite{Tze:lics07} appears as $\tilde{\,\Rightarrow}$). The latter can be seen as a function-space constructor merging the contravariant part of its
RHS with its LHS. For example, for any $A,B,C$, we have
\[ A\him\GN=\GN\qd\text{and}\qd A\him(B\impl C)=(A\ten B)\impl C \]
In the first equality we see that $\GN$ which appears on the RHS of $\him$ has no contravariant part, and hence $A$ is redundant. In the second
equality $B$, which is the contravariant part of $B\impl C$, is merged with $A$. This construction will be of great use when considering a monadic
semantics for store.

We move on to describe how are nominal games played. Plays of a game consist of sequences of moves from some prearena. These moves are attached with
name-lists to the effect of capturing name-environments.%
\bdefn%
A \boldemph{move-with-names} of a (pre)arena $A$ is a pair, written $m^\all$,
where $m$ is a move of $A$ and $\all$ is a finite list of distinct names (\emph{name-list}).%
\edefn%
If $x$ is a move-with-names then its name-list is denoted
by $\nlist(x)$ and its underlying move by $\ul{x}$\,; therefore,
\[ x=\ul{x}^{\nlist(x)}. \]
We introduce some notation for sequences (and lists).%
\benotn[\textbf{Sequences}]%
A sequence $s$ will be usually denoted by $xy\dots$, where $x,y,...$ are the elements of $s$. For sequences $s,t$,
\begin{DFNitemize}
  \item $s\leq t$ denotes that $s$ is a prefix of $t$, and then $t=s(t\pln s)$,
  \item $s\sublist t$ denotes that $s$ is a (not necessarily initial or contiguous) subsequence of $t$,
  \item $s^-$ denotes $s$ with its last element removed,
  \item if $s=s_1\dots s_n$ then $s_1$ is the first element of $s$ and $s_n$ the last. Also,
  \begin{aDescription}[$\circ$]{}
  \item $n$ is the {length} of $s$, and is denoted by $|s|$,
  \item $s.i$ denotes $s_i$ and $s.\arnt i$ denotes $s_{n+1-i}$\,, that is, the $i$-th element from the end of $s$
  (for example, $s.\arnt1$ is $s_n$),
  \item $s_{\leq s_i}$ denotes $s_1\dots s_i$\,, and so does $s_{<s_{i+1}}$,
  \item if $s$ is a sequence of moves-with-names then, by extending our previous notation, we have $s=\ul{s}^{\nlist(s)}$, where
        $\nlist(s)$ is a list of length $|s|$ of lists of names. \deq
  \end{aDescription}
\end{DFNitemize}
\ennotn%
A \emph{\bfseries justified sequence} over a prearena $A$ is a finite sequence $s$ of OP-alternating moves such that,
except for $s.1$ which is initial, every move $s.i$ has a \boldemph{justification pointer} to some $s.j$ such that $j<i$ and $s.j\vdash_A s.i$\,; we
say that $s.j$ \boldemph{(explicitly) justifies} $s.i$\,. A move in $s$ is an \boldemph{open question} if it is a question and there is no answer
inside $s$ justified by it.

There are two standard technical conditions that one may want to apply to justified sequences: \boldemph{well-bracketing} and \boldemph{visibility}. We
say that a justified sequence $s$ is \emph{well-bracketed} if each answer $s.i$ appearing in $s$ is explicitly justified by the last open question in
$s_{<i}$ (called the \boldemph{pending question}). For visibility, we
need to introduce the notions of \boldemph{Player- and Opponent-view}. For a justified sequence $s$, its P-view $\pv{s}$ and its O-view $\ov{s}$ are defined as follows.
\[\begin{array}{c | c}
\begin{aligned}
\pv{\ee} &\defn \ee\\
\pv{sx} &\defn \pv{s}x && \text{if $x$ a P-move}\\
\pv{x} &\defn x && \text{if $x$ is initial}\\
\pv{sxs'y} &\defn \pv{s}xy && \text{if $y$ an O-move}\\ &&& \text{expl. justified by $x$}
\end{aligned}\qd\text{}&\text{}\qd
\begin{aligned}
\ov{\ee} &\defn \ee\\
\ov{sx} &\defn \ov{s}x && \text{if $x$ an O-move}\\
\ov{sxs'y} &\defn \ov{s}xy && \text{if $y$ a P-move}\\ &&& \text{expl. justified by $x$} \\\text{}
\end{aligned}
\end{array}\]
The visibility condition states that any O-move $x$ in $s$ is justified by a move in $\ov{s_{<x}}$\,, and any P-move $y$ in $s$ is justified by a move in $\pv{s_{<y}}$. We can now define plays. %
\bdefn%
Let $A$ be a prearena. A legal sequence on $A$ is sequence of moves-with-names $s$ such that $\ul{s}$ is a justified sequence satisfying Visibility and
Well-Bracketing. A legal sequence $s$ is a \boldemph{play} if $s.1$ has empty name-list and $s$ also satisfies the following Name Change Conditions (cf.~\cite{Ong:lics02}):
\begin{Description}{asdasd}
  \item[\,(NC1)\,] The name-list of a P-move $x$ in $s$ contains as a prefix the name-list of the move preceding it. It possibly contains some other names,
  all of which are fresh for $s_{<x}$.
  \item[\,(NC2)\,] Any name in the support of a P-move $x$ in $s$ that is fresh for $s_{<x}$ is contained in the name-list of $x$.
  \item[\,(NC3)\,] The name-list of a non-initial O-move in $s$ is that of the move justifying it.
\end{Description}
\nada\\[-3mm]
The set of plays on a prearena $A$ is denoted by $P_A$. \edefn
It is important to observe that plays have strong support, due to the tagging of moves with lists of names (instead of sets of names~\cite{AGMOS}).
Note also that plays are the $\ee$-plays of \cite{Tze:lics07}. Now, some further notation.%
\benotn[Name-introduction]\label{n:plays} A name $\al$
{is introduced} (by Player) in a play $s$, written $\al\in\LL(s)$, if there exist consecutive moves $yx$ in $s$ such that $x$ is a P-move and
$\al\in\supp(\nlist(x)\pln\nlist(y))$. \deq \ennotn
From plays we move on to strategies. Recall the notion of name-restriction we introduced in definition~\ref{NS:d:NomAbstr}; for any nominal set $X$ and
any $x\in X$, \ $[x]=\{\,\pact x\,|\,\pi\in\perm(\A)\,\}$\,.
\bdefn%
Let $A$ be a prearena. A \boldemph{strategy} $\sigma$ on $A$ is a set of equivalence classes $[s]$ of plays in $A$, satisfying:
\begin{DFNitemize}
  \item[ \bf Prefix closure:] If $[su]\in\sigma$ then $[s]\in\sigma$.
  \item[ \bf Contingency completeness:] If even-length $[s]\in\sigma$ and $sx$ is a play then $[sx]\in\sigma$.
  \item[ \bf Determinacy:] If even-length $[s_1x_1],[s_2x_2]\in\sigma$ and $[s_1]=[s_2]$ then $[s_1x_1]=[s_2x_2]$.
\end{DFNitemize}
We write $\sigma:A$ whenever $\sigma$ is a strategy on $A$. \edefn%
By convention, the empty sequence $\ee$ is a play and hence, by prefix closure and contingency completeness, all strategies contain $[\ee]$ and $[i_A]$'s. Some basic strategies are the following\HY note that we give definitions \emph{modulo prefix closure}.
\bdefn\label{d:verybasicmorphisms} For any $\all',\all\in\A^\#$ with $\supp(\all')\ypo\supp(\all)$, $n\in\N$ and any arena $B$, define the following strategies.
\begin{DFNitemize}
\item $\n:1\paei\GN\defn\{[*\,n]\}$\\[-3mm]
\item $!_B : B\paei 1\defn\{[i_B\, *]\}$\\[-3mm]
\item $\pit{\all}{\all'}:\GA[\all]\paei\GA[\all']\defn\{[\all\,\all']\}$\\[-3mm]
\item $\id_B : B\paei B\defn\{\,[s]\,|\,s\in P_{B\indx{1}\tote B\indx{2}}\land \forall t\leq^{\text{even}}\! s.\ t\hrp B\indx{1}=t\hrp B\indx{2}\,\}$ \deq
\end{DFNitemize}
\edefn[q]%
It is easy to see that the aforedefined are indeed strategies. That definitions are given modulo prefix closure means that e.g.~$\n$ is in fact:
\[ \n = \{\,[\ee],\, [*],\, [*\,n]\,\}\,. \]
We proceed to composition of plays and strategies.
In ordinary games, plays are composed by doing ``parallel composition plus hiding" (v.~\cite{Abramsky_Jagadeesan:MLL}); in nominal games we need also take some extra care for names.
\bdefn%
Let $s\in P_{A\tote B}$ and $t\in P_{B\tote C}$. We say that:
\begin{Description}[$\bullet$]{.}
\item $s$ and $t$ are \boldemph{almost composable}, $s\smallsmile t$, if $\ul{s}\upharpoonright B=\ul{t}\upharpoonright B$.
\item $s$ and $t$ are \boldemph{composable}, $s\asymp t$, if $s\smallsmile t$ and, for any $s'\leq s$, $t'\leq t$ with $s'\smallsmile t'$:
\begin{Description}{qwe}
  \item[\,(C1)\!\!] If $s'$ ends in a (Player) move in $A$ introducing
  some name $\al$ then $\al\4 t'$. 

            Dually, if $t'$ ends in a move in $C$ introducing some name $\al$ then $\al\4 s'$.
  \item[\,(C2)\!\!] If both $s',t'$ end in $B$ and $s'$ ends in
  a move introducing some name $\al$ then $\al\4
  t'^-$.

            Dually, if $t'$ ends in a move introducing some name $\al$ then $\al\4 s'^-$. \deq
\end{Description}
\end{Description}
\edefn[q]%
The following lemma is taken verbatim from \cite{Honda:CBV}, adapted from~\cite{BDER}.
\belem[Zipper lemma] If $s\in P_{A\tote B}$ and $t\in P_{B\tote C}$ with $s\smallsmile t$ then either $\ul{s}\hrp B=\ul{t}=\ee$, or $s$ ends in $A$ and
$t$ in $B$, or $s$ ends in $B$ and $t$ in $C$, or both $s$ and $t$ end in $B$. \qed \enlem
Note that in the sequel we will use some standard \emph{switching condition} results (see e.g.~\cite{Honda:CBV,AJM:PCF}) without further mention. Composable plays are composed as below. Note that we may tag a move $m$ as $m\indx{O}$ (or $m\indx{P}$) to specify it is an O-move (a P-move). %
\bdefn%
Let $s\in P_{A\tote B}\,$ and $t\in P_{B\tote C}\,$ with $s\asymp t$\,. Their \boldemph{parallel interaction} $s\comp t$ and their \boldemph{mix}
$s\mix t$, which returns the final name-list in $s\comp t$, are defined by mutual recursion as follows. We set $\ee\comp\ee\defn \ee$\,, \
$\ee\mix\ee\defn \ee$\,, and:
\begin{align*}
sm_A^\bee\comp t            &\defn (s\comp t)m_A^{sm_A^\bee\mix t}\!\!\!\!\!\!\!\!
& sm_B^\bee\comp tm_B^\gaa  &\defn (s\comp t)m_B^{sm_B^\bee\mix tm_B^\gaa}\!\!\!\!\!\!
& s\comp tm_C^\gaa          &\defn (s\comp t)m_C^{s\mix tm_C^\gaa}
\\
sm_{A(P)}^{\bee_s\bee}\mix t                  &\defn (s\mix t)\,\bee
& sm_{B(P)}^{\bee_s\bee}\mix tm_{B(O)}^\gaa   &\defn (s\mix t)\,\bee
& s\mix tm_{C(P)}^{\gaa_t\gaa}                &\defn (s\mix t)\,\gaa
\\
sm_{A(O)}^\bee\mix t                          &\defn \bee'
& sm_{B(O)}^\bee\mix tm_{B(P)}^{\gaa_t\gaa}   &\defn (s\mix t)\,\gaa
& s\mix tm_{C(O)}^\gaa                        &\defn \gaa\,',
\end{align*}
where $\bee_s$ is the name-list of the last move in $s$, and $\bee'$ is the name-list of $m_{A(O)}$'s justifier inside $s\comp t$\,; similarly for $\gaa_t$ and $\gaa\,'$.\\
The \boldemph{composite} of $s$ and $t$ is:
\[s\cmp t \defn (s\comp t)\upharpoonright AC\,. \]
The set of \boldemph{interaction sequences} of $A,B,C$ is defined as:
$$\iseq(A,B,C) \defn \{\,s\comp t\,|\,s\in\play{A\tote B}\Land t\in\play{B\tote C}\Land s\asymp t\,\}\,.\eqno{\blacktriangle}$$
\edefn[a]%
When composing compatible plays $s$ and $t$, although their parts appearing in the common component ($B$) are hidden, the names appearing in (the support of) $s$ and $t$ are not lost but rather propagated to the output components ($A$ and $C$). This is shown in the following lemma (the proof of which is tedious but not difficult, see~\cite{Tze_PhD}).
\belem\label{l:asymp}%
Let $s\asymp t$ with $s\in P_{A\tote B}$ and $t\in P_{B\tote C}$.
\begin{cEnumerate}[\rm(\alph{enumi})\,]{123}
  \item If $s\comp t$ ends in a generalised P-move $m^\bee$ then $\bee$ contains as a prefix the name-list of $(s\comp t).\arnt2$\,.
  It possibly contains some other names, all of which are fresh for $(s\comp t)^-$.
  \item If $s\1t$ ends in a P-move $m^\bee$ then $\bee$ contains as a prefix the name-list of $(s\1t).\arnt2$\,.
  It possibly contains some other names, all of which are fresh for $(s\1t)^-$.
  \item If $s\comp t$ ends in a move $m^\bee$ then $\bee$ contains as a prefix the name-list of the move explicitly justifying $m^\bee$.
  \item If $s=s'm^\bee$ ends in $A$ and $t$ in $B$ then $\bee\sublist s\mix t$,\\
             if $s=s'm^\bee$ and $t=t'm^\gaa$ end in $B$ then $\bee\sublist s\mix t$ and $\gaa\sublist s\mix t$,\\
             if $s$ ends in $B$ and $t=t'm^\gaa$ in $C$ then $\gaa\sublist s\mix t$.
  \item $\supp(s)\cup\supp(t)=\supp(s\comp t)=\supp(s\cmp t)\cup\supp(s\mix t)$\,.
\qed
\end{cEnumerate}
\enlem
\beprop[Plays compose]\label{p:playscompose} If $s\in P_{A\tote B}\,$ and $t\in P_{B\tote C}\,$ with $s\asymp t$, then
$s\cmp t\in P_{A\tote C}\,$.
\enprop
\proof We skip visibility and well-bracketing, as these follow from ordinary CBV game analysis. It remains to show that the name change conditions hold
for $s\cmp t$. (NC3) clearly does by definition, while (NC1) is part (b) of previous lemma.\\
For (NC2), let $s\1t$ end in some P-move $m^{s\mix t}$ and suppose $\al\in\supp(m^{s\mix t})$ and $\al\4(s\1t)^-$. Suppose wlog that $s=s'm^\bee$, and
so $(s\cmp t)^-=s'\cmp t$. Now, if $\al\4s'\mix t$ then, by part (e) of previous lemma, $\al\4s',t$ and therefore $\al\in\bee$\,, by (NC2) of $s$. By
part (d) then, $\al\in\supp(s\mix t)$. Otherwise, $\al\in\supp(s'\mix t)$ and hence, by part (a), $\al\in\supp(s\mix t)$. \qed\medskip%
\noindent
We now proceed to composition of strategies. Recall that we write
$\sigma:A\paei B$ if $\sigma$ is a strategy on the prearena $A\tote
B$.\newpage

\bdefn%
For strategies $\sigma:A\tote B$ and $\tau:B\tote C$, their {composition} is defined as
\[ \sigma\cmp\tau \defn \{\,[s\cmp t]\,|\,[s]\in\sigma\land[t]\in\tau\land s\asymp t\,\}\,, \]
and is a candidate strategy on $A\tote C$.
\edefn%
Note that, for any sequence $u$, if $[u]\in\st$ then $u=\pi\actn(s\cmp t)=(\pi\actn s)\cmp(\pi\actn t)$ for some $[s]\in\sig,[t]\in\tau,s\asymp t$ and
$\pi$. Therefore, we can always assume $u=s\cmp t$ with $[s]\in\sig,[t]\in\tau$ and $s\asymp t$. Our next aim is to
show that composites of strategies are indeed strategies. Again, the proofs of the following technical lemmata are omitted for economy (but see~\cite{Tze_PhD}).%
\belem%
For plays $s_1\asymp t_1$ and $s_2\asymp t_2$\,, if $s_1\comp t_1=s_2\comp t_2$ then $s_1=s_2$ and $t_1=t_2$\,. Hence, if $s_1\comp t_1\leq s_2\comp
t_2$ then $s_1\leq s_2$ and $t_1\leq t_2$\,. \qed
\enlem %
%
\belem\label{l:somelemma}%
Let $\sigma:A\tote B$ and $\tau:B\tote C$ be strategies with $[s_1],[s_2]\in\sigma$ and $[t_1],[t_2]\in\tau$. If $|s_1\comp t_1|\leq|s_2\comp t_2|$ and
$[s_1\cmp t_1]=[s_2\cmp t_2]$ then there exists some $\per$ such that $\per\actn (s_1\comp t_1)\leq s_2\comp t_2$. \qed%
\enlem
\beprop[Strategies compose]\label{p:stratcomp} If $\sigma:A\tote B$ and $\tau:B\tote C$ are strategies then so is $\st$.%
\enprop%
\proof By definition and proposition \ref{p:playscompose}, $\sigma\1\tau$ contains equivalence classes of plays. We need also check prefix closure, contingency completeness and determinacy. The former two are rather straightforward, so we concentrate on the latter.\\
Assume even-length $[u_1 x_1],[u_2 x_2]\in\st$ with $[u_1]=[u_2]$, say $u_ix_i=s_i\1 t_i$, $[s_i]\in\sig$ and $[t_i]\in\tau$,
$i=1,2$\,.
By prefix-closure of $\sig,\tau$ we may assume that $s_i,t_i$ don't both end in $B$, for $i=1,2$.\\
If $s_i$ end in $A$ then $s_i=s_i'n_i^{\bee_i}$ and $s_i\1 t_i=(s_i'\1 t_i)n_i^{\bee_i'}$, $i=1,2$\,. Now, $[s_1'\1 t_1]=[u_1]=[u_2]=[s_2'\1 t_2]$, so,
by lemma~\ref{l:somelemma} and assuming wlog that $|s_1'\comp t_1|\leq|s_2'\comp t_2|$, we have $\pact(s_1'\comp t_1)\leq(s_2'\comp t_2)$, $\ara\pact
s_1'\leq s_2'$\,, say $s_2'=s_2''s_2'''$ with $s_2''=\pact s_1'$ and $s_2'''$ in $B$. Then $[s_2'']=[s_1']$, $\ara
[s_2''(s_2'''n_2^{\bee_2}).1]=[s_1'n_1^{\bee_1}]$, by determinacy of $\sig$, and hence $|s_2'''|=0$, $s_2'=\pact s_1'$ and $t_2=\pact t_1$\,. Moreover,
$\pact[']s_1'n_1^{\bee_1}=s_2'n_2^{\bee_2}$, some permutation $\per'$. Now we can apply the Strong Support Lemma, as (C1) implies
$(\supp(n_i^{\bee_i})\plhn\supp(s_i'))\cap\supp(t_i)=\keno$. Hence, there exists a permutation $\per''$ such that $\pact['']s_1=s_2$ and
$\pact['']t_1=t_2$, $\ara [s_1\1 t_1]=[s_2\1 t_2]$\,, as required.
\\
If $s_i$ end in $B$ and $t_i$ in $C$, then work similarly as above. These are, in fact, the only cases we need to check. Because if, say, $s_2,t_1$ end
in $B$, $s_1$ in $A$ and $t_2$ in $C$ then $t_1,s_2$ end in P-moves and $[s_1^-\1 t_1]=[s_2\1t_2^-]$ implies that $s_1^-,t_2^-$ end in O-moves in $B$.
If, say, $|s_1^-\comp t_1|\leq|s_2\comp t_2^-|$ then we have, by lemma~\ref{l:somelemma}, $\pact s_1^-\leq s_2$\,, some permutation $\per$. So if
$\pact s_1^-=s_2'$ and $s_2=s_2's_2''$, determinacy of $\sig$ dictates that $s_2''.1$ be in $A$, \contra to $|s_1\1t_1|=|s_2\1t_2|$ and $s_2\1t_2$
ending in $C$. \qed\noindent%
In order to obtain a category of nominal games, we still need to show that strategy composition is associative. We omit the (rather long) proof and
refer the interested reader to~\cite{Tze_PhD}.
\beprop%
For any $\sig:A\paei B$, $\sig_1:A'\paei A$ and $\sig_3:B\paei B'$,
\[ \id_A\1\sig = \sig = \sig\1\id_B \qd\land\qd  (\sig_1\1\sig)\1\sig_3=\sig_1\1(\sig\cmp\sig_3)\,. \]
\myqed\enprop
%
\bdefn The \boldemph{category $\GG$ of nominal games} contains nominal arenas as objects and nominal strategies as arrows. \edefn
In the rest of this section let us examine closer the proof of proposition~\ref{p:stratcomp} in order identify where exactly is strong support needed, and for which reasons is the nominal games model of~\cite{AGMOS} flawed.
%
\berem[The need for strong support]\label{rem:SS} The nominal games presented here differ from those of \cite{AGMOS} crucially in one aspect; namely, the
modelling of local state. In~\cite{AGMOS} local state is modelled by finite sets of names, so a move-with-names is a
move attached with a finite set of names, and other definitions differ accordingly. The problem is that thus determinacy is not preserved by strategy
composition: information separating freshly created names may be hidden by composition and hence a composite strategy may break determinacy by
distinguishing between composite plays that are equivalent.

In particular, in the proof of determinacy above we first derived from $[s'_1\1t_1]=[s'_2\1t_2]$ that there exists some $\per$ so that $\pact s'_1=s_2$
and $\pact t_1=t_2$, by appealing to lemma~\ref{l:somelemma}; in the (omitted) proof of that lemma, the Strong Support Lemma needs to be used several
times. In fact, the statement
\[ |s'_1\comp t_1|=|s'_2\comp t_2| \Land [s'_1\1t_1]=[s'_2\1t_2]\implies \exists\per.\,\pact s'_1=s'_2\land \pact t_1=t_2 \]
does not hold in a weak support setting such that of~\cite{AGMOS}. For take some $i\in\om$ and consider the following AGMOS-strategies (i.e.~strategies of~\cite{AGMOS}).
\begin{lefteqn*}\tag{\ref{rem:SS}:A}\label{e:SS:A}
\begin{aligned}
\sig&:1\paei\GA[]_i\defn\{\, [*\,\al^{\{\al,\be\}}]\,|\,\al,\be\in\A_i\land\al\neq\be\,\}\,, \\
\tau&:\GA[]_i\paei\GA[]_i\impl\GA[]_i\defn\{\,[\al * \ga\,\al]\,|\,\al,\ga\in\A_i\,\}\,.
\end{aligned}
\end{lefteqn*}%
Then,
\[ [*\,\al^{\{\al,\be\}}\1\,\al*\be]=[**^{\{\al,\be\}}\be^{\{\al,\be\}}]=[**^{\{\al,\be\}}\al^{\{\al,\be\}}]
= [*\,\al^{\{\al,\be\}}\1\,\al*\al]\,,
\]
yet for no $\per$ do we have $\pact(*\,\al^{\{\al,\be\}})=*\,\al^{\{\al,\be\}}$ and
$\pact(\al*\be)=\al*\al$. As a result, determinacy fails for $\sig\1\tau$ since both
$[**^{\{\al,\be\}}\be^{\{\al,\be\}}\al^{\{\al,\be\}}],[**^{\{\al,\be\}}\al^{\{\al,\be\}}\al^{\{\al,\be\}}]\in\sig\1\tau$.

Another point where we used the Strong Support Lemma in the proof of determinacy was in showing (the dual of):
\begin{align*}
&\exists\per,\per'.\, \pact (s_1,t'_1)=(s_2,t'_2)\Land\pact[']t'_1n_1^{\bee_1}=t'_2n_2^{\bee_2}\implies\exists\per''.\,
\pact[''](s_1,t'_1n_1^{\bee_1})=(s_2,t'_2n_2^{\bee_2})\\
&\text{i.e.}\qd[1] [s_1,t'_1]=[s_2,t'_2] \Land [t'_1n_1^{\bee_1}]=[t'_2n_2^{\bee_2}]\implies[s_1,t'_1n_1^{\bee_1}]=[s_2,t'_2n_2^{\bee_2}]\,.
\end{align*}
The above statement does not hold for AGMOS-games. To show this, we need to introduce\footnote{%
This is because our presentation of nominal games does not include plays and strategies with non-empty initial local state. In the AGMOS setting we
could have used to the same effect the $\{\al,\be\}$-strategies:
\begin{center}
$\sig:\GA[]_i\ten\GA[]_i\paei 1\defn\{\,[\,(\al,\be)^{\{\al,\be\}}*^{\{\al,\be\}}]_{\{\al,\be\}} \,\} \,,\qd[1] \tau:1\paei\GA[]_i\defn\{\,
[*^{\{\al,\be\}}\al^{^{\{\al,\be\}}}]_{\{\al,\be\}} \,\}\,.$
\end{center}} the flat arena
$\GA[]_i\mathop{\odot}\GA[]_i$ with $M_{\GA[]_i\odot\GA[]_i}\defn\PP_{2}(\A_i)$ (the set of 2-element subsets of $\A_i$). This is not a legal arena in our setting, since its moves are not strongly supported, but it is in the AGMOS setting. Consider the following strategies.
\[ \tag{\ref{rem:SS}:B}\begin{aligned}
\sig&:\GA[]_i\ten\GA[]_i\paei\GA[]_i\mathop{\odot}\GA[]_i\defn\{\, [\,(\al,\be)\,\{\al,\be\}]\,|\,\al,\be\in\A_i\land\al\neq\be \,\} \\
\tau&:\GA[]_i\mathop{\odot}\GA[]_i\paei\GA[]_i\defn\{\, [\{\al,\be\}\,\al]\,|\,\al,\be\in\A_i\land\al\neq\be \,\}
\end{aligned}\]
We have that $[\,(\al,\be)\,\{\al,\be\},\{\al,\be\}]=[\,(\al,\be)\,\{\al,\be\},\{\al,\be\}]$ and $[\{\al,\be\}\,\al]=[\{\al,\be\}\,\be]$\,, yet
\[ [\,(\al,\be)\,\{\al,\be\},\{\al,\be\}\,\al]\neq[\,(\al,\be)\,\{\al,\be\},\{\al,\be\}\,\be]\,. \]
In fact, determinacy is broken since $[\,(\al,\be)\,\al],[\,(\al,\be)\,\be]\in\sigma\1\tau$. \deq\enrem
\subsection{Arena and strategy orders in \texorpdfstring{$\GG$}{GG}}\label{sec:Orderings_G}
$\GG$ is the raw material from which several subcategories of nominal games will emerge. Still, though, there is structure in $\GG$ which will be
inherited to the refined subcategories we will consider later on. In particular, we consider (subset) orderings for arenas and strategies, the latter
enriching $\GG$ over Cpo.\footnote{%
By cpo we mean a partially ordered set with least element and least upper bounds for increasing $\om$-sequences. Cpo is the category of cpos and
continuous functions.} These will prove useful for solving domain equations in categories of nominal games.

\bdefn\label{d:Enrich}%
For any arenas $A,B$ and each $\sig,\tau\in\GG(A,B)$ define $\sig\ypoo\tau\iff\sig\ypo\tau$\,.\\
For each $\ypoo$-increasing sequence $(\sig_i)_{i\in\om}$ take $\bigsqcup_i\sig_i\defn\bigcup_i\sig_i$.
\edefn%
It is straightforward to see that each such $\bigsqcup_i\sig_i$ is indeed a strategy: prefix closure, contingency completeness and determinacy easily follow from the fact that the sequences we consider are $\ypoo$-increasing.
Hence, each $\GG(A,B)$ is a cpo with least element the empty strategy (i.e.~the one containing only $[\ee]$). More than that, these cpo's enrich $\GG$.%
\beprop\label{p:CpoEnrich} $\GG$ is
Cpo-enriched wrt $\ypoo$.
\enprop%
\proof Enrichment amounts to showing the following straightforward assertions.
\begin{align*}
  \sig\ypoo\sig' \land \tau\ypoo\tau' &\implies \sig\1\tau\ypoo\sig'\1\tau' \\
  (\sig_i)_{i\in\om}\text{ an $\om$-chain} &\implies (\bigsqcup_{i\in\om}\sig_i)\1\tau\ypoo \bigsqcup_{i\in\om}(\sig_i\1 \tau) \\
  (\tau_i)_{i\in\om}\text{ an $\om$-chain} &\implies \sig\1(\bigsqcup_{i\in\om}\tau_i) \ypoo \bigsqcup_{i\in\om}(\sig\1\tau_i)
\end{align*}
\myqed[-3]%
On the other hand, arenas are structured sets and hence also {ordered} by a `subset relation'. %
\bdefn%
For any $A,B\in Ob(\GG)$ define
\[ A\ypop B \iff M_A \ypo M_B \Land I_A\ypo I_B \Land \la_A \ypo \la_B \Land \mathord{\vdash_A} \ypo \mathord{\vdash_B}\,, \]
and for any $\ypop$-increasing sequence $(A_i)_{i\in\om}$ define
\[ \bigsqcup_{i\in\om}A_i\defn\bigcup_{i\in\om}A_i\,. \]
If $A\ypop B$ then we can define an \boldemph{embedding-projection pair} of arrows by setting:
\begin{myalign}
    \incl{A,B}&:A\paei B \defn \{\,[s]\in[\play{A\tote B}]\,|\,[s]\in \id_A\lor (odd(|s|)\land[s^-]\in\id_A)\,\}\,, \\
    \proj{B,A}&:B\paei A \defn \{\,[s]\in[\play{B\tote A}]\,|\,[s]\in \id_A\lor (odd(|s|)\land[s^-]\in\id_A)\,\}\,.
\end{myalign}%
There is also an indexed version of $\ypop$, for any $k\in\N$,
\[
    A\ypop[k]B \iff A\ypop B \Land \{\,m\in M_B\,|\,level(m)<k\,\} \ypo M_A\,.
\]
\deq[-1]\edefn[1]%
It is straightforward to see that $\bigsqcup_{i\in\om}A_i$ is well-defined, and that $\ypop$ forms a cpo on $Ob(\GG)$ with least element the empty arena $0$. By $\incl{A,B}$ and $\proj{B,A}$ being an
embedding-projection pair we mean that:
\begin{equation}
    \incl{A,B}\1\proj{B,A}=\id_A \qd\land\qd \proj{B,A}\1\incl{A,B}\ypoo\id_B
\end{equation}
Note that in essence both $\incl{A,B}$ and $\proj{B,A}$ are equal to $\id_A$, the latter seen as a partially defined strategy on prearenas $A\tote B$
and $B\tote A$. Finally, it is easy to show the following.
\[ A\ypop B\ypop C \implies \incl{A,B}\1\incl{B,C} = \incl{A,C} \tag{TRN}\]

\subsection{Innocence: the category \texorpdfstring{$\VV$}{VV}}
In game semantics for pure functional languages (e.g. PCF~\cite{HO:PCF}), the absence of computational effects corresponds to innocence of the strategies.
\parpic[r]{\fbox{ \ \xymatrix@C=5mm@R=2mm{
1\ar[rr] && 1_\bot\otimes \GA[]_i \\
{*} &&& \GOQ\\
        && (\rnode{A1}{*},\rnode{A2}{*}) &\GPA\\
        && \rnode{B}{*}\qd[2] &\GOQ\\
        && \rnode{C}{*^\al}\qd[2] &\GPA\\
        && \qd[2]\rnode{D}{*} &\GOQ\\
        && \qd[2]\rnode{E}{\al} &\GPA
\nccdu{B}{A1}\nccdu{D}{A2}\nccdu{C}{B}\nccdu{E}{D} }}}\noindent
Here, although our aim is to model a language with effects, our model will use innocent strategies: the effects
will still be achieved, by using monads.

Innocence is the condition stipulating that the strategies be completely determined by their behaviour on P-views. In our current setting the manipulation of P-views presents some difficulties, since P-views of plays need not be plays themselves. For example, the P-view of the play on the side (where curved lines represent justification pointers) is $*\,(*,*)*\,\al$ and violates (NC2).
Consequently, we need to explicitly impose innocence on plays.

%
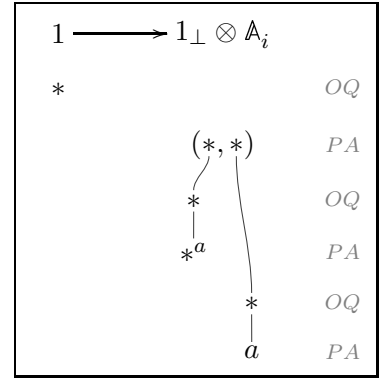
\bdefn%
A legal sequence $s$ is an \boldemph{innocent play} if $s.1$ has empty name-list and $s$ also satisfies the following Name Change Conditions:
\begin{Description}{qqqqq.}
  \item[\,(NC1)\,] The name-list of a P-move $x$ in $s$ contains as a prefix the name-list of the move preceding it. It possibly contains some other names,
  all of which are fresh for $s_{<x}$.
  \item[\,(NC2$'$)\,] Any name in the support of a P-move $x$ in $s$ that is fresh for $\pv{s_{<x}}$ is contained in the name-list of $x$.
  \item[\,(NC3)\,] The name-list of a non-initial O-move in $s$ is that of the P-move justifying it.
\end{Description}
\nada\\[-3mm]
The set of innocent plays of $A$ is denoted by $\iplay{A}$.
\edefn%
It is not difficult to show now that a play $s$ is {innocent} iff, for any $t\leq s$, $\pv{t}$ is a play.
We can obtain the following characterisation of {name-introduction} in innocent plays.%
\beprop[Name-introduction]\label{p:name-intro}%
Let $s$ be an innocent play. A name $\al$ is introduced by Player in $s$ iff there exists a P-move $x$ in $s$ such that $\al\in\supp(x)$ and
$\al\4\pv{s_{<x}}$.%
\enprop%
\proof If $\al$ is introduced by a P-move $x$ in $s$ then $\al\in\nlist(x)$ and $\al\4\nlist(s_{<x}.\arnt1)$, hence, by (NC1), $\al\4s_{<x}$ so
$\al\4\pv{s_{<x}}$. Conversely, if $\al\in\supp(x)$ and $\al\4\pv{s_{<x}}$ then, by (NC2$'$), $\al\in\nlist(x)$, while $\al\4\pv{s_{<x}}$ implies
$\al\4\nlist(s_{<x}.\arnt1)$. \qed\noindent%
%
%
%
Innocent plays are closed under composition (proof omitted, v.~\cite{Tze_PhD}).
\beprop\label{p:innplays}%
If $s\in\play{A\tote B}$\,, $t\in\play{B\tote C}$ are innocent and $s\asymp t$ then $s\cmp t$ is innocent.%
\qed
\enprop%
%
We now move on to innocent strategies and show some basic properties.%
\bdefn%
A strategy $\sig$ is an \boldemph{innocent strategy} if $[s]\in\sig$ implies that $s$ is innocent, and if even-length $[s_1x_1]\in\sig$ and odd-length
$[s_2]\in\sig$ have $[\pv{s_1}]=[\pv{s_2}]$ then there exists $x_2$ such that $[s_2x_2]\in\sig$ and
$[\pv{s_1x_1}]=[\pv{s_2x_2}]$.%
\edefn%
%
\belem\label{l:innstrat}\label{NomGames:l:innstrat}
Let $\sig$ be an innocent strategy. 
\begin{cEnumerate}[\normalfont(\arabic{enumi})\,]{aae}
\item If $[s]\in\sig$ then $[\pv{s}]\in\sig$.
\item If $sy$ is an even-length innocent play and $[s],[\pv{sy}]\in\sig$ then $[sy]\in\sig$.
\item If $\pv{sy}$ is even-length with $\nlist(y)=\nlist(s.\arnt1)$ and $[s],[\pv{sy}]\in\sig$ then $[sy]\in\sig$.
\item If $s$ is an even-length innocent play and, for any $s'\leq^{\text{even}}s$, $[\pv{s'}]\in\sig$ then $[s]\in\sig$.
\end{cEnumerate}
\enlem
\proof%
For (1) we do induction on $|s|$. The base case is trivial. Now, if $s=s'y$ with $y$ a P-move then $\pv{s}=\pv{s'}y$ and
$[\pv{s'}]\in\sig$ by prefix closure and IH. By innocence, there exists $y'$ such that $[\pv{s'}y']\in\sig$ and $[\pv{s'}y']=[\pv{sy}]$, so done. If
$s=s_1ys_2x$ and $x$ an O-move  justified by $y$ then $[\pv{s_1y}]\in\sig$ by prefix closure and IH, hence $[\pv{s_1y}x]\in\sig$ by
contingency completeness.

For (2) note that by innocence we have $[sy']\in\sig$ for some $y'$ such that $[\pv{sy}]=[\pv{sy'}]$. Then,
\[ [\pv{s},y]=[\pv{s},y']\land[\pv{s},s]=[\pv{s},s]\land(\supp(y)\setminus\supp(\pv{s}))\cap\supp(s)= (\supp(y')\setminus\supp(\pv{s}))\cap\supp(s)=\keno\,.\]
Thus we can apply the strong support lemma and get $[sy]=[sy']$, as
required.

For (3) it suffices to show that $sy$ is an innocent play. As
$s,\pv{s}y$ are plays, it suffices to show that $sy$ satisfies the
name conditions at $y$. (NC3) and (NC2$'$) hold because $\pv{sy}$ a
play. (NC1) also holds, as $y$ is non-introducing.

For (4) we do induction on $|s|$. The base case is encompassed in $\pv{s}=s$, which is trivial. For the inductive step, let $s=s^-x$ with $\pv{s}\neq
s$. By IH and contingency completeness we have $[s^-]\in\sig$, and since $[\pv{s}]\in\sig$, by (2), $[s]\in\sig$. \qed\noindent%
We can now show that innocent strategies are closed under composition (details in~\cite{Tze_PhD}). %
\beprop If $\sig:A\tote B,\tau:B\tote C$ are innocent strategies then so is $\sig\cmp\tau$. \qed\enprop
\bdefn $\VV$ is the lluf subcategory of $\GG$ of innocent strategies. \edefn%
Henceforth, when we consider plays and strategies we presuppose them being innocent.

\paragraph{\it Viewfunctions}
We argued previously that innocent strategies are specified by their behaviour on P-views. We formalise this argument by representing
innocent strategies by \emph{viewfunctions}.
\bdefn Let $A$ be a prearena. A \boldemph{viewfunction} $f$ on $A$ is a set of equivalence classes of innocent plays of $A$ which are even-length
P-views, satisfying:
\begin{DFNitemize}
  \item[ \bf Even-prefix closure:] If $[s]\in f$ and $t$ is an even-length prefix of $s$ then $[t]\in f$.
  \item[ \bf Single-valuedness:] If $[s_1x_1],[s_2x_2]\in f$ and $[s_1]=[s_2]$ then $[s_1x_1]=[s_2x_2]$.
\end{DFNitemize}
Let $\sig$ be an innocent strategy and let $f$ be a viewfunction. Then, we can define a corresponding viewfunction and a strategy by:
\begin{align*}
    \viewf(\sig) &\defn\{\,[s]\in\sig\,|\,|s|\text{ even}\land \pv{s}=s\,\}\,, \\
    \strat(f)    &\defn\bigcup\nolimits_n\strat_n(f)\,,
\end{align*}%
where $\strat_0(f)\defn\{[\ee]\}$ and:
\begin{leftalign}
\strat_{2n+1}(f)&\defn \{\,[sx]\,|\,sx\in\iplay{A}\land [s]\in\strat_{2n}(f)\,\}\,,\\
\strat_{2n+2}(f)&\defn \{\,[sy]\,|\,sy\in\iplay{A}\land [s]\in\strat_{2n+1}(f)\land [\pv{sy}]\in f\,\}\,.
\end{leftalign}
\deq[-1]
\edefn[q]%
Note in the above definition that, for any even-length $s$, $[s]\in\strat(f)$ implies $[\pv{s}]\in f$. We can show that the conversion functions are well-defined inverses.
\beprop%
For any innocent strategy $\sig$, $\viewf(\sig)$ is a viewfunction. Conversely, for any viewfunction $f$, $\strat(f)$ is an innocent strategy.
Moreover,
\[ f=\viewf(\strat(f)) \Land \sig=\strat(\viewf(\sig))\,. \]
\myqed \enprop%
Recall the subset ordering $\ypoo$ of strategies given in definition~\ref{d:Enrich}. It is easy to see that the ordering induces a cpo on innocent strategies and that $\VV$ is Cpo-enriched. We can also show the following.
\becor\label{cor:Venriched}\label{c:viewfstrat}%
For all viewfunctions $f,g$ and innocent strategies $\sig,\tau$,
\begin{cEnumerate}[\rm(\arabic{enumi})\,]{123}
  \item $f\ypo\strat(f)$\,,
  \item $\sig\ypo\tau \iff \viewf(\sig)\ypo\viewf(\tau)\,,\qd f\ypo g\iff \strat(f)\ypo\strat(g)$\,,
  \item $\viewf(\sig)\ypo\tau\Land\viewf(\tau)\ypo\sig\implies \sig=\tau$\,.
\end{cEnumerate}
Moreover, $\ypoo$ yields a cpo on viewfunctions, and $\viewf$ and $\strat$ are continuous with respect to $\ypoo$. \qed
\encor

\benotn[Diagrams of viewfunctions]\label{not:Diagr}
We saw previously that innocent strategies can be represented by their viewfunctions.
A viewfunction is a set of (equivalence classes of) plays, so the formal way to express such a construction is explicitly as a set. For example,
we have that
\[ \viewf(\id_A) = \{\,[sm\indx{1}m\indx{2}]\,|\,[s]\in\viewf(\id_A)\land (m\in I_A\lor(s.\arnt1\vdash_A m\indx{1}\land s.\arnt2\vdash_A m\indx{2}))\,\}\,. \]
The above behaviour is called \boldemph{copycat} (v.~\cite{Abramsky_Jagadeesan:MLL}) and is perhaps the most focal notion in game semantics.

A more convenient way to express viewfunctions is by means of diagrams. For example, for $\id_A$ we can have the following depiction.
\[
\xymatrix@R=0pt@C=8pt{\str[5mm]{\id_A:\quad}
A\ar[rr] && A \\
\rnode{A}{i_A}      &&&\GOQ \\
&& \rnode{B}{i_A}    &\GPA\\
\cc{A}{B}}
\]
The polygonal line in the above depiction stands for a \boldemph{copycat link},
meaning that the strategy copycats between the two $i_A$'s. A more advanced example of this notation is the strategy in the middle below.
{\small\[
\begin{array}{c l} A\impl B \\\cline{1-2}
\rnode{A}{*} & \GPA\\
\rnode{Z}{i_A}\qd[3] & \GOQ\\
\rnode{EXX}{\pspolygon[fillstyle=solid,fillcolor=white](0,0)(-0.4,-1.5)(0.4,-1.5)(0,0)}{\rput[bl]{0}(-0.2,-1.4){A^-}}
\qd[3] \rnode{B}{\pspolygon[fillstyle=solid,fillcolor=white](0,0)(-0.4,-1.5)(0.4,-1.5)}{\rput[bl]{0}(-0.2,-1.4){B}}
\nccdu{Z}{A}\nccdu{B}{Z}\nccdu{EXX}{Z}
\\\\\\\\\\\\\\\\\\\\
\end{array}\qd
\begin{array}{|c|c}
\xymatrix@R=0pt@C=4pt{\str[17mm]{h_{A,B}:}
(A\impl B) \ten A \ar[rrr] &&& B_\bot \\
\qd[3]\rnode{A}{(\rnode{A1}{*},\rnode{A2}{i_A})}    &&&&\GOQ\\
&&& \rnode{B}{*}                                    &\GPA\\
&&& \rnode{C}{*}                                    &\GOQ\\
\xinode[i_A]{D}\qd[3]                               &&&&\GPQ\\\\
\nccdu{D}{A1}\nccdu{C}{B}\xiccl[10pt]{D}{A2}\xxiccl{D}{C}}
&
\xymatrix@R=2pt@C=4pt{\str[17mm]{h_{A,B}:}
(A\impl B) \ten A \ar[rrr] &&& B_\bot \\
\qd[3]\rnode{A}{(\rnode{A1}{*},\rnode{A2}{i_A})}    &&&&\GOQ \\
&&& \rnode{B}{*}                                    &\GPA\\
&&& \rnode{C}{*}                                    &\GOQ\\
\rnode{D}{i_A}\qd[3]                                &&&&\GPQ\\
\biglinehere{-1.2,0}{3.4,0}\\
\rnode{E}{j_A}\qd[4]                                &&&&\GOQ\\
\qd[3]\rnode{F}{j_A}                                &&&&\GPQ\\
\biglinehere{-1.2,0}{3.4,0}\\
\qd\rnode{G}{i_B}                                   &&&&\GOA\\
&&& \rnode{H}{i_B}                                  &\GPA
\nccdu{D}{A1}\nccdu{C}{B}\nccdu{E}{D}\nccdu{F}{A2}\nccdu{G}{D}\nccdu{H}{C}\cc[.75mm]{G}{H}\cc[.75mm]{F}{E}}
\end{array}
\]}%
Note first that curved lines (and also the line connecting the two $*$'s) stand for justification pointers. Moreover, recall that the arena $A\impl B$ has the form given on the left above, so the leftmost $i_A$ (l-$i_A$) in the diagram of $h_{A,B}$ has two child components, $A^-$ and $B$. Then, the copycat links starting from the l-$i_A$ have the following meaning. $h_{A,B}$ copycats between the $A^-$-component of $\text{l-}i_A$ and the other $i_A$, and copycats also between the $B$-component of l-$i_A$ and the lower $*$. That is (modulo prefix-closure),
\[
h_{A,B}\defn\strat\{\,[\,(*,i_A)\,*\,*\,i_A\,s\,]\,|\,[\,i_A\,i_A\,s\,]\in\viewf(\id_A)\lor [s]\in\viewf(\id_B)\,\}\,.
\]
Another way to depict $h_{A,B}$ is by cases with regard to Opponent's next move after l-$i_A$, as seen on the right diagram above.

Finally, we will sometimes label copycat links by strategies (e.g.~in the proof of proposition~\ref{p:exponentials}). Labelling a copycat link by a strategy $\sigma$ means that the specified strategy plays like $\sigma$ between the linked moves, instead of doing copycat. In this sense, ordinary copycat links can be seen as links labelled with identities.
\ennotn

\subsection{Totality: the category \texorpdfstring{$\Vt$}{Vt}}
We introduce the notion of total strategies, specifying those strategies which
immediately answer initial questions without introducing fresh names. We extend this type of reasoning level-1 moves, yielding several subclasses of
innocent strategies. Note that an arena $A$ is \boldemph{pointed} if $I_A$ is singleton.
\bdefn\label{NomGames:d:tl4etal} %
An innocent strategy $\sig:A\paei B$ is \boldemph{total} if for any $[i_A]\in\sig$ there exists $[i_A\, i_B]\in\sig$.
A total strategy $\sig:A\paei B$ is:
\begin{DFNitemize}
\item \boldemph{l4} if whenever $[s]\in\sig$ and $\ul{s}.\arnt1\in J_A$ then $\trnn{\pv{s}}=4$,
\item \boldemph{t4} if for any $[i_A\, i_B\, j_B]\in\sig$ there exists $[i_A\, i_B\, j_B\, j_A^{\bee}]\in\sig$,
\item \boldemph{tl4} if it is both t4 and l4,
\item \boldemph{ttotal} if it is tl4 and for any $[i_A\, i_{B}\, j_{B}]\in\sig$ there exists $[i_A\, i_{B}\, j_{B}\, j_A]\in\sig$.
\end{DFNitemize}
A total strategy $\tau:C\ten A\paei B$ is:
\begin{DFNitemize}
\item \boldemph{l4*} if whenever $[s]\in\tau$ and $\ul{s}.\arnt1\in J_A$ then $\trnn{\pv{s}}=4$,
\item \boldemph{t4*} if for any $[\,(i_C,i_A) i_B\, j_B]\in\tau$ there exists
    $[\,(i_C,i_A) i_B\, j_B\, j_A^{\bee}]\in\tau$,
\item \boldemph{tl4*} if it is both t4* and l4*.
\end{DFNitemize}
We let $\Vt$ be the lluf subcategory of $\VV$ of total strategies, and $\Vtt$ its lluf subcategory of ttotal strategies.
$\Vts$ and $\Vtts$ are the full subcategories of $\Vt$ and $\Vtt$ respectively containing pointed arenas.
\edefn%
The above subclasses of strategies will be demystified in the sequel. For now, we show a technical lemma. Let us define, for each arena $A$, the diagonal strategy $\Delta_A$ as follows.
\begin{equation}\label{e:Delta}
  \Delta_A:A\paei A\ten A\defn\strat\{\,[\,i_A\,(i_A,i_A)\, s\,]\,|\,[\,i_A\,i_A\,\,s\,]\in\viewf(\id_A)\,\}
\end{equation}
\belem[Separation of Head Occurrence]\label{l:SepHead}\label{NomGames:l:SepHead}%
Let $A$ be a pointed arena and let $f:A\paei B$ be a t4 strategy.
There exists a tl4* strategy $\tilde{f}:A\ten A\paei B$ such that $f=\Delta\1\tilde{f}$.
\enlem %
\proof Let us tag the two copies of $A$ in $A\otimes A$ as $A\indx1$ and $A\indx2$, and take
\[ \tilde{f}\defn\strat\{\,[\,(i_A,i_A) i_B\, j_B\, j_{A\indx{2}}^{\bee}s\,]\,|\,
    [\,i_A\, i_B\, j_B\, j_{A\indx2}^{\bee}s\,]\,\tilde\in\,\viewf(f)\Land \forall i.\,\ul{s}.i\notin J_{A\indx2}\,\}\,, \]
where $\tilde\in$ is the composition of de-indexing from $M_{A\indx1}$ and $M_{A\indx2}$ to $M_A$ with $\in$.
Intuitively, $\tilde f$ plays the first $J_A$-move of $f$ in $A\indx2$, and then mimics $f$ until the next $J_A$-move of $f$, which is played in
$A\indx1$. All subsequent $J_A$-moves are also played in $A\indx1$. Clearly, $\tilde f$ is tl4* and $f=\De\1\tilde f$. %
\qed\noindent%
We proceed to examine $\Vt$. Eventually, we will see that it contains finite products and that it contains \emph{some} exponentials, and that lifting promotes to a functor.
\paragraph{\it Lifting and product}
We first promote the lifting and tensor arena-constructions to functors. In the following definition recall $\LL$ from notation~\ref{n:plays} and note that we write $\LL(m)\4 m'$ for $\LL(m)\cap\supp(m')=\keno$.
\bdefn%
Let $f:A\paei A'$, $g:B\paei B'$  in $\Vt$. 
Define the arrows
\begin{align*}
  f\ten g
&\defn\strat\{\,[\,(i_A,i_B)\,(i_{A'},i_{B'})\, s\,]\,|\,\\
&\phantom{{}\defn\strat\{\,}(\,[\,i_A\, i_{A'}\,
  s\,]\in\viewf(f)\land[i_B\,i_{B'}]\in g\land
  \LL(i_A\, i_{A'}\, s)\4i_B)\\
&\phantom{{}\defn\strat\{\,}\lor(\,[\,i_B\, i_{B'}\, s\,]\in\viewf(g)\land [i_A\, i_{A'}]\in f\land \LL(i_B\, i_{B'}\, s)\4i_A)\,\}\,,\\
f_\bot&\defn\strat\{\,[* *' *' * s]\,|\,[s]\in\viewf(f)\,\}\,,
\end{align*}
of types $f\ten g:A\ten B\paei A'\ten B'$ and $f_\bot:A_\bot\paei A'_\bot$.
\edefn%
Let us give an informal description of the above constructions:
\begin{aDescription}[$\bullet$]{}
\item $f_\bot:A_\bot\paei A_\bot'$ initially plays a sequence of asterisks $[* *' *' *]$ and then continues playing like $f$.
\item $f\ten g:A\ten B\paei A'\ten B'$ answers initial moves $[\,(i_A,i_B)\,]$ with $f$'s answer to $[i_A]$ and $g$'s answer to
    $[i_B]$. Then, according to whether Opponent plays in $J_{A'}$ or in $J_{B'}$\,, Player plays like $f$ or like $g$ respectively.
\end{aDescription}
Note that $f_\bot$ is always ttotal. We can show the following.
\beprop%
$\uscore\ten\uscore:\Vt\times\Vt\tote\Vt$ and $(\uscore)_\bot:\Vt\tote\Vtts$ are functors.
\qed\enprop
%
%
%
Moreover, $\ten$ yields products and hence $\Vt$ is cartesian.
\beprop\label{p:products}\label{NomGames:p:products}%
$\Vt$ is cartesian: $1$ is a terminal object and $\ten$ is a product constructor. \enprop %
\proof %
Terminality of $1$ is clear. Moreover, it is straightforward to see that $\ten$ yields a symmetric monoidal structure on $\Vt$\,, with its unit being
$1$ and its associativity, left-unit, right-unit and symmetry isomorphisms being the canonical ones. Hence, it suffices to show that there exists a
natural coherent diagonal, that is, a natural transformation $\Delta:Id_{\Vt}\paei\otimes\circ\ang{Id_{\Vt},Id_{\Vt}}$ (where $\ang{Id_{\Vt},Id_{\Vt}}$ is the diagonal functor on $\Vt$) such that the following diagrams commute for any $A,B$ in $\Vt$.
\[
\xymatrix@C=2cm{A\ten B\ar[dr]_{\Delta_{A\ten B}}\ar[r]^-{\Delta_A\ten\Delta_B} & (A\ten \rnode{1}{A})\ten(\rnode{2}{B}\ten B)\\
    & (A\ten \rnode{3}{B})\ten(\rnode{4}{A}\ten B) \nccdu[,linecolor=lightgray]{3}{2}\nccdu[,linecolor=lightgray]{4}{1} \ar@{<-}[u]_{\isom}}
\qd[3]
\xymatrix@C=13mm{& A\ar[dl]_{\cong}\ar[d]^{\Delta_A}\ar[dr]^{\cong} \\ 1\ten A & A\ten A\ar[l]^{!_A\ten\id_A}\ar[r]_{\id_A\ten!_A} & A\ten1}
\]
But it is easy to see that the diagonal of~\eqref{e:Delta}
makes the above diagrams commute. Naturality follows from the single-threaded nature of strategies (v.~\cite{Harmer_PhD}).
\qed\noindent%
Products are concretely given by triples $A\Rnza{\pi_1}A\ten B\Lnza{\pi_2}B$, where
\[ \pi_1=\strat\{\,[\,(i_A,i_B)\, i_A\,s\,]\,|\,[i_A\,\,i_A\,\,s]\in\viewf(\id_A)\,\} \]
and $\pi_2$ similarly, while for each $A\Rnza{f}C\Lnza{g}B$ we have
\begin{align*}
\ang{f,g}:C\paei A\ten B=\strat\{\,&[\,i_C\,(i_A,i_B)\, s\,]\,|\,\\
&(\,[i_C\,\,i_A\,\,s]\in\viewf(f)\land[i_C\,\,i_B]\in\viewf(g)\,) \\
    &\lor(\,[i_C\,\,i_A]\in\viewf(f)\land[i_C\,\,i_B\,\,s]\in\viewf(g)\,)\,\}\,.
\end{align*}%
Finally, we want to generalise the tensor product to a version applicable to countably many arguments. In arenas, the construction comprises of gluing
countably many arenas together at their initial moves. The problem that arises then is that the product of infinitely many (initial) moves need not
have finite support, breaking the arena specifications. Nevertheless, in case we are interested only in pointed arenas, this is easily
bypassed: a pointed arena has a unique initial move, which is therefore equivariant, and the product of equivariant moves is of course also
equivariant.
\bepropdefn For pointed arenas $\{A_i\}_{i\in\om}$ define $\bigotimes_{i}A_i$ by:
\begin{align*}
M_{\Ten{A}}&\defn\{*\}+\biguplus\nolimits_{i}\, \BI{A_i}\,, &
\la_{\Ten{A}}&\defn[\,(*\mapsto PA),[\la_{A_i}\,^{i\in\om}]]\,, \\
I_{\Ten{A}}&\defn\{*\}\,, &
\vdash_{\Ten{A}}&\defn\{(\dagger,*)\}\cup\{\,(*,j_{A_i})\,|\,i\in\om\,\}\cup\bigcup\nolimits_{i}(\vdash_{A_i}\hrp\BI{A_i}\,^2)\,.
\end{align*}
For $\{f_i:A_i\tote B_i\}_{i\in\om}$ with $A_i$'s and $B_i$'s pointed define:
\[ \Ten{f}\defn\strat\{\,[*\, *\, s]\,|\,\exists k.\,[i_{A_k}\, i_{B_k}\, s]\in \viewf(f_k)\,\}\,. \]
Then, \ $\bigotimes\_:\prod\Vts\paei\Vts$ \ is a functor. \qed%
\enpropdefn%
In fact, we could proceed and show that the aforedefined tensor yields general products of pointed objects, but this will not be of use here.

\paragraph{\it Partial exponentials}
We saw that $\Vt$ has products, given by the tensor functor $\ten$. We now show that the arrow constructor yields appropriate partial
exponentials, which will be sufficient for our modelling tasks.

Let us introduce the following transformations on strategies.

\bdefn\label{d:exponentials} For all arenas $A,B,C$ with $C$ pointed, define a bijection
\[ \La^B_{A,C}:\Vt(A\ten B,C)\lred{\isom}\Vt(A,B\him C) \]
by taking, for each $h:A\ten B\paei C$ and $g:A\paei B\him C$\,,\footnote{Note the reassignment of pointers that takes place implicitly in the definitions of $\La,\La^{-1}$, in order e.g.~for $(i_A,i_B)\,i_C\,j_C\,s$ to be a play of $\viewf(h)$.}
\begin{align*}
\La^B_{A,C}(h):A\paei B\him C &\defn \strat\{\,[i_A\,i_C\,(i_B,j_C)\, s]\,|\,[\,(i_A,i_B)\,i_C\,j_C\,s\,]\in\viewf(h)\,\}\,, \\[5pt]
\La^{B\ -1}_{A,C}(g):A\ten B\paei C &\defn \strat\{\,[\,(i_A,i_B)\,\,i_C\,\,j_C\,\,s\,]\,|\,[i_A\,\,i_C\,\,(i_B,j_C)\, s]\in\viewf(g)\,\}\,.
\end{align*}
For each $(f,g):(A,B)\paei(A',B')$, define the arrows
\begin{align*}
  \ev_{A,B}&:(A\him B)\ten A\paei B\defn \La^{A\ -1}_{A\him B,B}(\id_{A\him B})\,, \\
  f\him g&:A'\him B\paei A\him B'\defn \La^{A'}_{A\him B,A'\him B'}(\id \ten f\1\ev \1g)\,.
\end{align*}
\mindeq[-1]\edefn[1]%
It is not difficult to see that $\La$ and $\La^{-1}$ are well-defined and mutual inverses. What is more, they supply us with exponentials.

\beprop\label{p:exponentials}%
$\Vt$ has partial exponentials wrt to $\ten$, in the following sense. For any object $B$, the functor $\uscore\ten B:\Vt\paei\Vt$ has a partial right
adjoint $B\him\uscore:\Vts\paei\Vt$, that is, for any object $A$ and any pointed object $C$ the bijection
$\La^B_{A,C}$ is natural in $A$.
\enprop%
\parbox{.53\linewidth}{%
\proof It suffices to show that, for any\linebreak $f:A\ten B\paei C$ and $g:A\paei B\him C$,
\\[1.5mm]
\centerline{$\La(f)\ten\id\1\ev = f\,,\qd g\ten\id\1\ev=\La^{-1}(g)$\,.}
\\[1.52mm]
These equalities are straightforward. For example, the viewfunction of $\La(f)\ten\id\1\ev$ is given by the diagram on the side, which also gives the viewfunction of $f$. \qed}%
\parbox{.45\linewidth}{\small\qd\ 
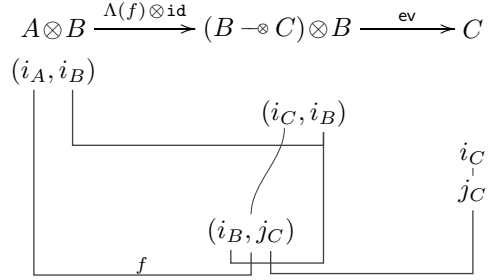
\xymatrix@C=12mm@R=0mm{
A\ten B\ar[r]^-{\La(f)\ten\id} & (B\him C)\ten B\ar[r]^-{\ev} & C \\
\rnode{A1}{(\rnode{A1a}{i_{A}},\rnode{A1b}{i_B})} \\
    & \qd[2]\rnode{B2}{(\rnode{B2a}{i_C},\rnode{B2b}{i_B})} \\
    &   & \rnode{C3}{i_C} &  \\
    &   & \dunode{C6}{j_C}{C3} \\
    &   \dunode{B7}{(\rnode{B7a}{i_B},\rnode{B7b}{j_C})}{B2a}\qd[2]
\cc{A1b}{B2b} \cc{B2b}{B7a} \cc[3mm]{C6}{B7b} \cc[3mm]{A1a}{B7}\bput[-2.6mm]{:U}{^f} \\\\\\ }}
\\%
A consequence of partial exponentiation is that $\him$ naturally upgrades to a functor:
\[ \uscore\him\uscore:(\Vt)^{\mathrm{op}}\times\Vts\paei\Vt\,. \]
Now, in case $g$ is ttotal, the strategy $f\him g:A'\him B\paei A\him B'$ is given concretely by $\strat(\phi)$, where
\begin{align*}
  \phi
&=\{\,[i_B\, i_{B'}\,(i_{A},j_{B'})\,(i_{A'},j_B)\, s]\,|\,\\
&\phantom{{}=\{\,}([i_{A}\, i_{A'}\, s]\in\viewf(f)\land[i_B\, i_{B'}\, j_{B'}\, j_{B}]\in g \land \LL(i_{A}\, i_{A'}\, s)\#i_B,j_{B'})\\
&\phantom{{}=\{\,}\,\lor\,([i_B\, i_{B'}\, j_{B'}\, j_{B}\, s]\in\viewf(g)\,\land\,[i_{A}\, i_{A'}]\in f\,\land\,\LL(i_B\, i_{B'}\, j_{B'}\, j_{B}\, s)\#i_{A})\,\}.
\end{align*}
That is, $f\him g$ answers initial moves $[i_B]$ like $g$ and then responds to $[i_B\, i_{B'}\,(i_{A},j_{B'})\,]$ with $f$'s answer to $[i_{A}]$ and
$g$'s response to $[i_B\, i_{B'}\, j_{B'}]$ (recall $g$ ttotal). It then plays like $f$ or like $g$, according to Opponent's next move. Note that $\phi$ is a viewfunction even if $B,B'$ are not pointed.

A special case of ttotality in the second argument arises in the defined functor:
\begin{equation}
\uscore\impl\uscore:(\Vt)^\mathrm{op}\times\Vt\paei\Vtts\defn \uscore\him(\uscore)_\bot\,.
\end{equation}

\berem In the work on CBV games of Honda \& Yoshida~\cite{Honda:CBV} the following version of partial exponentiation is shown.
\begin{equation}\label{e:HO_expon}
\VV(A\ten B,C) \cong \Vt(A,B\impl C)
\end{equation}
Interestingly, that version can be derived from ours (using also another bijection shown in~\cite{Honda:CBV}),
\[ \VV(A\ten B,C) \overset{}{\cong} \Vt(A\ten B,C_\bot) \overset{}{\cong} \Vt(A,B\him C_\bot) = \Vt(A,B\impl C)\,. \]
But also vice versa, if $C$ is pointed then $C\cong C_2\impl C_1$, for some arenas $C_1,C_2$,\footnote{%
~In fact, for $C$ to be expressed as $C_2\impl C_1$ we need a stronger version of condition~(f), namely:
\begin{aDescription}{asd}
  \item[\,(f')\,] For each $m\in M_A$, there exists unique $k\geq0$ and a unique sequence $x_1\dots x_n\in\{Q,A\}^*$ such that
  $I_A\ni m_1\vdash_A\cdots\vdash_A m_k\vdash_A m$\,, for some $m_l$'s in $M_A$ with $\la_C^{QA}(m_l)=x_l$\,.
\end{aDescription}
In such a case, $C_1$ and $C_2$ are given by taking \ $K_C^A\defn \{\,m\in M_C\,|\,\exists j_C.\,j_C\vdash_C m\land \la_C(m)=PA\,\}$ \ and
\begin{align*}
  M_{C_1} &\!\defn\! K_C^A + \{\,m\in M_C\,|\, \exists k\in K_C^A.\, k\vdash_C\cdots\vdash_C m\,\} &
  \!\!I_{C_1} &\!\defn\! K_C^A &
  \!\vdash_{C_1} &\!\defn\!\,\vdash_C\restriction (M_{C_1}\times\BI{C_1}) &
  \!\!\la_{C_1} &\!\defn\! \la_C\!\restriction\! M_{C_1} \\
  M_{C_2} &\!\defn\! \BI{C}\plhn M_{C_1}  \qd \la_{C_2} \!\defn\! [i_{C_2}\mapsto PA, m\mapsto\bar{\la}_{C}(m)\,]
  &\!\!I_{C_2} &\!\defn\! J_C
  &\!\vdash_{C_2} &\!\defn\!\,\vdash_C\restriction (M_{C_2}\times\BI{C_2})\,.
\end{align*}} %
and
\[ \Vt(A\ten B,C_2\impl C_1) \overset{\eqref{e:HO_expon}}{\cong} \VV(A\ten B\ten C_2,C_1) \overset{\eqref{e:HO_expon}}{\cong}
    \Vt(A,(B\ten{C_2})\impl C_1) = \Vt(A,B\him({C_2}\impl C_1))\,. \]
\enrem

\paragraph{\it Strategy and arena orders}
Recall the orders defined for strategies ($\ypoo$) and arenas ($\ypop$) in section~\ref{sec:Orderings_G}. These being subset orderings are
automatically inherited by $\Vt$. Moreover, by use of corollary~\ref{cor:Venriched} we can easily show that the aforedefined functors are continuous.
Note that, although the strategy order $\ypoo$ is inherited from $\VV$, the least element (the empty strategy) is lost, as it is not total.
\beprop\label{p:unionofstrat}\label{NomGames:p:unionofstrat}
$\Vt$ and $\Vtt$ are PreCpo-enriched wrt $\ypoo$.\footnote{%
~By precpo we mean a cpo which may not have a least element. PreCpo is the category of precpos and continuous functions.} Moreover,
\[ \begin{myaligned}
\botf &: \Vt\paei\Vtts\,,  & \tenf &: \Vt\times\Vt\paei\Vt\,,  & \bigtenf&:\prod\Vt\!_{*}\paei\Vt\!_{*}\,, \\
\himf &: \Vt[\mathrm{op}]\times\Vtts\paei\Vtts\,,  & \implf&: \Vt[\mathrm{op}]\times\Vt\paei\Vtts
\end{myaligned} \]
are locally continuous functors. \qed%
\enprop
%
The order of arenas in $\Vt$ is the same as in $\GG$, and therefore $Ob(\Vt)$ is a cpo with least element $0$. Note that $A\ypop B$ does not imply that the corresponding projection is a total strategy\HY but $A\ypop[1]B$ does imply it. In fact,
\[
    A\ypop[1]B \implies \proj{B,A}\in\Vtt(B,A)\qd\land\qd
    A\ypop[2]B \implies \incl{A,B}\in\Vtt(A,B)\,.
\]%
Moreover, we have the following.
\beprop\label{p:CpoObj}\label{NomGames:p:CpoObj} All of the functors of proposition~\ref{p:unionofstrat} are continuous wrt $\ypop$\,. Moreover,
\begin{align*}
A\ypop A'\land B\ypop B' &\implies \incl{A,A'}\tenn\incl{B,B'}=\incl{A\tenn B,A'\tenn B'} \\
A\ypop[1]A'\land B\ypop[1] B' &\implies \proj{A',A}\tenn\proj{B',B}=\proj{A'\tenn B',A\tenn B} \\
\forall i\in\om.\, A_i\ypop A'_i &\implies \bigotimes\nolimits_i\incl{A_i,A'_i} = \incl{\Ten{A},\Ten{A'}} \\
\forall i\in\om.\, A_i\ypop A'_i &\implies \bigotimes\nolimits_i\proj{A'_i,A_i} = \proj{\Ten{A'},\Ten{A}} \\
A\ypop[1]A'\land B\ypop B' &\implies \proj{A',A}\impl\incl{B,B'}=\incl{A\impl B,A'\impl B'} \\
A\ypop A'\land B\ypop[1]B' &\implies \incl{A,A'}\impl\proj{B',B}=\proj{A'\impl B',A\impl B} \\
A\ypop[1]A'\land B\ypop[2]B' &\implies \proj{A',A}\him\incl{B,B'}=\incl{A\him B,A'\him B'} \\
A\ypop A'\land B\ypop[1]B' &\implies \incl{A,A'}\him\proj{B',B}=\proj{A'\him B',A\him B} \,.
\end{align*}
\enprop \proof All the clauses are in effect functoriality statements, since the underlying sets of inclusions and
projections correspond to identity strategies. \qed

\subsection{A monad, and some comonads}
We now proceed to construct a monad and a family of comonads on $\Vt$ that will be of use in later sections. Specifically, we will upgrade lifting to a monad and introduce a family of product comonads for initial state.
\paragraph{\it Lifting monad}
It is a more-or-less standard result that the lifting functor induces a monad.
\bdefn\label{NomGames:d:Lift} %
Define the natural transformations $\upp,\dnn,\stt$ as follows.
\begin{align*}
    \upp[A] &:A\paei A_\bot = \strat\{\,[i_A *_1 *_2\,i_A \,s]\,|\,[i_A\,i_A\,s]\in\viewf(\id_A)\,\} \\[4pt]
\dnn[A] &:A_{\bot\bot}\paei A_\bot \defn \strat\{\,[*_1 *_1' *_2' *_2 *_3 *_4  s]\,|\,[s]\in\viewf(\id_A)\,\} \\[4pt]
\stt[A,B] &:A\ten B_\bot\paei (A\ten B)_\bot \defn \strat\{\,[\,(i_A,*_1) *_1' *_2' *_2\,i_B\,(i_A,i_B)\,s]\\
    &\qd[19]|\,[\,(i_A,i_B)\,(i_A,i_B)\,s]\in\viewf(\id_{A\ten B})\,\}
\end{align*}
{\footnotesize(primed asterisks are used for arenas on the RHS, where necessary).} \edefn
\beprop\label{NomGames:p:Lift} The quadruple $((\uscore)_\bot,\upp,\dnn,\stt)$ is a strong monad on $\Vt$. Moreover, it yields monadic exponentials
by taking $(C_\bot)^B$ to be $B\impl C$, for each $B,C$. \enprop
\proof It is not difficult to see that $((\uscore)_\bot,\upp,\dnn,\stt)$ is a strong monad.
Moreover, for each $B,C$ we have that $B\impl C=B\him C_\bot$ is a $(\uscore)_\bot$-exponential, because of exponentiation properties of $\him$.
\qed\noindent%
Although finding a canonical arrow from $A$ to $A_\bot$ is elementary ($\upp[A]$), finding a canonical arrow in the inverse direction is not always
possible. In some cases, e.g.~$A=\GAA{i}$\,, there is no such arrow at all, let alone canonical. An exception occurs when $A$ is pointed, by setting:
\begin{equation}\label{e:up-pu}
 \puu[A]:A_\bot\paei A\defn \strat\{\,[*\, i_A\,j_A\,*\, i_A\,j_A\,s]\,|\,[i_A\,i_A\,j_A\,j_A\,s]\in\viewf(\id_A)\,\}\,.
\end{equation}%
\belem\label{l:puu}%
$\puu[A]$ yields a natural transformation $\puu:(\uscore)_{\bot(\Vtts)}\paei Id_{\Vtts}$. Moreover, for any arenas $A,B$ with $B$ pointed,
$\upp[A]\1\puu[A]=\id_A$\,,
$\puu[A_\bot] = \dnn[A]$\, and
\[ \puu[A\him B] = \La\Big((A\him B)_\bot\ten A\lred{\stt'}((A\him B)\ten A)_\bot\lred{\ev\!_\bot}B_\bot\lred{\puu[B]}B\Big).\]
\myqed[-1]\enlem%
\paragraph{\it Initial-state comonads}\label{NomGames:c:Comon}
Our way of modelling terms-in-local-state will be by using initial state comonads, in the spirit of intensional program modelling of Brookes \&
Geva~\cite{BrookesGeva:91:ComputationalComonads}. In our setting, the initial state can be any list $\all$ of distinct names; we define a comonad for
each one of those lists.
\bdefn[Initial-state comonads] For each $\all\in\A^\#$ define the triple $(\prn,\eps,\de)$ by taking $\prn:\Vt\paei\Vt \defn\GA\ten\uscore$ \, and
\begin{lefteqn*}\begin{aligned}
    \eps &:\prn\paei Id_{\Vt} \defn\{\,\eps_A:\GA\ten A\lred{\pi_2}A\,\}\,, \\[1mm]
    \de &:\prn\paei(\prn)^2 \defn\{\,\de_A:\GA\ten A\lred{\De\ten\id}\GA\ten\GA\ten A\,\}\,.
\end{aligned}\end{lefteqn*}%
For each $\supp(\all')\ypo\supp(\all)$ define the natural transformation $\pit{\all}{\all'}:\prn\paei\prn[\all']$ by taking
\[    (\pit{\all}{\all'})_A :\GA\ten A\paei\GA[\all']\ten A \defn (\pit{\all}{\all'})_1\ten\id_A\,, \]
where $(\pit{\all}{\all'})_1$ is $\pit{\all}{\all'}$ of definition~\ref{d:verybasicmorphisms}, that is, $(\pit{\all}{\all'})_1\defn \{\,[\,(\all,*)\,(\all',*)\,]\,\}$\,.
\edefn%
Note that $\prn[\ee]$, the comonad for empty initial state, is the identity comonad. Note also that we have suppressed indices $\all$ from
transformations $\eps,\de$ for notational economy.

Clearly, each triple $(\prn,\eps,\de)$ forms a product comonad on $\Vt$. Moreover, it is straightforward to show the following.
\beprop[Chain rule] For each $\supp(\all')\ypo\supp(\all)\in\A^\#$, the transformation $\pit{\all}{\all'}$ is a comonad morphism. Moreover,
$\pit{\all}{\ee}=\eps:\prn\paei Id_{\Vt}$\,, $\pit{\all}{\all}=\id:\prn\paei\prn$ and, for each $\supp(\all')\ypo\supp(\all'')\ypo\supp(\all)$, \ $\pitt{\all}{\all''}\1\pitt{\all''}{\all'} = \pitt{\all}{\all'}$\,.
\qed\enprop%
Finally, for each name-type $i$, we can define a name-test arrow:
\begin{lefteqn}
  \eq[i]:\GAA{i}\ten\GAA{i}\paei\GN\defn\{\,[\,(\al,\al)\,0]\,\}\cup\{\,[\,(\al,\be)\,1]\,|\,\al\neq\be\,\}\,,
\end{lefteqn}%
which clearly makes the (N1) diagram (definition~\ref{d:laNR}) commute.
\paragraph{\it Fresh-name constructors}
Combining the monad and comonads defined previously we can obtain a monadic-comonadic setting $\MODEL[]{\Vt,\Lift,\prn[]}$, where $\prn[]$
denotes the family $\MODEL{\prn}$. This setting, which in fact yields a sound model of the $\nu$-calculus~\cite{AGMOS,Tze_PhD}, will be used as the
basis of our semantics of nominal computation in the sequel. Nominal computation of type $A$, in
name-environment $\all$ and variable-environment $\Gamma$, will be translated into the set of strategies
\[ \{\,\sig:\prn\trn{\Gamma}\paei \trn{A}_\bot\,\}\,. \]
The lifting functor, representing the monadic part of our semantical setting, will therefore incorporate the computational effect of fresh-name
creation.

We describe in this section the semantical expression of fresh-name creation. Fresh names are created by means of natural transformations which
transform a comonad $\prn$, say, to a monad-comonad composite $(\prn[\all\al]\uscore)_\bot$.
\bdefn\label{NomGames:d:new}%
Consider the setting $\MODEL[]{\Vt,\Lift,\prn[]}$. We define natural transformations
$\frs^{\allal}:\prn\paei (\prn[\all\al]\uscore)_\bot$ \, by
\begin{lefteqn*}\begin{aligned}
    \frs^\allal_A&\defn\GA\ten A\lred{\frs_1^\allal\ten\id_A}(\GA[\allal])_\bot\ten A\lred{\stt'}(\GA[\allal]\ten A)_\bot\,, \\[1mm]
    \frs^\allal_1&:\GA\ten1\paei (\GA[\allal]\ten1)_\bot\defn\strat\{\,[\,(\all,*)\,*\,*\,(\all\al,*)^\al]\,\}\,,
\end{aligned}\end{lefteqn*}%
for each $\all\al\in\NAA[\#]$.
\edefn%
That $\frs$ is a natural transformation is straightforward: for any $f:A\paei B$ we can form the following commutative diagram.
\[
\xymatrix@C=17mm{ \GA\ten A\ar[d]_{\id\ten f}\ar[r]^-{\frs_1\ten\id} & (\GA[\all\al])_\bot\ten A\ar[d]_{\id\ten f}\ar[r]^{\stt'}
    & (\GA[\all\al]\ten A)_\bot\ar[d]^{(\id\ten f)_\bot} \\
    \GA\ten B\ar[r]_-{\frs_1\ten\id} & (\GA[\all\al])_\bot\ten B\ar[r]_{\stt'} & (\GA[\all\al]\ten B)_\bot }
\]
Moreover, we can show the following.%
\beprop\label{NomGames:p:N2}%
In the setting $\MODEL[]{\Vt,\Lift,\prn[]}$ with $\frs$ defined as above, the {\rm(N2)} diagrams (definition~\ref{d:laNR}) commute.\qed
\enprop%
%
%
The fresh-name constructor allows us to define name-abstraction on strategies by taking:
\begin{equation}\label{e:abs}
  \abs\sig\defn\prn B\lred{\frs_B^{\all\al}}(\prn[\all\al]B)_\bot\lred{\sig\!_\bot}C_\bot\lred{\puu[C]}C\,.
\end{equation}
Name-abstraction can be given an explicit description as follows. For any sequence of moves-with-names $s$ and any name $\al\4\nlist(\al)$, let $s^{\al}$ be $s$ with $\al$ in the head of all of its name-lists. Then, for $\sig$ as above, we can show that:
\begin{equation}\label{e:abs_expl}
\viewf(\abs\sig)= \{\,[\,(\all,i_B)\,i_C\,j_C\,m^{\al\bee}\,s^\al]\,|\,[\,(\all\al,i_B)\,i_C\,j_C\,m^{\bee}\,s]\in\viewf(\sig)\land\al\4i_B,j_C\,\}
\end{equation}
We end our discussion on fresh-name constructors with a technical lemma stating that name-abstraction and currying commute.

\belem\label{NomGames:l:LambdaAbs}%
Let $f:\prn[\allal](A\ten B)\paei C$ with $C$ a pointed arena. Then,
\[ \abs\La(\zet'\1f)=\La(\zet'\1\abs f):\prn A\paei B\him C\,. \]
\enlem%
\proof
As follows.
\begin{align*}
  \abs\La(\zet'\1f) &= \frs_{A}^{\allal}\1(\La(\zet'\1f))_\bot\1\puu[B\him C]=\frs_{A}^{\allal}\1(\La(\zet'\1f))_\bot\1\La(\stt'\1\ev_\bot\1\puu[C]) \\
    &= \La(\frs_{A}^{\allal}\ten\id_B\1(\La(\zet'\1f))_\bot\ten\id_B\1\stt'\1\ev_\bot\1\puu[C]) \\
    &= \La(\frs_{A}^{\allal}\ten\id_B\1\stt'\1(\La(\zet'\1f)\ten\id_B)_\bot\1\ev_\bot\1\puu[C]) \\
    &= \La(\frs_{A}^{\allal}\ten\id_B\1\stt'\1(\zet'\1f)_\bot\1\puu[C])
    \overset{\text{(N2)}}{=} \La(\zet'\1\frs_{A\ten B}^{\allal}\1f_\bot\1\puu[C])
\end{align*}
and the latter equals $\La(\zet'\1\abs f)$. \qed\noindent%
Note that the above result does \emph{not} imply that $\nu$- and $\lambda$-abstractions commute in our semantics of nominal languages, i.e.~that we obtain identifications of the form $\trn{\new[\al]\la x.M}=\trn{\la x.\new[\al]M}$. As we will see in the sequel, $\lambda$-abstraction is not simply currying, because of the use of monads.

\subsection{Nominal games \`a la Laird} As aforementioned, there have been two independent original presentations of nominal games, one due to Abramsky, Ghica, Murawski, Ong and Stark (AGMOS)~\cite{AGMOS} and another one due to Laird~\cite{Laird:fossacs04,Laird:Names_Pointers}. Although Laird's constructions are are not explicitly based on nominal sets (natural numbers are used instead of atoms), they constitute nominal constructions nonetheless. In this section we highlight the main differences between our nominal games, which follow AGMOS, and those of~\cite{Laird:fossacs04,Laird:Names_Pointers}.

Laird's presentation concerns the $\nu$-calculus with pointers, i.e.~with references to names.
The main difference in his presentation is in the treatment of name-introduction. In particular, a name does not appear in a play at the point of evaluation of its $\nu$-constructor, but rather at the point of its first \emph{use}; let us refer to this condition as \emph{name-frugality} (cf.~\cite{Tze_Murawski:RML}).
An immediate result is that strategies are no longer innocent, as otherwise e.g.~$\nu\al.\la x.\al$ and $\la x.\nu\al.\al$ would have the same denotation.\footnote{Non-innocence can be seen as beneficial in terms of simplicity of the model, since strategies then have one condition less. On the other hand, though, innocent strategies are specified by means of their viewfunctions, which makes their presentation simpler. Moreover, non-innocence diminishes the power of definability results, as finitary behaviours are less expressive in absence of innocence.} More importantly, name-frugality implies that strategies capture the examined nominal language more \emph{accurately}: Opponent is not expected to guess names he is not supposed to know and thus, for example, the denotations of $\nu\al.\sskip$ and $\sskip$ are identical. In our setting, Player is not frugal with his names and therefore the two terms above are identified only at the extensional level (i.e.~after quotienting).\footnote{Note here, though, that the semantics being too explicit about the created names can prove beneficial: here we are able to give a particularly concise proof adequacy for $\nurho$ (see section~\ref{s:Adequacy} and compare e.g.~with respective proof in~\cite{Abramsky+:GamesReferences}) by exploiting precisely this extra information!}

The major difference between~\cite{Laird:fossacs04} and ~\cite{Laird:Names_Pointers} lies in the modelling of (ground-type, name-storing) store. In~\cite{Laird:fossacs04} the store is modelled by attaching to strategies a global, top-level (non-monadic), store arena. Then, a good-store-discipline is imposed on strategies via extra conditions on strategy composition which enforce that hidden store-moves follow the standard read/write pattern. As a result (and in contrast to our model), the model relies heavily on quotienting by the intrinsic preorder in order for the store to work properly.

The added accuracy obtained by using frugality conditions is fully exploited in~\cite{Laird:Names_Pointers}, where a carefully formulated setting of moves-with-store\footnote{Inter alia, frugality of names implies that sequences of moves-with-store have strong support even if stores are represented by sets!} allows for an \emph{explicit characterisation} result, that is, a semantic characterisation of operational equality at the intensional level. The contribution of using moves-with-store in that result is that thus the semantics is relieved from the (too revealing) internal workings of store: for example, terms like $(\al:=\be)\1\la x.\bang\al\1 0$ and $(\al:=\be)\1\la x.0$ are equated semantically at the intensional level, in contrast to what happens in our model.\footnote{In our model they correspond to the strategies (see also section~\ref{S:model}):
\[ \sig_1 \defn\{\,[\,(\al,\be)*\ixi(*,\ixi)(n,\ixi)\,\al\,\ga\, 0]\,\}\,,\qd\sig_2\defn\{\,[\,(\al,\be)*\ixi(*,\ixi)(n,\ixi)\,0]\,\}\,. \]
Thus, the inner-workings of the store revealled by $\sig_1$ (i.e.~the moves $\al\,\ga$) differentiate it from $\sig_2$. In fact, in our attempts to obtain an explicit characterisation result from our model, we found store-related innaccuracies to be the most stubborn ones.} Note, though, that in  a setting with higher-order store such that of $\nurho$, moves-with-store would not be as simple since stores would need to store higher-order values, that is, strategies.

Laird's approach is therefore advantageous in its use of name-frugality conditions, which allow for more accurate models. At the same time, though, frugality conditions are an extra burden in constructing a model: apart from the fact that they need to be dynamically preserved in play-composition by garbage collection, they presuppose an appropriately defined notion of \emph{name-use}. In~\cite{Laird:fossacs04,Laird:Names_Pointers}, a name is considered as used in a play if it is accessible through the store (in a reflexive transitive manner) from a name that has been \emph{explicitly played}. This definition, however, does not directly apply to languages with different nominal effects (e.g.~higher-order store). Moreover, frugality alone is not enough for languages like Reduced ML or the $\nu$-calculus: a name may have been used in a play but may still be inaccessible to some participant (that is, if it is outside his view~\cite{Tze_Murawski:RML}). On the other hand, our approach is advantageous in its simplicity and its applicability on a wide rage of nominal effects (see~\cite{Tze_PhD}), but suffers from the accuracy issues discussed above. 

\section{The nominal games model}\label{S:model}%
\noindent
We embark on the adventure of modelling $\nurho$ in a category of nominal arenas and strategies. Our starting point is the category $\Vt$ of nominal
arenas and total strategies. Recall that $\Vt$ is constructed within the category $\nom$ of nominal sets so, for each type $A$, we have an arena $\GAA{A}$ for references to type $A$.

The semantics is monadic in a \boldemph{store monad} built around a store arena $\xi$,
and co\-mo\-na\-dic in an initial state comonad. The store monad is defined on top of the lifting monad
(see definition~\ref{NomGames:d:Lift}) by use of a side-effect monad constructor, that is,
\[ TA \defn \xi\him(A\otimes\xi)_\bot \qd[2]\text{i.e.}\ TA=\xi\impl A\otimes\xi\,. \]
Now, $\xi$ contains the values assigned to each name (reference), and thus it is of the form
\[ \Tenn[A\in\TY]{(\GAA{A}\impl\trn{A})} \]
where $\trn{A}$ is the translation of each type $A$. Thus, a recursive (wrt type-structure) definition of the type-translation is not possible because of the following cyclicity.
\begin{equation} \label{e:storeEq}\tag{SE}
\begin{aligned}
\trn{A\tote B}&=\trn{A}\him(\xi\impl \trn{B}\otimes\xi)\\
\xi&=\Tenn[A]{(\GAA{A}\impl\trn{A})}
\end{aligned}
\end{equation}
Rather, both $\xi$ and the type-translation have to be \emph{computed} as the least solution to the above domain equation. By the way, observe that $\trn{A\paei B}=\trn{A}\otimes\xi\impl\trn{B}\otimes\xi$\,.
\subsection{Solving the Store Equation}
The full form of the store equation~\eqref{e:storeEq} is:
\[ \begin{array}{c@{\qd[2]}c@{\qd[2]}c@{\qd[2]}c}
\trn{\ena}=1\,, & \trn{\N}=\GN\,, & \trn{[A]}=\GAA{A}\,, & \trn{A\tote B}=\trn{A}\ten\trn{B}\,, \\[5pt]
\multicolumn{3}{c}{\trn{A\tote B}=\trn{A}\him(\xi\impl \trn{B}\otimes\xi)\,,} & \xi=\Tenn[A]{(\GAA{A}\impl\trn{A})}\,.
\end{array} \]
This can be solved either as a fixpoint equation in the cpo of nominal arenas or as a domain equation in the PreCpo-enriched category $\Vt$.
We follow the latter approach, which provides the most general notion of canonical solution (and which incorporates the solution in the cpo of nominal arenas, analogously to~\cite{McCusker:PhD_book}). It uses the categorical constructions of~\cite{Smyth_Plotkin:RDE,Freyd:90a} for solving recursive domain equations, as adapted to games in~\cite{McCusker:PhD_book}.

\bdefn Define the category
\begin{lefteqn*}
    \CC\defn\Vt\times\prod_{A\in\text{TY}}\Vt
\end{lefteqn*}%
with objects $D$ of the form $(D_\xi,D_A\,^{A\in\text{TY}})$ and arrows $f$ of the form $(f_\xi,f_A\,^{A\in\text{TY}})$.

Now take $F:(\CC)^\mathrm{op}\times\CC\tote\CC$ to be defined on objects by
$F(D,E)\defn(\xi_{D,E},\trn{A}_{D,E}\,^{A\in\TY})$, where:
\[\begin{array}{@{}l|@{\qd\,}l|@{\qd\,}l}
\trn{\ena}_{D,E}\defn1 & \trn{A\times B}_{D,E}\defn\trn{A}_{D,E}\ten \trn{B}_{D,E} & \trn{[A]}_{D,E}\defn\GAA{A} \\[7pt]
\trn{\N}_{D,E}\defn\GN & \trn{A\tote B}_{D,E}\defn D_A\him(\xi_{E,D}\impl E_B\ten \xi_{D,E}) & \xi_{D,E}\defn\Tenn[A\in\TY]{(\GAA{A}\impl E_A)}
\end{array}\]
and similarly for arrows, with $F(f,g)\defn(\xi_{f,g},\trn{A}_{f,g}\,^{A\in\TY})$\,.%
\edefn%
Now~\eqref{e:storeEq} has been reduced to:
\[ \tag{SE$^*$}\label{e:SEF} D= F(D,D) \]
where $F$ is a locally continuous functor wrt the strategy ordering (proposition~\ref{NomGames:p:unionofstrat}), and continuous wrt the arena ordering
(proposition~\ref{NomGames:p:CpoObj}). The solution to~\eqref{e:SEF} is given via a \emph{local bilimit} construction to the following $\om$-chain in $\CC$.%
\footnote{Recall that we call an arrow $e:A\paei B$ an \boldemph{embedding}
if there exists $e^R:B\paei A$ such that
\[ e\1e^R=\id_A\ \land\ e^R\1e\ypoo\id_B\,. \]
Given an $\om$-chain $\Delta=(D_i,e_i)_{i\in\om}$ of objects and embeddings, a \boldemph{cone} for $\Delta$ is an object $D$ together with a family
$(\eta_i:D_i\paei D)_{i\in\om}$ of embeddings such that, for all $i\in\om$, \ $\eta_i=e_i\1\eta_{i+1}$. Such a cone is a \boldemph{local bilimit} for $\Delta$ if, for all $i\in\om$,
\[ \eta_i^R\1\eta_i\ypoo\eta_{i+1}^R\1\eta_{i+1}\ \land\ \bigsqcup\nolimits_{i\in\om}(\eta_i^R\1\eta_i)=\id_D\,. \]}

\bdefn In $\CC$ form the sequence $(D_i)_{i\in\om}$ taking $D_0$ as below and \ $D_{i+1}\defn F(D_i,D_i)$.
\begin{align*}
D_{0,\ena}&\defn 1 & D_{0,\N}&\defn\GN & D_{0,[A]}&\defn\GAA{A}\\
D_{0,A\tote B}&\defn 1 & D_{0,A\times B}&\defn D_{0,A}\ten D_{0,B} & D_{0,\xi}&\defn\Tenn[A]{(\GAA{A}\impl0)}
\end{align*}
Moreover, define arrows $e_i:D_i\paei D_{i+1}$ and $e_i^R:D_{i+1}\paei D_i$ as:
\begin{align*}
e_0 &\defn\incl{D_0,D_1} & e_0^R &\defn\proj{D_1,D_0} &\qd[2] e_{i+1} &\defn F(e_i^R,e_i) & e_{i+1}^R &\defn F(e_i,e_i^R)\,.
\end{align*}
\deq[-1] \edefn[q]%
The above inclusion and projection arrows are defined componentwise. In fact, there is a hidden lemma here which allows us to define the
projection arrow, namely that $D_0\ypop[1]D_1$ (which means $D_{0,\xi}\ypop[1]D_{1,\xi}$ and $D_{0,A}\ypop[1]D_{1,A}$ for all $A$).
\[ (\Delta) \qd[2] \xymatrix@C=13mm{D_0\ar[r]^{e_0}&D_1\ar[r]^{e_1}&D_2\ar[r]^{e_2}&D_3\ar[r]^{e_3}&\cdots} \]
Thus, we have formed the $\omega$-chain $\Delta$.
We show that $\Delta$ is a $\ypop$-increasing sequence of objects and embeddings, and proceed to the main result.

\belem For $(e_i,e_i^R)_{i\in\om}$ as above and any $i\in\om$,
\[ e_i=\incl{D_i,D_{i+1}}\qd\land\qd e_i^R=\proj{D_{i+1},D_i}\,. \]
\enlem%
\proof It is easy to see that $D_i\ypop[1]D_{i+1}$, all $i\in\om$, so the above are well-defined. We now do induction on $i$; the base case is true by
definition. The inductive step follows easily from proposition~\ref{NomGames:p:CpoObj}. \qed
\bethm We obtain a local bilimit $(D^*,\eta_i\,^{i\in\om})$ for $\Delta$ by taking:
\[ D^*\defn\Lub D_i\,,\qd[2]\eta_i\defn\incl{D_i,D^*}\qd\text{(each $i\in\om$)}. \]
Hence, $\id_{D^*}:F(D^*,D^*)\paei D^*$ is a minimal invariant for $F$.
\enthm%
\proof First, note that $D_0\ypop[1] D_i$, for all $i\in\om$, implies that all $D_i$'s share the same initial moves, and hence
$D_i \ypop[1] D^*$.
Thus, for each $i\in\om$, we can define
$\eta_i^R\defn\proj{D^*,D_i}$\,,
and hence each $\eta_i$ is an embedding. We now need to show the following.
\begin{enumerate}
  \item $(D^*,\eta_i\,^{i\in\om})$ is a cone for $\Delta$,
  \item for all $i\in\om$, \ $\eta_i^R\1\eta_i\ypoo\eta_{i+1}^R\1\eta_{i+1}$\,,
  \item $\bigsqcup_{i\in\om}(\eta_i^R\1\eta_i)=\id_{D^*}$.
\end{enumerate}
For 1, we nts that, for any $i$, \ $\incl{D_1,D^*}=\incl{D_i,D_{i+1}}\1\incl{D_{i+1},D^*}$\,, which follows from (TRN).
For 2 we essentially nts that $\id_{D_i}\ypo\id_{D_{i+1}}$, and for 3 that $\bigcup_i\id_{D_i}=\id_{D^*}$\,; these are both straightforward.

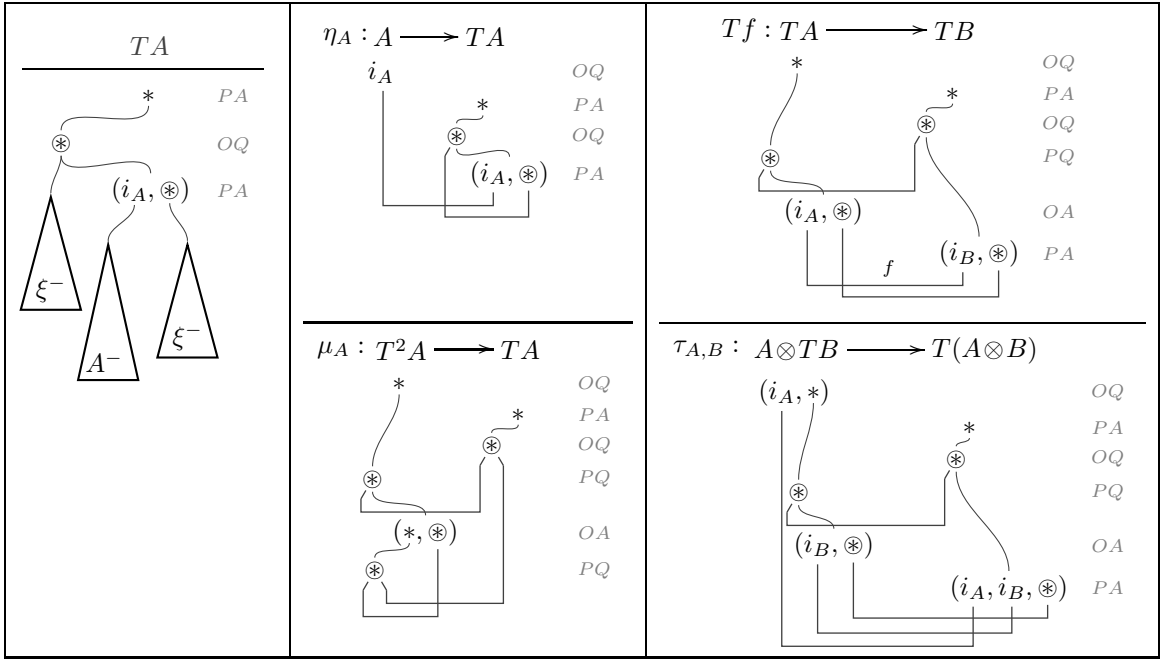
\begin{figure}[t]\renewcommand\arraystretch{1.5}\centerline{\small$\begin{array}{|l|l|l|}%
\hline
\begin{array}{c}\!\!
{\begin{array}{l l} \qd[3]\;\;\color{darkgray}TA \\\cline{1-2}
\qd[4]\rnode{A}{*} & \GPA\\
\;\;\rnode{Z}{\ixi} & \GOQ\\
\;\;\rnode{EXX}{\pspolygon[fillstyle=solid,fillcolor=white](0,0)(-0.4,-1.5)(0.4,-1.5)(0,0)}{\rput[bl]{0}(-0.2,-1.4){\xi^-}}
\qd[2]\; \rnode{B}{(\rnode{IB}{i_{A}},\rnode{IX}{\ixi})} & \GPA\\
\qd[2]\;\;\,\rnode{CC}{\pspolygon[fillstyle=solid,fillcolor=white](0,0)(-0.4,-1.8)(0.4,-1.8)}{\rput[bl]{0}(-0.3,-1.7){A^-}}\qd[3]
\rnode{C}{\pspolygon[fillstyle=solid,fillcolor=white](0,0)(-0.4,-1.5)(0.4,-1.5)(0,0)}{\rput[bl]{0}(-0.2,-1.4){\xi^-}}
\nccdu{C}{IX}\nccdu{CC}{IB}\nccdu{Z}{A}\nccdu{B}{Z}\nccdu{AX}{Z}\nccdu{AA}{ZA}\nccdu{EXX}{Z}
\\\\\\\\\\\\\\\\\\
\end{array}}\end{array}
&
\begin{array}{c}
\xymatrix@R=0pt@C=1pt{\str[5mm]{\eta_A:}
A\ar[r]   & T A \\
\rnode{A}{i_A\,}                &                               & \GOQ\\
                                   & \rnode{a}{*}\,             & \GPA\\
                                   & \xinode{B}\,\qquad    & \GOQ\\
                                   & \qquad\rnode{b}{(\rnode{C}{i_A},\rnode{D}{\ixi})}\,
                                                                & \GPA
\\\\\\\\\\\\\\\\
\cc[5pt]{A}{C} \xiccl[10pt]{B}{D} \nccdu{B}{a} \nccdu{b}{B} }%
\\\hline
\xymatrix@R=0pt@C=1pt{\str[8mm]{\mu_A:}
\TT A\ar[r]   & T A \\
\rnode{a}{*}\,                           &                      & \GOQ\\
                                   & \rnode{b}{*}\,             & \GPA\\
                                   & \xinode{A}\,\qquad    & \GOQ\\
\xinode{B}\,\qquad         &                               & \GPQ\\
\\
\qquad\rnode{c}{(\rnode{cc}{*},\rnode{D}{\ixi})}\,               &          & \GOA\\
\xinode{E}\qd[2]             &                              & \GPQ\\
\\\\\\
\xicc[5pt]{B}{A}\xiccl[10pt]{E}{D}\xxicc[5pt]{E}{A} 
\nccdu{A}{b} \nccdu{B}{a} \nccdu{c}{B} \nccdu{E}{cc}} 
\end{array}&
\begin{array}{c}
\xymatrix@R=0pt@C=1pt{\str[7mm]{T f:}
T A\ar[r]   & T B                                               \\
\rnode{a}{*}\,                           &                      & \GOQ\\
                                   & \rnode{b}{*}\,             & \GPA\\
                                   & \xinode{A}\,\qquad    & \GOQ\\
\xinode{B}\,\qquad         &                               & \GPQ\\
\\
\qquad\rnode{c}{(\rnode{C}{i_A},\rnode{D}{\ixi})}\,            && \GOA\\
                                   & \qquad\rnode{d}{(\rnode{E}{i_B},\rnode{F}{\ixi})}\, & \GPA
\\\\\\
\xicc{B}{A} \cc[10pt]{D}{F}\aput[7pt](1.3){_f} \cc[5pt]{C}{E} & \nccdu{A}{b} \nccdu{B}{a} \nccdu{c}{B} \nccdu{d}{A}}%
\\\hline
\xymatrix@R=0pt@C=1pt{\str[12mm]{\tau_{A,B}:}
A\ten T B\ar[r]   & T(A\ten B) \\
\rnode{a}{(\rnode{A}{i_A},\rnode{b}{*})}\,                &     & \GOQ\\
                                   & \rnode{c}{*}\,\quad        & \GPA\\
                                   & \xinode{B}\,\qd[2]    & \GOQ\\
\,\xinode{C}                &                              & \GPQ\\
\\
\qd[3]\rnode{d}{(\rnode{D}{i_B},\rnode{E}{\ixi})}\, &           & \GOA\\
                                   & \qquad\rnode{e}{(\rnode{F}{i_A},\rnode{G}{i_B},\rnode{H}{\ixi})}\,
                                                                & \GPA
\\\\\\
\xicc[5pt]{B}{C} \cc[10pt]{D}{G} \cc[5pt]{E}{H} \cc[15pt]{A}{F} %
\nccdu{B}{c} \nccdu{C}{b} \nccdu{d}{C} \nccdu{e}{B} }%
\end{array}
\\\hline
\end{array}$}
\caption{The store monad.}\label{NR:f:Monad}
\end{figure}

From the local bilimit $(D^*,\eta_i\,^{i\in\om})$ we obtain a minimal invariant $\alpha:F(D^*,D^*)\paei D^*$ by taking (see e.g.~\cite{Abramsky:DT}):
\[ \alpha\defn\Lub\alpha_i\,,\qd \alpha_i\defn F(\eta_i,\eta_i^R)\1\eta_{i+1}
  \overset{\text{prop.~\ref{NomGames:p:CpoObj}}}{=}\proj{F(D^*,D^*),D_{i+1}}\1\incl{D_{i+1},D^*}\,. \]
Moreover, $D^*=F(D^*,D^*)$ by the Tarski-Knaster theorem, and therefore $\alpha_i=\eta_{i+1}^R\1\eta_{i+1}$\,,
which implies $\alpha=\id_{D^*}$.
\qed\noindent%
Thus, $D^*$ is the canonical solution to $D=F(D,D)$, and in particular it solves:
\[
D_{A\tote B}=D_{A}\him(D_\xi\impl D_{B}\ten D_\xi)\,,\qd
D_\xi=\Tenn[A]{(\GAA{A}\impl D_A)}\,.
\]
\bdefn\label{d:xi} Taking $D^*$ as in the previous theorem define, for each type $A$,
\[ \xi \defn D^*_\xi\,, \qd \trn{A}\defn D_A^*\,. \]
\mindeq[-1]\edefn[a]%
The arena $\xi$ and the translation of compound types are given explicitly in the following figure. $\xi$ is depicted by means of unfolding it to
$\Tenn[A]{(\GAA{A}\impl\trn{A})}$\,: it consists of an initial move $\ixi$ which justifies each name-question $\al\in\NA{A}$, all types $A$, with the answer to the latter being the denotation of $A$ (and modelling the stored value of $\al$). Note that we reserve the symbol ``$\ixi$" for the initial move of $\xi$. $\ixi$-moves in type-translations can be seen as \emph{opening a new store}.
\begin{center}\fbox{\small\qd
\parbox[t][47mm]{26mm}{\nada\\$\begin{array}{c l}%
\xi \\\hline
\rnode{A}{\ixi} & \GPA\\
\rnode{B}{\al}\qd[2]\nada & \parbox[t]{12mm}{$\GOQ$\\\scriptsize $(\al\in\NA{A})$}\\
\qd\rnode{C}{\pspolygon[fillstyle=solid,fillcolor=white](0,0)(-0.5,-1.3)(0.5,-1.3)(0,0)}{\rput[bl]{0}(-0.275,-1.23){\trn{A}}} \nccdu{C}{B}\nccdu{B}{A}
\end{array}$}\qd[2]%
\parbox[t]{29mm}{\nada\\$\begin{array}{c l}%
\trn{A\times B} \\\hline
\rnode{AB}{(\rnode{A}{i_{\trn{A}}},\rnode{B}{i_{\trn{B}}})} & \GPA \\
\rnode{AA}{\pspolygon[fillstyle=solid,fillcolor=white](0,0)(-0.6,-1.6)(0.6,-1.6)(0,0)}{\rput[bl]{0}(-0.4,-1.5){\trn{A}^-}}\qd[4]
\rnode{BB}{\pspolygon[fillstyle=solid,fillcolor=white](0,0)(-0.6,-1.6)(0.6,-1.6)(0,0)}{\rput[bl]{0}(-0.4,-1.5){\trn{B}^-}}\nccdu{AA}{A}\nccdu{BB}{B}
\end{array}$}\qd[2]%
\parbox[t]{49mm}{\nada\\$\begin{array}{c@{\qd[3]}c l}%
\multicolumn{2}{c}{\trn{A\tote B}} \\\hline
\multicolumn{2}{c}{\rnode{A}{*}} & {\scriptstyle\color{gray}PA}\\
\rnode{Z}{(\rnode{ZA}{i_{\trn{A}}},\rnode{ZB}{\ixi})} && {\scriptstyle{\color{gray}OQ}}\\
\rnode{AA}{\pspolygon[fillstyle=solid,fillcolor=white](0,0)(-0.63,-1.6)(0.63,-1.6)}{\rput[bl]{0}(-0.455,-1.5){\trn{A}^-}}\qd[3]
\rnode{AX}{\pspolygon[fillstyle=solid,fillcolor=white](0,0)(-0.4,-1.3)(0.4,-1.3)(0,0)}{\rput[bl]{0}(-0.2,-1.2){\xi^-}}
& \rnode{B}{(\rnode{IB}{i_{\trn{B}}},\rnode{IX}{\ixi})} & {\scriptstyle{\color{gray}PA}}\\
& \;\,\rnode{CC}{\pspolygon[fillstyle=solid,fillcolor=white](0,0)(-0.63,-1.6)(0.63,-1.6)}{\rput[bl]{0}(-0.455,-1.5){\trn{B}^-}}\qd[3]
\rnode{C}{\pspolygon[fillstyle=solid,fillcolor=white](0,0)(-0.4,-1.3)(0.4,-1.3)(0,0)}{\rput[bl]{0}(-0.2,-1.2){\xi^-}}
\nccdu{C}{IX}\nccdu{CC}{IB}\nccdu{Z}{A}\nccdu{B}{ZB}\nccdu{AA}{ZA}\nccdu{AX}{ZB}
\end{array}$}
}
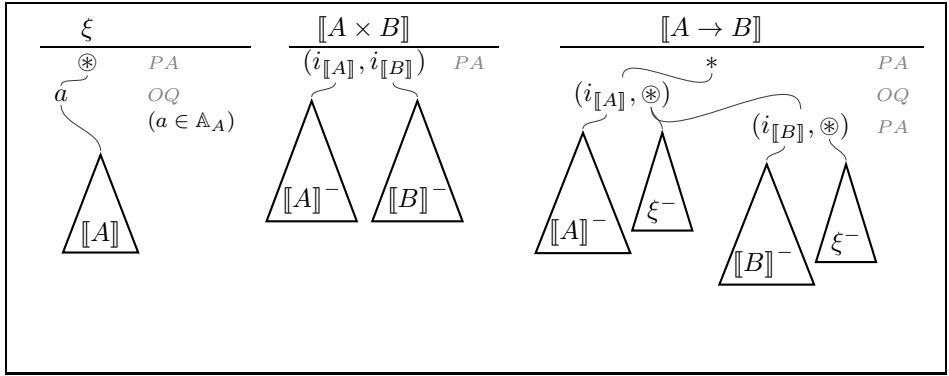
\captionof{figure}{The store arena and the type translation.}\label{NR:f:trans}\label{f:TA}
\end{center}
\paragraph{\it The store monad $T$}
There is a standard construction (v.~\cite{Moggi:88}) for defining a monad of $A$-side-effects (any object $A$) starting from a given strong monad with exponentials. Here we define a store monad, i.e.~a $\xi$-side-effects monad, from the lifting monad as follows.
\begin{gather}\raisetag{18mm}\begin{aligned}
&T  :\CC\paei\CC \defn \xi\impl(\uscore\otimes \xi) \\
&\eta_A :A\paei T A\defn\La\left(A\otimes \xi\nzaa{\upp}(A\otimes\xi)_\bot\right)\\
&\mu_A :\TT A\paei T A \defn \La\left(\TT A\otimes\xi\nzaa{\ev}(T A\otimes\xi)_\bot\nzaa{\ev_\bot}
    (A\otimes\xi)_{\bot\bot}\nzaa{\dnn}(A\otimes\xi)_\bot\right) \\
&\tau_{A,B} :A\otimes T B\paei T(A\otimes B)\defn\La\left(A\otimes T B\otimes\xi\nzaa{\id\otimes\ev}
    A\otimes(B\otimes\xi)_\bot\nzaa{\stt}(A\otimes B\otimes\xi)_\bot\right)
\end{aligned}\end{gather}
A concrete description of the store monad is given in figure~\ref{NR:f:Monad} (the diagrams of strategies depict their viewfunctions, as described in notation~\ref{not:Diagr}).
For the particular case of $\ixi$-moves which appear as second moves in $TA$'s, let us recall the convention we are following. Looking at the diagram for $TA$ (figure~\ref{NR:f:Monad}), we see that $\ixi$ justifies a copy of $\xi^-$ (left) and a copy of $A\ten\xi$ (right). Thus, a copycat link connecting to the lower-left of a $\ixi$ expresses a copycat concerning the $\xi^-$ justified by $\ixi$ (e.g.~the link between the first two $\ixi$-moves in the diagram for $\mu_A$), and similarly for copycat links connecting to the lower-right of a $\ixi$. Thus, for example, $\mu_A$ is given by:
\begin{align*}
\mu_A = \strat(\ &\{\,[**\ixi\ixi s]\,|\,[\ixi\ixi s]\in\viewf(\id_\xi)\,\} \\
                 &\cup\{\,[**\ixi\ixi(*,\ixi')\ixi's]\,|\,[\ixi'\ixi' s]\in\viewf(\id_\xi)\lor[s]\in\viewf(\id_{A\ten\xi})\,\}\ ) \,.
\end{align*}

\noindent%
A consequence of lifting being a strong monad with exponentials is that the store monad is also a strong monad with exponentials.
$T$-exponentials are given by:
\begin{equation}
  \eistin{TB}{A}\defn A\him TB\,,\qd[2]\LaT(f:A\ten B\paei TC)\defn\La(f)\,.
\end{equation}
Moreover, for each arena $A$ we can define an arrow:
\begin{equation}
\alpha_A\defn A_\bot\lred{(\eta_A)_\bot}(TA)_\bot\lred{\puu[TA]}TA\,.
\end{equation}
The transformation $\puu$ was introduced in~\eqref{e:up-pu}. Using lemma~\ref{l:puu} we obtain $\alpha_A = \La(\stt[A,\xi]')$.
Moreover, we can show that \ $\alpha : (\uscore)_\bot\paei T$ \ is a monad morphism.
\subsection{Obtaining the \texorpdfstring{$\laNR$}{laNR}-model}
Let us recapitulate the structure that we have constructed thus far to the effect of obtaining a $\laNR$-model in $\Vt$. Our numbering below follows
that of definition~\ref{d:laM}.
\begin{cEnumerate}[\normalfont\Roman{enumi}.]{123}
 \item $\Vt$ is a category with finite products (proposition~\ref{NomGames:p:products}).
 \item The store monad $T$ is a strong monad with exponentials.
 \item $\Vt$ contains adequate structure for numerals.
 \item There is a family $\MODEL{\prn,\eps,\de,\zet}$ of product comonads, with each $\prn$ having basis $\GA$ (see section~\ref{NomGames:c:Comon}),
     which fulfils specifications (a,b). There are also fresh-name constructors, $\frs^{\all\al}:\prn\paei(\prn[\allal])_\bot$\,,
     which satisfy (N2).
 \item There are name-equality arrows, $\eq[A]$ for each type $A$, making the (N1) diagram commute (section~\ref{NomGames:c:Comon}).
\end{cEnumerate}
From $\frs$ we can obtain a fresh-name transformation for the store monad.

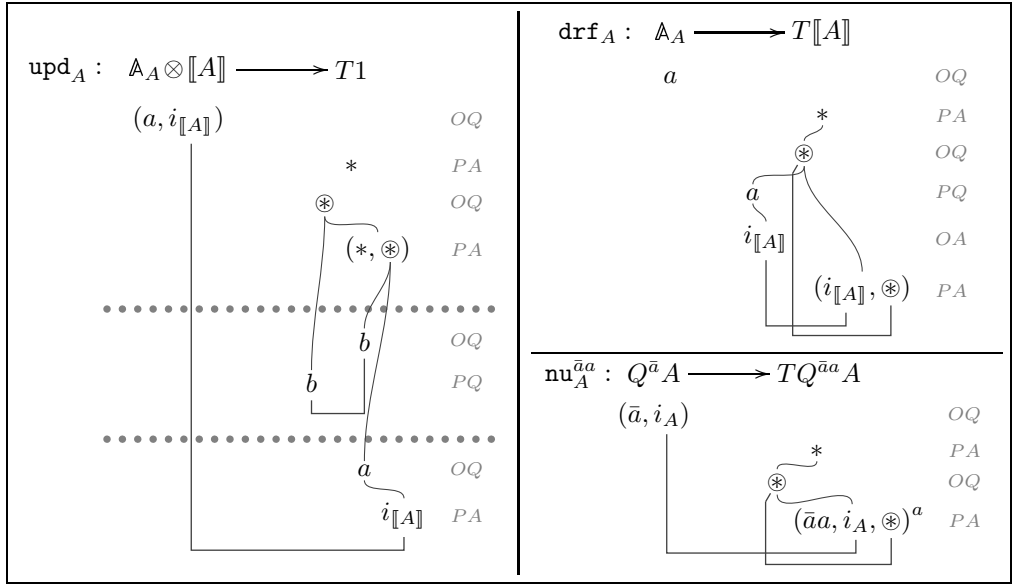
\begin{figure}[t]
\fbox{\small$\begin{array}{@{}l|l@{}}
\begin{array}{c}\xymatrix@R=1mm@C=1mm{\str[15mm]{\upd[A]:}
\GAA{A}\ten\trn{A}\ar[rr]   && T1 \\
\rnode{a}{(\al,\rnode{A}{i_{\trn{A}}})}            &&            & \GOQ\\
                                   && \rnode{b}{*}               & \GPA\\
                                   && \rnode{c}{\ixi}\qquad      & \GOQ\\
                                   && \qquad\rnode{d}{(*,\rnode{e}{\ixi})}
                                                                 & \GPA\\
\\\biglinehere{-1,0}{4.2,0}\\
                                   && \quad\rnode{B}{\be}        & \GOQ\\
                                   && \rnode{C}{\be}\qd[3]       & \GPQ\\
\\\biglinehere{-1,0}{4.2,0}\\
                                   && \quad\rnode{f}{\al}        & \GOQ\\
                                   && \qd[4]\rnode{D}{i_{\trn{A}}}&\GPA 
\cc{A}{D} \cc{B}{C} \nccdu{d}{c} \nccdu{B}{e} \nccdu{C}{c} \nccdu{f}{e} \nccdu{D}{f} }
\end{array} &
\begin{array}{c}
\xymatrix@R=1mm@C=1mm{\str[10mm]{\drf[A]:}
\GAA{A}\ar[rr] && T\trn{A} \\
\rnode{a}{\al}                                                 &&& \GOQ\\
                                   && \rnode{b}{*}               & \GPA\\
                                   && \;\ \xinode{A}\qquad      & \GOQ\\
                                   && \rnode{c}{\al}\qd[5]\;       & \GPQ\\
                                   && \rnode{B}{i_{\trn{A}}}\qd[4]\;&\GOA\\
                                   && \qquad\quad\rnode{d}{(\rnode{C}{i_{\trn{A}}},\rnode{D}{\ixi})}
                                                                & \GPA
\xiccl[4mm]{A}{D}\cc{B}{C}\nccdu{A}{b}\nccdu{c}{A} \nccdu{B}{c} \nccdu{d}{A} }
\\\\[3pt]\hline
\xymatrix@R=0pt@C=1mm{\str[10mm]{\nw^{\allal}_A:}
\prn A\ar[r]   & T\prn[\allal] A \\
\rnode{A}{(\all,\rnode{AA}{i_A})}                && \GOQ \\
                                   & \rnode{a}{*}\, & \GPA \\
                                   & \xinode{B}\,\qd[3] & \GOQ \\
                                   & \qd[3]\rnode{b}{(\allal,\rnode{C}{i_A},\rnode{D}{\ixi})}^\al & \GPA\\\\\text{}
\cc[5pt]{AA}{C} \xiccl[10pt]{B}{D} \nccdu{B}{a} \nccdu{b}{B} }
\end{array}
\end{array}$}%
\captionof{figure}{Strategies for update, dereferencing and fresh-name creation.}\label{f:nu}
\end{figure}

\bdefn\label{NR:d:nw} For each $\allal\in\NAA[\#]$, define a natural transformation $\nw^{\allal}:\prn\paei T\prn[\allal]$ by:
\[ \nw^{\allal}_A\defn \prn A\lred{\frs_A}(\prn[\allal]A)_\bot\lred{\alpha_{{\prn[\allal]}A}}T\prn[\allal]A\,. \]
Moreover, for each $f:\prn[\all\al]A\paei TB$, take  $\abs f\defn\prn A\lred{\nw_A}T\prn[\all\al]A\lred{Tf}\TT B\lred{\mu_B}TB$\,.
\edefn%
Each arrow $\nw^{\allal}_A$ is explicitly given by (note we use the same conventions as in~\eqref{e:abs_expl}):
\begin{align*}
\nw^{\allal}_A
= \strat\{\,&[(\all,i_A)\,*\,\ixi\,(\allal,i_A,\ixi)^\al s^\al]\,|\\
            &\al\4i_A\land([i_A i_A s]\in\viewf(\id_A)\lor[\ixi\ixi s]\in\viewf(\id_\xi))\,\}
\end{align*}
and diagrammatically as in figure~\ref{f:nu}. Moreover, using the fact that $\alpha$ is a monad morphism and lemma~\ref{l:puu} we can show that, in fact, $\abs f$ is given exactly as in~\eqref{e:abs}, that is,
\[
\abs f  =\frs_A\1f_\bot\1\puu[TB]\,.
\]
Finally, $\alpha$ being is a monad morphism implies also the following.

\beprop The $\nw$ transformation satisfies the {\rm(N2)} diagrams of definition~\ref{d:laNR}.\qed \enprop%


\noindent%
What we are only missing for a $\laNR$-model is update and dereferencing maps.

\bdefn\label{d:upddrf} For any type $A$ we define the following arrows in $\Vt$\,, %
\begin{align*}%
\drf[A]&\defn\strat\{\,[\,\al * \ixi\,\al\,i_{\trn{A}}\,(i_{\trn{A}},\ixi)\, s\,]\,|\\
       &\phantom{{}\defn\strat\{\,}[\ixi\ixi s]\in\viewf(\id_\xi)\lor[i_{\trn{A}} i_{\trn{A}} s]\in\viewf(\id_{\trn{A}})\,\}\,, \\
\upd[A]&\defn\strat\bigl(\{\,[\,(\al,i_{\trn{A}})* \ixi\,\be\,\be\, s\,]\,|\,
       [\ixi\,\ixi\,\be\,\be\, s]\in\viewf(\id_\xi)\land\be\#\al\,\}\\
       &\qd[5]\cup\{\,[\,(\al,i_{\trn{A}}) * \ixi\,\al\, i_{\trn{A}}\, s\,]\,|\,
       [i_{\trn{A}}\, i_{\trn{A}}\, s]\in\viewf(\id_{\trn{A}})\,\}\bigr)\,,
\end{align*}
depicted also in figure~\ref{f:nu}. \edefn%
These strategies work as follows. $\upd[A]$ responds with the answer $(*,\ixi)$ to the initial sequence $(\al,i_{\trn{A}})*\ixi$ and then:
\begin{Itemize}
  \item for any name $\be\4\al$ that is asked by $O$ to $(*,\ixi)$ (which is a store-opening move), it copies $\be$ under the store $\ixi$
    (opened by $O$) and establishes a copycat link between the two $\be$'s;
  \item if $O$ asks $\al$ to $(*,\ixi)$, it answers $i_{\trn{A}}$ and establishes a copycat link between the two $i_{\trn{A}}$'s.
\end{Itemize}
On the other hand, $\drf[A]$ does not immediately answer to the initial sequence $\al *\ixi$ but rather asks (the value of) $\al$ to $\ixi$. Upon receiving $O$'s answer $i_{\trn{A}}$\,, it answers $(i_{\trn{A}},\ixi)$ and establishes two copycat links.
We can show by direct computation the following.
\beprop\label{NR:p:NR_SNR}
The {\rm(NR)} and {\rm(SNR)} diagrams of definition~\ref{d:laNR} commute. \qed
\enprop%
We have therefore established the following.
\bethm $(\Vt,T,\prn[])$ is a \laNR-model. \qed
\enthm%
We close this section with a discussion on how the store-effect is achieved in our innocent setting, and with some examples of translations of $\nurho$-terms in $\Vt$.

\berem[Innocent store]\label{r:InnStore}
The approach to the modelling of store which we have presented differs fundamentally from previous such approaches in game semantics.
Those approaches, be they for basic or higher-order store~\cite{Abramsky_McCusker:IA97,Abramsky+:GamesReferences}, are based on the following methodology. References are modelled by read/write product types, and fresh-reference creation is modelled by a ``cell" strategy which creates the fresh cell and imposes a good read/write discipline on it. In order for a cell to be able to return the last stored value, innocence has to be broken
since each read-request hides previous write-requests from the P-view. Higher-order cells have to also break visibility in order to establish copycat links between read- and write-requests.

\piccaption{A dialogue in innocent store.}
\parpic(97mm,47mm)(0mm,25mm)[br]{%
\fbox{\itshape\small\parbox{.62\linewidth}{%
P -- What's the value of $\al$? \\
\nada\qd[1] O -- I don't know, you tell me: what's the value of $\al$? \\
\nada\qd[2] P -- I don't know, you tell me: what's the value of $\al$? \\
\nada\qd[4] $\vdots$ \\
\nada\qd[3] O -- I don't know, you tell me: what's the value of $\al$? \\
\nada\qd[3] P -- I know it, it is $v$. \\
\nada\qd[4] $\vdots$ \\
\nada\qd[2] O -- I know it, it is $v$. \\
\nada\qd[1] P -- I know it, it is $v$. \\
O -- I know it, it is $v$.}}%
}

  Here instead we have only used innocent strategies and a monad on a
  store $\xi$. Because of the monad, an arena $\trn{A}$ contains
  several copies of $\xi$, therefore several stores are opened inside
  a play. The read/ write discipline is then kept in
  an \emph{interactive} way: when a participant asks (the value of) a
  name $\al$ at the last (relevant) store,\footnote{i.e.~at the last
  store-opening move played by the other participant.} the other
  participant either answers with a value or asks himself $\al$ at the
  penultimate store, and so on until one of the participants answers
  or the first store in the play is reached. At each step, a
  participant answers the question $\al$ only if he updated the value
  of $\al$ before opening the current store (of that step, i.e.~the
  last store in the participant's view)\HY note that this behaviour
  does \emph{not} break innocence. If no such update was made by the
  participant then he simply passes $\al$ to the previous store and
  establishes a copycat link between the two $\al$'s. These links
  ensure that when an answer is eventually obtained then it will be
  copycatted all the way to answer the original question $\al$. Thus,
  we innocently obtain a read/write discipline: at each question
  $\al$, the last update of $\al$ is returned.
\enrem

\beexam Consider the typed terms:
\[ \seq{\ee}{\keno}{\new[\al]\al := \ang{\fst\bang\al,\snd\bang\al}}\,,\qd \seq{\be}{\keno}{\be:=\la x.(\bang\be)\sskip}\,, \qd
\seq{\be}{\keno}{(\bang\be)\sskip}\]
with $\al\in\NA{\N\times\N}$ and $\be\in\NA{\ena\tote B}$. Their translations in $\Vt$ are as follows.
{\small
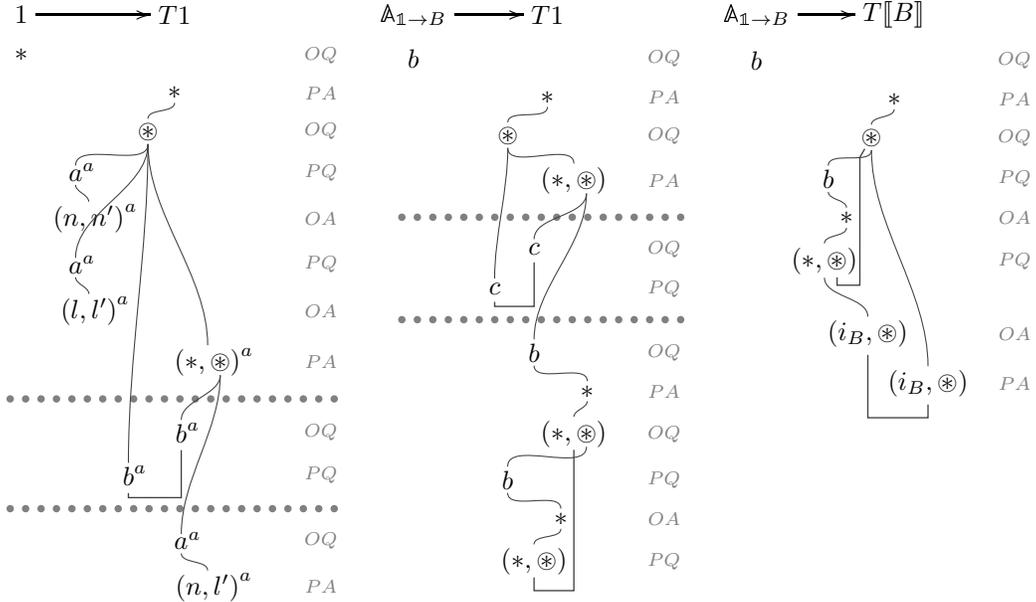
\[
\xymatrix@R=1mm@C=-1mm{%
1\ar[rrr]   &&& T1 \\
\rnode{a}{*}                 &&&                                            & \GOQ\\
                                   &&& \rnode{b}{*}                               & \GPA\\
                                   &&& \rnode{c}{\ixi}\qquad                      & \GOQ\\
                                   &&& \rnode{d}{\al}^\al\qd[7]                       & \GPQ\\
                                   &&& \rnode{e}{(n,n')}^\al\qd[6]                    & \GOA\\
                                   &&& \rnode{dd}{\al}^\al\qd[7]                      & \GPQ\\
                                   &&& \rnode{ee}{(l,l')}^\al\qd[6]                   & \GOA\\
                                   &&& \qd[3]\rnode{f}{(*,\rnode{ff}{\ixi})}^\al      & \GPA\\
\biglinehere{-.2,0}{4.2,0}\\
                                   &&& \quad\rnode{g}{\be}^\al                        & \GOQ\\
                                   &&& \rnode{h}{\be}^\al\qd[3]                       & \GPQ\\
\biglinehere{-.2,0}{4.2,0}\\
                                   &&& \quad\rnode{gg}{\al}^\al                       & \GOQ\\
                                   &&& \qd[3]\rnode{hh}{(n,l')}^\al                   & \GPA\\
\cc[0.61mm]{g}{h} \nccdu{c}{b} \nccdu{d}{c} \nccdu{e}{d} \nccdu{dd}{cc} \nccdu{dd}{c} \nccdu{ee}{dd} \nccdu{f}{c} \nccdu{g}{ff} \nccdu{h}{c}
\nccdu{gg}{ff} \nccdu{hh}{gg} } 
\qd
\xymatrix@R=1mm@C=-2mm{%
\GAA{\ena\tote B}\ar[rr]   && T1 \\
\rnode{a}{\be}                       &&                                            & \GOQ\\
                                   && \rnode{b}{*}                               & \GPA\\
                                   && \rnode{c}{\ixi} \qd[3]                     & \GOQ\\
                                   && \qd[2]\rnode{d}{(*,\rnode{e}{\ixi})}   & \GPA\\
\biglinehere{-.2,0}{3.6,0}\\
                                   && \rnode{B}{\ga}\qd                      & \GOQ\\
                                   && \rnode{C}{\ga}\qd[4]                   & \GPQ\\
\biglinehere{-.2,0}{3.6,0}\\
                                   && \rnode{f}{\be}\qd                      & \GOQ\\
                                   && \qd[5]\rnode{g}{*}\qd[2]               & \GPA\\
                                   && \qd[2]\rnode{h}{(*,\rnode{hh}{\ixi})}  & \GOQ\\
                                   && \rnode{i}{\be}\qd[3]                   & \GPQ\\
                                   && \qd[1]\rnode{j}{*}                     & \GOA\\
                                   && \rnode{k}{(*,\rnode{D}{\ixi})}\qd[1]   & \GPQ\\
\cc[0.61mm]{B}{C} \cc[5pt,nodesep=1pt]{k}{h} \nccdu{c}{b} \nccdu{d}{c} \nccdu{B}{e} \nccdu{C}{c} \nccdu{f}{e} \nccdu{g}{f} \nccdu{h}{g} \nccdu{i}{hh} \nccdu{j}{i} \nccdu{k}{j} }
\qd
\xymatrix@R=1mm@C=-2mm{%
\GAA{\ena\tote B}\ar[rr]   && T\trn{B} \\
\rnode{a}{\be}                       &&                                            & \GOQ\\
                                   && \rnode{b}{*}                               & \GPA\\
                                   && \;\xinode{c}\qd[2]                     & \GOQ\\
                                   && \rnode{d}{\be}\qd[5]                   & \GPQ\\
                                   && \rnode{e}{*}\qd[3]\;\                     & \GOA\\
                                   && \rnode{f}{(*,\rnode{ff}{\ixi})}\qd[5]\;        & \GPQ\\\\
                                   && \rnode{g}{(i_B,\ixi)}\qd[2]                  & \GOA\\
                                   && \qd[2]\;\;\rnode{h}{(i_B,\ixi)}                  & \GPA\\
\xiccr[3pt]{ff}{c}\cc{g}{h} \nccdu{c}{b} \nccdu{d}{c} \nccdu{e}{d} \nccdu{f}{e} \nccdu{g}{f} \nccdu{h}{c}  } \]}%
In the first example we see that, although the strategy is looking up the fresh (and therefore uninitialised) reference $\al$, the play does not deadlock: if Opponent answered the question $\al^\al$ then the play would proceed as depicted. In practice, however, Opponent will never be able to answer that question and the play will halt indeed (this is because Opponent must play \emph{tidily}, see section~\ref{s:tidy}).
Moreover, from the latter two examples we can compute
$\trn{\sstop_B}:1\paei T\trn{B}=\{\,[*\,*\,\ixi]\,\}$\,.
\qed\enexam
\subsection{Adequacy}\label{s:Adequacy}
We proceed to show that $\Vt$ is adequate (v.~definition~\ref{d:adeq}). First we characterise non-reducing terms as follows.

\belem\label{NR:l:Values}%
Let $\Aseq{\keno}{M\o A}$ be a typed term. $M$ is a value iff there exists a store $S$ such that $S\modl M$ has no reducts
and $[(\all,*)\,*\,\ixi\,(i_A,\ixi)^\bee]\in\trn{\bS\1 M}$\,, for some $i_A,\bee$.
\enlem%
\proof The ``only if"-part is straightforward. For the ``if"-part assume that $M$ is a non-value and take any $S$ such that $S\modl M$ has no reducts.
We show by induction on $M$ that there exist no $i_A,\bee$ such that $[(\all,*)\,*\,\ixi\,(i_A,\ixi)^\bee]\in\trn{\bS\1 M}$. The base case follows trivially from $M$ not being a value. Now, for the inductive step, the specifications of $S\modl M$ (and $M$) imply that either $M\equiv\bang\al$ with $\al$ not having a value in $S$, or $M\equiv\ctxE[K]$ with $\ctxE$ an evaluation context and $K$ a non-value typed as $\Aseq{\keno}{K\o B}$ and such that $S\modl K$ non-reducing.

In case of $M\equiv\bang\al$, we have that $[(\all,*)*\ixi\,\al]\in\trn{\bS\1 M}$, which proves the claim because of determinacy. On the other hand, if $M\equiv\ctxE[K]$ then, as in proof of proposition~\ref{p:Correctness}, we have:
\begin{align*}
\trn{\bS\1M} &= \ang{\La(\zet'\1\trn{\ctxE[x]}),\trn{\bS\1K}}\1\tau\1T\ev\1\mu
             = \ang{\id,\trn{\bS\1K}}\1\tau\1T(\zet'\1\trn{\ctxE[x]})\1\mu
\end{align*}
By IH, there are no $i_B,\gaa$ such that $[(\all,*)*\ixi\,(i_B,\ixi)^\gaa]\in\trn{\bS\1K}$, which implies that there are no $i_A,\bee$ such that
$[(\all,*)*\ixi\,(i_A,\ixi)^\bee]\in\trn{\bS\1 M}$. \qed\noindent%
Because of the previous result, in order to show adequacy it suffices to show that, whenever $\trn{M}=\abs[\bee]\trn{\bS\1\nm{0}}$, there is no
infinite reduction sequence starting from $\all\modl M$. We will carry out the following reasoning.
\begin{Itemize}
 \item Firstly, since the calculus without DRF reductions is strongly normalising\HY this is inherited from strong normalisation of the
       $\nu$-calculus\HY it suffices to show there is no reduction sequence starting from $\all\modl M$ and containing infinitely many DRF reduction
       steps.
 \item In fact, the problem can be further reduced to showing that, whenever $[(\all,*)*\ixi\,(0,\ixi)^\bee]\in\trn{M}$, there is no reduction
       sequence starting from $\all\modl M$ and containing infinitely many NEW reduction steps. The latter clearly holds, since $M$ cannot create
       more than $|\bee|$ fresh names in that case, because of
       correctness.

       The reduction to this simpler problem is achieved as follows. For each term $M$, we construct a term $M'$ by adding immediately before each
       dereferencing in $M$ a fresh-name construction. The result is that, whenever there is a sequence with infinitely many DRF's starting from
       $S\modl M$, there is a sequence with infinitely many NEW's starting from $S\modl M'$. The reduction is completed by finally showing that,
       whenever we have $[(\all,*)*\ixi\,(0,\ixi)^\bee]\in\trn{M}$, we also have $[(\all,*)*\ixi\,(0,\ixi)^{\bee'}]\in\trn{M'}$.
\end{Itemize}
The crucial step in the proof is the reduction to ``the simpler problem", and particularly showing the connection between $\trn{M}$ and $\trn{M'}$
described above. The latter is carried out by using the observational equivalence relation on strategies, defined later in this section. Note,
though, that a direct proof can also be given (see~\cite{Tze_PhD}).
\beprop[Adequacy]\label{p:Adequacy}%
$\MODEL[]{\Vt,T,\prn[]}$ is adequate.
\enprop%
\proof This follows from $O$-adequacy (lemma~\ref{l:OAdeq}), which is proved independently. \qed\noindent
Hence, $\MODEL[]{\Vt,T,\prn[]}$ is a sound model for $\nurho$ and thus, for all terms $M,N$,
\[ \trn{M}=\trn{N} \implies M\ypob N\,. \]

\subsection{Tidy strategies}\label{s:tidy}
Leaving adequacy behind, the route for obtaining a fully abstract model of $\nurho$ proceeds to \emph{definability}. That is, we aim for a model in
which elements with \emph{finite descriptions} correspond to translations of $\nurho$-terms.

However, $\Vt$ does not satisfy such a requirement: it includes (finitary) store-related behaviours that are disallowed in the
operational semantics of $\nurho$. In fact, our strategies treat the store $\xi$ like any other arena, while in $\nurho$ the treatment of store
follows some basic guidelines. For example, if a store $S$ is updated to $S'$ then the original store $S$ is not accessible any more (\emph{irreversibility}). In strategies we do not have such a condition: in a play there may be several $\xi$'s opened, yet there is no discipline on which of these are accessible to Player whenever he makes a move. Another condition involves the fact that a store either `knows' the value of a name or it doesn't know it. Hence, when a name is asked, the store either returns its value or it deadlocks: there is no third option. In a play, however, when Opponent asks the value of some name, Player is free to evade answering and play somewhere else!

To disallow such behaviours we will constrain total strategies with further conditions, defining thus what we call \emph{tidy strategies}. But
first, let us specify store-related moves inside type-translating nominal arenas. %

\bdefn%
Consider $\Vnr$\,, the full subcategory of $\Vt$~with objects given by:
\[ Ob(\Vnr)\ni A,B ::= 1\,|\,\GN\,|\,\GA\,|\,A\ten B\,|\,A\him TB \]
For each such arena $A$ we define its set of \boldemph{store-Handles}, $H_A$, as follows.
\begin{align*}
H_1&=H_{\GN}=H_{\GA}\defn\keno\,, \qd[1]
H_{A\ten B}\defn H_A\cup H_B\,,\\
H_{A\him TB}&\defn\{(i_A,\ixi_A),(i_B,\ixi_B)\}\cup H_A\cup H_B\cup H_{\xi_A}\cup H_{\xi_B}\qd\text{with }H_\xi\defn\bigcup\nolimits_{C}H_{\trn{C}}\,,
\end{align*}
where we write $A\him TB$ as $A\him(\xi_A\impl B\ten\xi_B)$, and $\xi$
as $\Tenn[C]{(\GAA{C}\impl\trn{C})}$. 

In an arena $A\in Ob(\Vnr)$, a store-Handle justifies (all) questions of the form $\al$, which we call \boldemph{store-Questions}. Answers to store-Questions are called \boldemph{store-Answers}. %
\edefn%
Note in particular that, for each type $A$, we have $\trn{A},\prn\trn{A},T\trn{A}\in Ob(\Vnr)$, assuming that $T\trn{A}$ is equated with $1\him
T\trn{A}$. Note also there is a circularity in $H_{A\him TB}$ in the above definition. In fact, it is a definition by induction: we take
$H_A\defn\bigcup_{i\in\om}H_A^i$ and,
\begin{align*}
H_1^i&=H_{\GN}^i=H_{\GA}^i=H_A^0\defn\keno\,,\qd[1]
H_{A\ten B}^i\defn H_A^i\cup H_B^i\,, \\
H_{A\him TB}^{i+1}&\defn \{(i_A,\ixi_A),(i_B,\ixi_B)\}\cup H_A^i\cup H_B^i\cup H_{\xi_A}^{i+1}\cup H_{\xi_B}^{i+1} \qd\text{with}~
H_\xi^{i+1}\defn\bigcup\nolimits_{C}H_{\trn{C}}^i\,.
\end{align*}
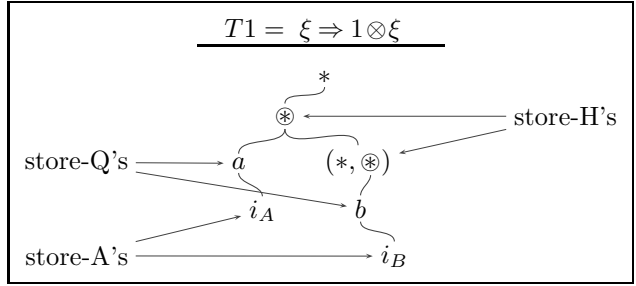
\piccaption{Store-H's -Q's -A's in arena $T1$.}\parpic[r]{%
\fbox{\small$\xymatrix@R=3pt@C-3.5 pt{
&\underline{\qd T1=\;\xi\impl 1\ten\xi\;\ \qd} \\
&\;\rnode{AA}{*} \\
&\;\rnode{C}{\ixi}\qd[3] & \rnode{H}{\text{store-H's}}\\
\rnode{Q}{\text{store-Q's}}&\;\rnode{D}{\al}\qquad\nccdu{D}{C}\qd[1]\rnode{B}{(*,\rnode{G}{\ixi})}{\quad}\nccdu{B}{C}\\
&\;\rnode{A}{i_A}\nccdu{A}{D}\qd[3]\rnode{E}{\be}\qd\;\nccdu{E}{G} \\
\rnode{AN}{\text{store-A's}} &\;\qd[5]\;\rnode{F}{i_B}\nccdu{F}{E}
\nccdu{C}{AA}
\ncline[linecolor=darkgray,linewidth=0.2pt,nodesep=3pt]{->}{Q}{D}\ncline[linecolor=darkgray,linewidth=0.2pt,nodesep=3pt]{->}{Q}{E}
\ncline[linecolor=darkgray,linewidth=0.2pt,nodesep=3pt]{->}{AN}{A}\ncline[linecolor=darkgray,linewidth=0.2pt,nodesep=3pt]{->}{AN}{F}
\ncline[linecolor=darkgray,linewidth=0.2pt,nodesep=3pt]{->}{H}{C}\ncline[linecolor=darkgray,linewidth=0.2pt,nodesep=3pt]{->}{H}{B}}$}}
\noindent%
Intuitively, store-H's are store-opening moves, while store-Q's and store-A's are obtained from unfolding the store structure. On the side we give examples of store-related moves in a simple arena.

From now on we work in $\Vnr$\,, unless stated otherwise. A first property we can show is that a move is exclusively either initial or an
element of the aforedefined move-classes.\picskip{0}

\beprop\label{p:storeH} For any $A\in Ob(\Vnr)$,
\[ M_{A}= I_{A} \uplus H_A \uplus \{\,m\in M_{A}\,|\,m\text{ a store-Q}\} \uplus \{m\in M_{A}\,|\,m\text{ a store-A}\,\}\,. \]
\enprop
\proof We show that any $m\in M_{A}$ belongs to exactly one of the above sets. We do induction on the level of $m$, $l(m)$, inside $A$ and on the size
of $A$, $|A|$, specified by the inductive definition of $Ob(\Vnr)$. If $m$ is initial then, by definition, it can't be a store-H. Neither can it be a
store-Q or store-A, as these moves presuppose non-initiality.

Assume $l(m)>0$. If $A$ is base then trivial, while if $A=A_1\ten A_2$ then use the IH on $(l(m),|A|)$. Now, if $A=A_1\him TA_2$ then let us write $A$
as ${A_1}\him (\xi_1\impl{A_2}\ten\xi_2)$; we have the following cases.
\begin{aDescription}[$\bullet$]{}
\item If $m=(i_{A_1},\ixi_1)\in H_A$ then $m$ a question and not a store-Q, as store-Q's are names.
\item If $m=(i_{A_2},\ixi_2)\in H_A$ then $m$ an answer and not a store-A as its justifier is $(i_{A_1},\ixi_1)$.
\item If $m$ is in ${A_1}$ or in ${A_2}$ then use the IH.
\item If $m$ is in $\xi_1$ then it is either some store-Q $\al$ to $(i_{A_1},\ixi_1)$ (and hence not a store-H or store-A), or it is in some $\trn{C}$.
    In the latter case, if $m$ initial in $\trn{C}$ then a store-A in $\trn{A}$ and therefore not a store-H, as $m$ not a store-H in $\trn{C}$ by IH
    (on $l(m)$). If $m$ is non-initial in $\trn{C}$ then use the IH and the fact that store-H's -Q's -A's of $\trn{C}$ are the same in $\trn{A}$.
\item Similarly if $m$ is in $\xi_2$. \qed
\end{aDescription}
The notion of store-handles can be straightforwardly extended to prearenas. %
\bdefn Let $A,B\in Ob(\Vnr)$. The set $H_{A\tote B}$ of store-handles in prearena
$A\tote B$ is $H_A\cup H_B$. Store-Q's and store-A's are defined accordingly. %
\edefn %
Using the previous proposition, we can see that, for any $A$ and $B$, the set $M_{{A}\tote{B}}$ can be decomposed as:
\begin{equation}
    I_{A} \uplus I_{B} \uplus H_{A\tote B} \uplus \{\,m\in M_{{A}\tote {B}}\,|\,m\text{ a store-Q}\,\} \uplus \{\,m\in M_{{A}\tote{B}}\,|\,m\text{ a store-A}\,\}
\end{equation}%
We proceed to define tidy strategies. We endorse the following notational convention. Since stores $\xi$ may occur in several places inside a
(pre)arena we may use parenthesised indices to distinguish identical moves from different stores. For example, the same store-question $q$ may be
occasionally denoted $q_{(O)}$ or $q_{(P)}$\,, the particular notation denoting the OP-polarity of the move. Moreover, by O-store-H's we mean
store-H's played by Opponent, etc.%
\bdefn[Tidy strategies] A total strategy $\sig$ is \boldemph{tidy} if whenever odd-length $[s]\in\sig$ then:
\begin{aDescription}{(TD1)}
\item[(TD1)] If $s$ ends in a store-Q $q$ then $[sx]\in\sig$\,, with $x$ being either a store-A to $q$ introducing no new names, or a copy of $q$.
    In particular, if $q=\al^\all$ with $\al\4\pv{s}^-$ then the latter case holds.
\item[(TD2)] If $[sq\indx{P}]\in\sig$ with $q$ a store-Q then $q\indx{P}$ is justified by last O-store-H in $\pv{s}$.
\item[(TD3)] If $\pv{s}=s'q\indx{O}q\indx{P}t\,y\indx{O}$ with $q$ a store-Q then $[sy\indx{P}]\in\sig$, where $y\indx{P}$ is justified by $\pv{s}.\arnt3$\,. \deq
\end{aDescription}
\edefn[q]%
(TD1) states that, whenever Opponent asks the value of a name, Player either immediately answers with its value or it copycats the question to the
previous store-H. The former case corresponds to Player having updated the given name lastly (i.e.~between the previous O-store-H and the last one).
The latter case corresponds to Player not having done so and hence asking its value to the previous store configuration, starting thus a copycat between the last and the previous store-H. Hence, the store is, in fact, composed by layers of stores\HY one on top of the other\HY and only when a name has not been updated in the top layer is Player allowed to search for it in layers underneath. We can say that this is the nominal games
equivalent of a \emph{memory cell} (cf.~remark~\ref{r:InnStore}).
(TD3) further guarantees the above-described behaviour. It states that when Player starts a store-copycat then he must copycat the store-A and all
following moves he receives, unless Opponent chooses to play elsewhere. (TD2) guarantees the multi-layer discipline in the store: Player can see one
store at each time, namely the last played by Opponent in the P-view.

The following straightforward result shows that (TD3), as stated, provides the intended copycat behaviour.
\beprop\label{p:tidy_CC} %
Let $\sig$ be a tidy strategy. If $[s'q\indx{O}q\indx{P}t]\in\sig$ is an even-length P-view and $q$ is a store-Q then $q\indx{O}q\indx{P}t$ is a
copycat.%
\enprop
\proof We do induction on $|t|$. The base case is straightforward. For the inductive step, let $t=t'xz$. Then, by prefix closure,
$[s'q\indx{O}q\indx{P}t'x]\in\sig$, this latter a P-view. By IH, $q\indx{O}q\indx{P}t'$ is a copycat. Moreover, by (TD3),
$[s'q\indx{O}q\indx{P}t'xx]\in\sig$ with last $x$ justified by $(q\indx{O}q\indx{P}t'x).\arnt3$, thus $s'q\indx{O}q\indx{P}t'xx$ a copycat. Now, by
determinacy, $[s'q\indx{O}q\indx{P}t'xx]=[s'q\indx{O}q\indx{P}t'xz]$, so there exists $\pi$ such that $\pi\actn x=x\Land\pi\actn x=z$, $\ara x=z$, as
required. \qed\noindent%
A \emph{good store discipline} would guarantee that store-Handles OP-alternate in a play. This indeed happens in P-views played by tidy strategies. In
fact, such P-views have canonical decompositions, as we show below.
\beprop[Tidy Discipline] \label{p:TidyDiscipline} Let $\sig:A\paei B$ be a tidy strategy and $[s]\in\sig$ with $\pv{s}=s$. Then, $s$ is decomposed as
in the following diagram.
{\tiny
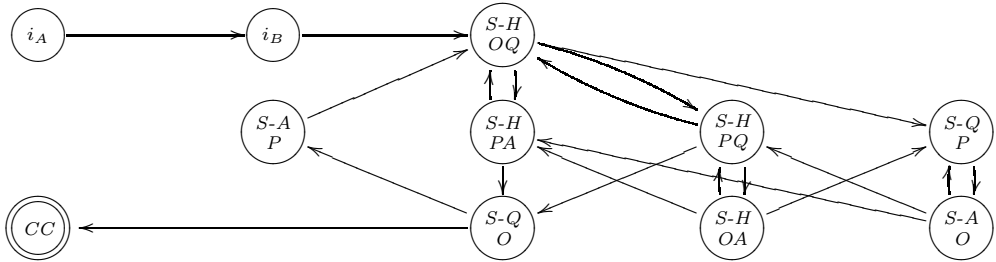
\[\xymatrix@C=60pt@R=10pt{%
*++[Fo]{i_A}\ar[r] & *++[Fo]{i_B}\ar[r] & \aaaa{S-H}{OQ}\ar@/^5pt/[d]\ar@/^5pt/[dr]\ar[drr] \\
& \aaaa{S-A}{P}\ar[ur] & \aaaa{S-H}{PA}\ar[d]\ar@/^5pt/[u] & \aaaa{S-H}{PQ}\ar@/^5pt/[d]\ar[dl]\ar@/^5pt/[ul] & \aaaa{S-Q}{P}\ar@/^5pt/[d] \\
*+++o[F=]{\text{CC}} && \aaaa{S-Q}{O}\ar[ul]\ar[ll] & \aaaa{S-H}{OA}\ar@/ ^5pt/[u]\ar[ul]\ar[ur] & \aaaa{S-A}{O}\ar[ull]\ar@/^5pt/[u]\ar[ul]}
\]}
{\scriptsize(by CC we mean the state that, when reached by a sequence $s=\pv{s}$, the rest of $s$ is copycat.)}
\enprop %
\proof The first two transitions are clear. After them neither P nor O can play initial moves, so all remaining moves in $s$ are store-H -Q
-A's. Assume now O has just played a question $x_0$ which is a store-H and the play continues with moves $x_1x_2x_3...$\,.

$x_1$ cannot be a store-A, as this would not be justified by $x_0$,
breaching well-bracketing. If $x_1$ is a store-Q then $x_2$ must be a store-A, by P-view. If $x_1$ is an answer-store-H then $x_2$
is an $OQ$, while if $x_1$ a question-store-H then $x_2$ is either a store-Q or a store-H.

If $x_2$ is a store-Q then, by (TD1), $x_3$ either a store-A or a store-Q, the latter case meaning transition to the CC state.
If $x_2$ is not a store-Q then $x_3$ can't be a store-A:
if $x_3$ were a store-A justified by $q\neq x_2$ then, as $q$ wouldn't have been immediately answered, $s_{\geq q}$ would be a copycat and therefore
we would be in the CC state right after playing $q$.

Finally, if $x_3$ is a store-A then $x_4$ must be justified by it, so it must be a Q-store-H. \qed
\becor[Good Store Discipline]\label{c:TidyDiscipline}%
Let $[s]\in\sig$ with $\sig$ tidy and $\pv{s}=s$. Then:
\begin{Itemize}
\item The subsequence of $s$ containing its store-H's is OP-alternating and $O$-starting.
\item If $s.\arnt1=q$ is a P-store-Q then either $q$ is justified by last store-H in $s$, or $s$ is in copycat mode at $q$. \qed
\end{Itemize}
\encor
\noindent%
Observe that strategies that mostly do copycats are tidy; in particular, identities are tidy. Moreover, tidy strategies are closed under composition (proof delegated to the appendix).
\beprop\label{p:tidy}%
If $\sig:A\paei B$ and $\tau:B\paei C$ are tidy strategies then so is $\st$. \qed %
\enprop
\bdefn
$\T$ is the lluf subcategory of $\Vnr$ of tidy strategies.
\edefn%
Finally, we need to check that all structure required for a sound $\laNR$-model pass from $\Vt$ to $\T$.
It is not difficult to see that all such structure which does not handle the store remains safely within the tidy universe. On the other hand, strategies for update and dereferencing are tidy by construction. (A fully formal proof is given in~\cite{Tze_PhD}.)
\beprop
$\T$ forms an adequate \laNR-model by inheriting all the necessary structure from $\Vt$\,. \qed
\enprop
Henceforth, by strategies we shall mean tidy strategies, unless stated otherwise.
\subsection{Observationality}
Strategy equality is \emph{too fine grained} to capture contextual equivalence in a complete manner.
For example, even simple contextual equivalences like
\[ \sskip\Eypob\new[\al]\sskip \]
are not preserved by the semantical translation, since strategies include in their name-lists all introduced names, even useless ones. For similar reasons, equivalences like
\[ \new[\al]\new[\be]M\Eypob\new[\be]\new[\al]M \]
are not valid semantically. In fact,
it is not only because of the treatment of name-creation that the semantics is not complete. Terms like
\[ \al:=1\1\la x.\bang\al\12 \ \Eypob \ \al:=1\1\la x.2 \]
are distinguished because of the `explicit' way in which the store works.

So there are many ways in which our semantics is too expressive for our language.
We therefore proceed to a quotienting by the intrinsic preorder and prove full-abstraction in the extensional model.
Following the steps described in section~\ref{s:CatSem}, in this section we introduce the intrinsic preorder on $\T$ and show that the resulting model is observational. Full-abstraction is then shown in the following section.
\bdefn%
Expand $\T$ to $\MODEL[]{\T,T,\prn[],\obs[]}$ by setting, for each $\all\in\A^{\#}$,
\begin{align*}
 & \obs\,\defn\{\,f\in\T(\prn1,T\GN)\,|\,\exists\bee.\, [(\all,*)\,*\,\ixi\,(0,\ixi)^{\bee}]\in f\,\}\,.
\intertext{Then, for each $f,g\in\T(\prn A,TB)$,\; $f\ypoc g$ if}
& \forall\rho:\prn(A\him TB)\paei T\GN.\ \ (\Laa(f)\1\rho\in\obs\implies \Laa(g)\1\rho\in\obs)\,.
\end{align*}\mindeq[-1]
\edefn[a]%
Thus, the observability predicate $\obs[]$ is a family $\MODEL{\obs}$, and the intrinsic preorder $\ypoc[]$ is a family $\MODEL{\ypoc}$. Recall that by
$\Laa(f)$ we mean $\LaQQT{\prn}(f)$, that is,
\[ \Laa(f)=\prn1\lred{\de}\prn\prn1\lred{\prn\La(\zet'\1f)}\prn(A\him TB)\,. \]
Note in particular that $f\ypoo g$ implies $\Laa(f)\1\rho\ypoo\Laa(g)\1\rho$, for any relevant $\rho$, and therefore:
\begin{equation} \label{e:ypooypoc}
 f\ypoo g\implies f\ypoc g
\end{equation}%
The intrinsic preorder is defined by use of \emph{test arrows} $\rho$, which stand for possible program contexts. As the following result shows, not
all such tests are necessary.
\belem[tl4 tests suffice] \label{l:tl4enough} %
Let $f,g\in\T(\prn1,B)$ with $B$ pointed. The following are equivalent (recall definition~\ref{NomGames:d:tl4etal}).
\begin{RmEnumerate}[xxxx]
\item $\forall\rho:\prn B\paei T\GN.\ \  \de\1\prn f\1\rho\in\obs \implies \de\1\prn g\1\rho\in\obs$
\item $\forall\rho:\prn B\paei T\GN.\ \ \rho\text{ is tl4}\implies(\de\1\prn f\1\rho\in\obs\implies\de\1\prn g\1\rho\in\obs)$
\end{RmEnumerate}
Hence, for each $\all$ and $f,g\in\T(\prn A,TB)$, $f\ypoc[\all]g$ iff
\[ \forall\rho:\prn(A\him TB)\paei T\GN.\ \ \rho\text{ is tl4}\implies(\Laa(f)\1\rho\in\obs\implies\Laa(g)\1\rho\in\obs)\,. \]
\enlem%
\proof $\text{I}\impl\text{II}$ is trivial. Now assume II holds and let $\rho:\prn B\paei T\GN$ be any strategy
such that $\de\1\prn f\1\rho\in\obs$. Then, there exist $[s]\in \de\1\prn f$ and $[t]\in\rho$ such that
$[s\1t]=[(\all,*)\,*\,\ixi\,(0,\ixi)^{\bee}]\in(\de\1\prn f)\1\rho$. We show by induction on the number of $J_B$-moves appearing in $s\comp t$ that
$\de\1\prn g\1\rho\in\obs$. 

If no such moves appear then $t=(\all,i_B)\,*\,\ixi\,(0,\ixi)^{\bee}$, so done. If $n+1$ such moves appear then $\rho$ is necessarily t4,
as $B$ is pointed, 
so by lemma~\ref{NomGames:l:SepHead} there exists tl4* strategy $\tilde\rho$ such that $\rho=\De\1\tilde\rho$. It is not difficult to see that $\rho$ being tidy implies that $\tilde\rho$ is tidy. Moreover,
$\de\1\prn f\1\rho=\de\1\prn f\1\De\1\tilde\rho 
=\de\1\prn f\1\ang{\id,\prn!\1\delta\1\prn f}\1\tilde\rho=\de\1\prn
f\1\rho'$\,, with $\rho'$ being $\ang{\id,\prn!\1\delta\1\prn
f}\1\tilde\rho$. Now, by definition of $\tilde\rho$,
$[(\all,*)\,*\,\ixi\,(0,\ixi)^{\bee}]=[s'\1t']\in \de\1\prn f\1\rho'$
with $s'\comp t'$ containing $n$ $J_B$-moves so, by IH, $\de\1\prn
g\1\rho'\in \obs$. But $\de\1\prn g\1\rho'=\de\1\prn
g\1\ang{\id,\prn!\1\delta\1\prn f}\1\tilde\rho=\de\1\prn
f\1\ang{\prn!\1\delta\1\prn g,\id}\1\tilde\rho=\de\1\prn f\1\rho''$\,,
where $\rho''$ is given by $\ang{\prn!\1\delta\1\prn g,\id}\1\tilde\rho$.
But $\rho''$ is tl4, thus, by hypothesis, $\obs\ni \de\1\prn g\1\rho''=\de\1\prn g\1\rho$\,, as required.
\qed\noindent%
We can now prove the second half of observationality. %
\belem\label{l:sempreorder}%
For any morphism $f:\prn[\all\al]1\paei B$, with $B$ pointed, and any tl4 morphism $\rho:\prn B\paei T\N$,
\[ \de\1\prn\abs f\1\rho\in\obs \iff \de\1\prn[\all\al]f\1\pit{\all\al}{\all}\1\rho\in\obs[\all\al] \]
Moreover, for each $\all$ and relevant $\al,\all',f,g$,
\[
    f\ypoc[\allal]g \implies \abs f\ypoc\abs g \,,\qd
    f\ypoc g \implies \pit{\all'}{\all}\1f \ypoc[\all'] \pit{\all'}{\all}\1g\,.
\]
\enlem %
\proof For the first part, $\rho$ being tl4 and $B$ being pointed imply that there exists some $\bee\4\all$ and a ttotal strategy $\rho'$ such that $\rho=\abs[\bee]\rho'$. Now let $\de\1\prn\abs f\1\rho\in\obs$, so there exists
$[s\1t]=[(\all,*)\,*\,\ixi\,(0,\ixi)^{\bee\al\gaa}]\in (\de\1\prn\abs f)\1\rho$, and let $s=(\all,*)\,(\all,i_B)\,j_B\,m^{\al\dee}s'$ and
$t=(\all,i_B)\,*\,\ixi\,j_B^{\bee}\,t'$. Letting $s\plin{\al}$ be $\ul{s}^{\nlist(s)\listplin\al}$, we can see that
$[(\allal,*)\,i_B\,j_B\,m^{\dee}s'\plin{\al}]\in f$ and thus
$[s'']\defn[(\all\al,*)\,(\all,i_B)\,j_B\,m^{\dee}s'\plin{\al}]\in \de\1\prn[\all\al]f\1\pit{\all\al}{\all}$\,.
Hence, $[s''\1t]=[(\all\al,*)\,*\,\ixi\,(0,\ixi)^{\bee\gaa}]\in \de\1\prn[\all\al]f\1\pit{\all\al}{\all}\1\rho$, as required. The converse
is shown similarly.

For the second part, suppose $f\ypoc[\allal]g:\prn[\all\al]A\paei TB$ and take any tl4 morphism $\rho:\prn(A\him TB)\paei T\GN$. Then,
\begin{align*}
\Laa(\abs f)\1\rho\in\obs &\iff \de\1\prn\La(\zet'\1\abs f)\1\rho\in\obs
    \overset{\text{lem~\ref{NomGames:l:LambdaAbs}}}{\iff}\de\1\prn\abs(\La(\zet'\1f))\1\rho\in\obs \\
    &\iff \de\1\prn[\all\al]\La(\zet'\1f)\1\pit{\all\al}{\all}\1\rho \in \obs[\all\al] \\
    &\overset{f\ypoc[\all\al]g}{\implies}\de\1\prn[\all\al]\La(\zet'\1g)\1\pit{\all\al}{\all}\1\rho \in \obs[\all\al]
    \iff \Laa(\abs g)\1\rho\in\obs\,.
\end{align*}
For the other claim, let us generalise the fresh-name constructors $\frs$ to:
\[
    \binom{\all}{\all'}:\GA\paei(\GA[\all'])_\bot\defn\{\,[\,(\all,*)\,*\,*\,(\all',*)^{\all'{\scriptscriptstyle\smallsetminus}\all}]\,\}
\]
for any $\supp(\all)\ypo\supp(\all')$. The above yields a natural transformation of type $\prn\paei\prn[\all']_\bot$. It is easy to see that, for any $h:\prn[\all']1\paei T\GN$, $h\in\obs[\all']$ iff $\tbinom{\all}{\all'}\1h_\bot\1\puu\in\obs$ and, moreover, that the diagram on the right below commutes.
Hence, if $f\ypoc g$ then
\[ \begin{array}{l@{\qd}|@{\qd}r}
\begin{aligned}
\de\1\prn[\all']&\La(\zet'\1\pit{\all'}{\all}\1f)\1\rho\in\obs[\all']\\
    &\iff \de\1\prn[\all']\pit{\all'}{\all}\1\prn[\all']\La(\zet'\1f)\1\rho\in\obs[\all'] \\
    &\iff \tbinom{\all}{\all'}\1(\de\1\prn[\all']\pit{\all'}{\all}\1\prn[\all']\La(\zet'\1f)\1\rho)_\bot\1\puu\in\obs \\
    &\iff \de\1\prn\La(\zet'\1f)\1\tbinom{\all}{\all'}\1\rho_\bot\1\puu\in\obs \\
    &\overset{f\ypoc g}{\implies}\de\1\prn\La(\zet'\1g)\1\tbinom{\all}{\all'}\1\rho_\bot\1\puu\in\obs \\
    &\iff \de\1\prn[\all']\La(\zet'\1\pit{\all'}{\all}\1g)\1\rho\in\obs[\all'],
\end{aligned}
&
\begin{array}{c}\!\!\!
\xymatrix@C=14mm@R=11mm{ \GA\ar[d]_{\tbinom{\all}{\all'}}\ar[r]^-{\ang{\tbinom{\all}{\all'},\id}} & (\GA[\all'])_\bot\otimes\GA\ar[d]^{\stt'} \\
                    (\GA[\all'])_\bot\ar[r]_-{\ang{\id,\pit{\all'}{\all}}_\bot} & (\GA[\all']\otimes\GA)_\bot}
\end{array}\end{array}\]
as required. \qed\noindent%
In order to prove that $\T$ is observational, we are only left to show that
\[ \trn{M}\in\obs \iff \exists\bee,S.\,\trn{M}=\abs[\bee]\trn{\bS\1 0} \]
for any $\Aseq{\keno}{M:\N}$. The ``$\Longleftarrow$" direction is trivial. For the converse, because of correctness, it suffices to show the following
generalisation of adequacy.
\belem[$O$-Adequacy] \label{l:OAdeq} %
Let $\Aseq{\keno}{M:\N}$ be a typed term. If $\trn{M}\in\obs[\ee]\,$ then there exists some $S$ such that $\all\modl M\rred S\modl 0$.%
\enlem%
\proof The idea behind the proof is given above proposition~\ref{p:Adequacy}.
It suffices to show that, for any such
$M$, there is a non-reducing sequent $S\modl N$ such that $\all\modl M\rred S\modl N$\,; therefore, because of Strong Normalisation in the
$\nu$-calculus, it suffices to show that there is no infinite reduction sequence starting from $\all\modl M$ and containing infinitely many DRF
reduction steps.

To show the latter we will use an operation on terms adding new-name constructors just before dereferencings. The operation yields, for each term $M$,
a term $\oted{M}$ the semantics of which is equivalent to that of $M$. On the other hand, $\all\modl\oted{M}$ cannot perform infinitely many DRF
reduction steps without creating infinitely many new names. For each term $M$, define $\oted{M}$ by induction as:
\[ \oted{\al}\defn\al\;,\quad \oted{x}\defn x\;,\quad ...\quad \oted{\la x.M}\defn\la x.\oted{M}\;,\quad \oted{M\, N}\defn\oted{M}\oted{N}\;,\quad... \]
and $\oted{\bang N}\defn\new[\al]\bang\oted{N}$\;, \ some $\al\4N$.

We show that $\trn{\oted{M}}\backsimeq\trn{M}$, by induction on $M$; the base cases are trivial. The induction step follows immediately from the IH and the fact that $\backsimeq$ is a congruence, in all cases except for $M$ being $\bang N$. In the latter case we have that
$\trn{\oted{M}}=\abs(\pit{\all\al}{\all}\1\trn{\bang\oted{N}})$\,, while the IH implies that $\trn{M}\backsimeq\trn{\bang\oted{N}}$. Hence, it
sts that for each $f:\prn A\paei TB$ we have $f\backsimeq\abs(\pit{\all\al}{\all}\1f)$\,. Indeed, for any relevant $\rho$ which is tl4,
\begin{align*}
\Laa(\abs(\pit{\all\al}{\all}\1 f))\1\rho\in\obs
    &\overset{\text{lem~\ref{l:sempreorder}}}{\iff} \de\1\prn[\all\al]\La(\zet'\1\pit{\all\al}{\all}\1f)\1\pit{\all\al}{\all}\1\rho\in\obs[\all\al] \\
    &\iff \de\1\prn[\all\al]\pit{\all\al}{\all}\1\pit{\all\al}{\all}\1\prn\La(\zet'\1f)\1\rho\in\obs[\all\al] \\
    &\iff \pit{\all\al}{\all}\1\Laa(f)\1\rho\in\obs[\all\al] \iff \Laa(f)\1\rho\in\obs.
\end{align*}
Now, take any $\Aseq{\keno}{M:\N}$ and assume $\trn{M}\in \obs$, and that $\all\modl M$ diverges using infinitely many DRF reduction steps. Then,
$\all\modl\oted{M}$ diverges using infinitely many NEW reduction steps. However, since $\trn{\oted{M}}\backsimeq\trn{M}$, we have $\trn{\oted{M}}\in
\obs$ and therefore $[(\all,*)\,*\,\ixi\,(\nm{0},\ixi)^{\bee}]\in\trn{\oted{M}}$ for some $\bee$. However, $\all\modl\oted{M}$ reduces to some $S\modl
M'$ using $|\bee|+1$ NEW reduction steps, so $\trn{\oted{M}}=\abs[\gaa]\trn{\bS\1M'}$ with $|\gaa|=|\bee|+1$, \contra to determinacy. %
\qed\noindent%
We have therefore shown observationality.
\beprop[Observationality]\label{p:p-obs} $\MODEL[]{\T,T,\prn[],\obs[]}$ is observational. \qed\enprop
\vskip-8 pt%

\subsection{Definability and full-abstraction}
We now proceed to show definability for $\T$, and through it ip-definability. According to the results of section~\ref{s:Compl}, this will suffice for full abstraction.

We first make precise the notion of \emph{finitary strategy}, that is, of (tidy) strategy with finite description, by introducing truncation functions that remove inessential branches from a strategy's description.
\bdefn\label{d:finitary} %
Let $\sig:{A}\paei{B}$ in $\T$ and let $[s]\in\viewf(\sig)$ be of even length. Define $\trunc(s)$ and $\trunc'(s)$ by induction as follows.
\begin{align*}
\trunc(\ee)=&\ \trunc'(\ee)\defn\ee \\
\trunc(x\indx{O}y\indx{P}s')\defn&\begin{cases}
    \ee &\text{ , if $x=y$ are store-Q's}\\
    xy\,\trunc(s') &\text{ , o.w.}\end{cases}\\
\trunc'(x\indx{O}y\indx{P}s')\defn&\begin{cases}
    \ee &\text{ , if $x=y$ are store-Q's}\\
    \ee &\text{ , if $x$ store-Q , $y$ a store-A and $s'=\ee$}\\
    \ee &\text{ , if $x\in I_A$, $y\in I_B$ and $s'=\ee$}\\
    xy\,\trunc'(s') &\text{ , o.w.}\end{cases}
\end{align*}
Moreover, say $\sig$ is \boldemph{finitary} if $\trunc(\sig)$ is finite, where
\[ \trunc(\sig)\defn\{\,[\trunc(s)]\,|\,[s]\in\viewf(\sig)\Land |s|>3\,\}\,. \]
Finally, for any $[t]\in\sig$ define:
$$ \sig\trnat{t}\defn\strat\{\,[s]\in\viewf(\sig)\,|\,\exists\, t'\leq t.\ \trunc'(s)=\pv{t'}\,\}\,.\eqno{\blacktriangle}$$
\edefn[a]
Hence, finitary are those strategies whose viewfunctions become finite if we delete all the store-copycats and all default initial
answers\HY the latter dictated by totality. Moreover, the strategy $\sig\trnat{t}$ is the strategy we are left with if we truncate $\viewf(\sig)$ by
removing all its branches of size greater than $3$ that are not contained in $t$, except for the store-copycats which are left intact and for the store-A's branches which are truncated to the point of leaving solely the store-A, so that we retain tidiness.
Note that, in general, $\trunc'(s)\leq\trunc(s)\leq s$. We can then show the following (proof in~\cite{Tze_PhD}). %
\beprop\label{NR:p:trunc} %
If $\sig$ is a strategy and $[t]\in\sig$ is even-length then $\sig\trnat{t}$ is a finitary strategy with $[t]\in\sig\trnat{t}$ and $\sig\trnat{t}\ypoo\sig$.\qed %
\enprop %

\noindent We proceed to show definability. The proof is facilitated by the following lemma, the proof of which is delegated to the appendix.
Note that for economy we define strategies by means of their viewfunctions modulo totality and even-prefix closure. Moreover, we write $\sigma\hrp i$ for the (total) restriction of a strategy $\sigma$ to an initial move $i$, and $s\plin{\bee}$ for $s$ with $\bee$ removed from all of its name-lists.

\belem[Decomposition Lemma]\label{l:Decomp}%
Let $\sig:\prn\trn{A}\paei T\trn{B}$ be a strategy. We can decompose $\sig$ as follows.
\begin{cEnumerate}{aa}
\item If there exists an $i_{A(0)}$ such that
    $\exists x_0.\,[(\all,i_{A(0)})\,*\ixi x_0]\in\sig$ then
\[\vcenter{\xymatrix{
  \prn\trn{A}
         \ar[d]_{\sig}
         \ar[dr]^{\quad\ang{[x\overset{\all}{=}i_{A(0)}],\ang{\sig_0,\sig'}}}
         \cr
   T\trn{B}
  &\GN\ten(T\trn{B})^2
         \ar[l]^-{\cnd}
  }}
\]         
  where:
\begin{align*}
  [x\overset{\all}{=}i_{A(0)}]
&:\prn\trn{A}\paei\Rlap{\GN}\phantom{T\trn{B}}
  \defn \{\,[(\all,i_{A(0)})\, 0\,]\}\cup\{\,[\,(\all,i_A)\,1\,]\,|\,
          [\,(\all,i_A)\,]\neq[\,(\all,i_{A(0)})\,]\,\}\,, \\
  \sig_0
&:\prn\trn{A}\paei T\trn{B}
  \defn\strat\{\,[\,(\all,i_{A(0)})\, s\,]\in\viewf(\sig)\,\}\,, \\
  \sig'
&:\prn\trn{A}\paei T\trn{B} 
  \defn\strat\{\,[\,(\all,i_A)\, s\,]\in\viewf(\sig)\,|\,
                 [\,(\all,i_A)\,]\neq[\,(\all,i_{A(0)})\,]\,\}\,.
\end{align*}

\item If there exists $i_{A(0)}$ such that $\forall i_A.\,(\exists x_0.\,
  [(\all,i_A)\,*\,\ixi\,x_0]\in\sig)\!\iff\![(\all,i_A)]=[(\all,i_{A(0)})]$\,, 
  then $\sig=\abs[\bee]\sig_\bee$ where:
\begin{align*}
  \sig_\bee:\prn[\all\bee]\trn{A}\paei T\trn{B}\defn\strat\{\,&
 [\,(\all\bee,i_{A(0)})*\,\ixi\, m_0\, s\plin{\bee}\,]\,|\,\cr
&[\,(\all,i_{A(0)})\,*\,\ixi\, m_0^{\bee}\,s\,]\in\viewf(\sig)\,\}\,.
\end{align*}

\item If there exist $i_{A(0)},m_0$ such that
    $\forall i_A,x.\, [(\all,i_A)\,*\,\ixi\, x]\in\sig\!\!\iff\!\! [(\all,i_A)\,x]=[(\all,i_{A(0)})\,m_0]$\,,
    then one of the following is the case.
\begin{Description}{w.}
\item[{\bf(a)\!}] $m_0=\al$, a store-Q of type $C$ under $\ixi$, in
  which case $\sig=\sig'\hrp(\all,i_{A(0)})$ where 
\begin{align*}
  \sig':\prn\trn{A}\paei T\trn{B}
 &\defn\ang{\id,\phi};\tau\1T\zet';T\sig_\al;\mu\\
  \sig_\al:\prn(\trn{A}\ten\trn{C})\paei T\trn{B}
 &\defn\strat\{\,[\,(\all,i_{A(0)},i_C)\,*\,\ixi\, s]\,|\,\\
 &\phantom{{}\defn\strat\{\,}[\,(\all,i_{A(0)})\,*\,\ixi\,\al\, i_C\,s\,]\in\viewf(\sig)\ \}\,, \\
  \phi:\prn\trn{A}\paei T\trn{C}
 &\defn
    \begin{cases}
        \prn!\1\pit{\all}{\al}\1\drf[C]     &\text{, if }\al\in\supp(\all)\\
        \prn\pi_j\1\pit{\all}{\ee}\1\drf[C] &\text{, if }\al\4\all\,.
    \end{cases}
    \end{align*}

\item[{\bf(b)\!}] $m_0=j_A\lor m_0=(i_B,\ixi)\,$, a store-H, in which
    case if $[\,(\all,i_{A(0)})\,*\,\ixi\, m_0\,\al\,i_C\,]\in\sig$, for
    some store-Q $\al$ and store-A $i_C$, then 
\[\vcenter{\xymatrix@C+12 pt{
  \prn\trn{A}
          \ar[d]_{\sig}
          \ar[r]^-{\ang{\De,\sig_\al}}
 &\prn\trn{A}\ten\prn\trn{A}\ten T\trn{C}
          \ar[d]^{\tau;T(\id\tenn\phi;\tau);\mu}
          \cr
  T\trn{B}
 &T\prn\trn{A}
          \ar[l]^{T\sig'\1\mu}
  }}
\]
  where:  

\begin{align*}  
  \sig_\al:\prn\trn{A}\paei T\trn{C}
 &\defn\strat\{\,[\,(\all,i_{A(0)})\,*\,\ixi\,(i_C,\ixi)\, s\,]\,|\,\\
 &\phantom{{}\defn\strat\{\,}
        [\,(\all,i_{A(0)})\,*\,\ixi\,m_0\,\al\,i_C\,s\,]\in\viewf(\sig)\\
 &\phantom{{}\defn\strat\{\,}\lor[\,\ixi\,\ixi\,s\,]\in\viewf(\id_\xi)\,\}\,,\\
  \sig':\prn\trn{A}\paei T\trn{B}
 &\defn\strat(\,\{\,[\,(\all,i_{A(0)})\,*\,\ixi\,m_0\,y\,s\,]%
                            \in\viewf(\sig)\,|\,y\neq\al\,\}  \\
 &\phantom{{}\defn\strat(\,}%
  \cup\{\,[\,(\all,i_{A(0)})\,*\,\ixi\,m_0\,\al\,s\,]\,|\\
 &\qd[6]\,\,\,[\,\ixi\,\ixi\,\al\, s\,]\in\viewf(\id_\xi)\}\,)\,, \\
  \phi:\prn\trn{A}\ten\trn{C}\paei T1
 &\defn
    \begin{cases}(\prn!\1\pit{\all}{\al})\ten\id_{\trn{C}}\1\upd[C]    &\text{, if }\al\in\supp(\all) \\
                 (\prn\pi_j\1\pit{\all}{\ee})\ten\id_{\trn{C}}\1\upd[C]&\text{, if }\al\4\all\,.
    \end{cases}
\end{align*}
\end{Description}
In both cases above, we take $j=\min\{\,j\,|\,(i_{A(0)})_j=\al\,\}$. \qed
\end{cEnumerate}\enlem%
The proof of definability is a nominal version of standard definability results in game semantics. In fact, using the Decomposition Lemma we reduce the problem of definability of a finitary strategy $\sig$ to that of definability of a finitary strategy $\sig_0$ of equal length, with $\sig_0$ having no
initial effects (i.e.~fresh-name creation, name-update or name-dereferencing). On $\sig_0$ we then apply almost verbatim the methodology of \cite{Honda:CBV}\HY itself based on previous proofs of definability.

\bethm[Definability]\label{t:Dfn}\label{NR:t:Dfn}%
Let $A,B$ be types and $\sig:\prn\trn{A}\paei T\trn{B}$ be finitary. Then $\sig$ is definable.%
\enthm%
\proof We do induction on $(|\trunc(\sig)|,\|\sig\|)$, where we let $\|\sig\|\defn\max\{\,|\LL(s)|\,|\,[s]\in\viewf(\sig)\,\}$, i.e.~the maximum number of names introduced in any play of $\trunc(\sigma)$. If $|\trunc(\sig)|=0$ then
$\sig=\trn{\sstop_B}$\,; otherwise, there exist $x_0,i_{A(0)}$ such that $[(\all,i_{A(0)})*\ixi\,x_0]\in\sig$\,. By Decomposition Lemma,
\[ \sig=\ang{[x\overset{\all}{=}i_{A(0)}],\ang{\sig_0,\sig'}};\cnd \]
with $|\trunc(\sig')|<|\trunc(\sig)|$ and $(0,0)<(|\trunc(\sig_0)|,\|\sig_0\|)\leq(|\trunc(\sig)|,\|\sig\|)$\,, so by IH there exists term $M'$ such
that $\trn{M'}=\sig'$. Hence, if there exist terms $M_0,N_0$ with $\text{$\trn{M_0}\hrp(\all,i_{A(0)})=\sig_0$}$ and $\trn{N_0}=[x\overset{\all}{=}i_{A(0)}];\eta$\,, then we can see that
\[ \sig=\trn{\ifo{N_0}{M_0}{M'}}\,. \]
We first construct $N_0$\,. Assume that $A=A_1\times
A_2\times\cdots\times A_n$ with $A_i$'s non-products, and similarly
$B=B_1\times\cdots\times B_m$. Moreover, assume without loss of generality that $A$ is segmented in four parts: each of $A_1,...,A_k$ is $\N$; each of $A_{k+1},...,A_{k+i},...,A_{k+k'}$ is $[A_i''']$;
each of $A_{k+k'+1},...,A_{k+k'+i},...,A_{k+k'+k''}$ is $A_i'\tote A_i''$; and the rest are all $\ena$. Take $\zz,\zz',\zz'',\zz'''$ to be
variable-lists of respective types. Define $\phi_0,\phi'_0$ by:
\begin{align*}
\phi_0 &\defn \kappa_1,...,\kappa_k\,,\text{ with $(\kappa_1,...,\kappa_k)$ being the initial $\N$-segment of $i_{A(0)}$}\,, \\[1.5mm]
\phi'_0&\defn \kappa_1',...,\kappa_{k'}'\,,\text{ with each }\kappa_i'\defn
\begin{cases}(i_{A(0)})_{k+i}\!\!\!\!&\text{, if }(i_{A(0)})_{k+i}\in\supp(\all)\\
                 z'_j       \!\!\!\!&\text{, if }(i_{A(0)})_{k+i}\4\all \\
                            \!\!\!\!&\qd\ \Land j=\min\{\,j<i\,|\,(i_{A(0)})_{k+i}=(i_{A(0)})_{k+j}\,\}\\
                 \mathtt{fresh}(i)\!\!\!\!&\text{, otherwise}\,. \end{cases}
\end{align*}
$\mathtt{fresh}(i)$ is a meta-constant denoting that Opponent has played a fresh name in $A_{k+i}$. If the same fresh name is played in several places
inside $i_{A(0)}$ then we regard its leftmost occurrence as introducing it\HY this explains the second item in the cases-definition above. Now, define
\begin{align*}
N_0 &\defn [\ang{\zz,\zz'}=\ang{\phi_0,\phi'_0}]\qd\text{where:}\\
[\ang{\zz,\zz'}=\ang{\vec\kappa,\vec\kappa'}]&\defn [z_1=\kappa_1]\land\cdots\land[z_k=\kappa_k]\land[z_1'=\kappa_1']\land\cdots\land
    [z_{k'}'=\kappa_{k'}']\,,\\
[z'=\mathtt{fresh}(i)]&\defn [z'\neq\al_1]\land\cdots\land[z'\neq\al_{|\all|}]\land[z'\neq z'_1]\land\cdots\land[z'\neq z'_{i-1}]\,,
\end{align*}
with the logical connectives $\land$ and $\lnot$ defined using $\iifo$'s, and $[z_i=\kappa_i]$ using $\pred$'s, in the standard way. It is not difficult to show that indeed $\trn{N_0}\overset{\all}{=}[x=i_{A(0)}];\eta$\,.

We proceed to find $M_0$\,. By second part of Decomposition Lemma, $\sig_0=\abs[\bee]\sig_\bee$ with $\bee=\nlist(x_0)$,
$|\trunc(\sig_\bee)|=|\trunc(\sig_0)|$\, and $\|\sig_\bee\|=\|\sig_0\|-|\bee|$\,. If $|\bee|>0$ then, by IH, there exists term $M_\bee$ such that
$\trn{M_\bee}=\sig_\bee$\,, so taking
\begin{equation*}
    M_0\defn\new[\bee]M_\bee
\end{equation*}%
we have $\sig_0=\trn{M_0}$\,.

Assume now $|\bee|=0$, so $x_0=m_0$. $\sig_0$ satisfies the hypotheses of the third part of the Decomposition Lemma. Hence, if $m_0=\al$, a store-Q of
type $C$ under $\ixi$, then
\[ \sig_0=(\ang{\id,\phi}\1\tau\1T\zet'\1T\sig_\al\1\mu)\hrp (\all,i_{A(0)}) \]
with $\trunc(\sig_\al)<\trunc(\sig_0)$\,. Then, by IH, there exists $\Gseq[y\o C]{\all}{M_\al:B}$ such that $\sig_\al=\trn{M_\al}$\,, and taking
\begin{equation*}
M_0\defn\begin{cases}   (\la y.M_\al)(\bang\al)  &\text{, if }\al\in\supp(\all)\\
                        (\la y.M_\al)(\bang z_j') &\text{, if }\al\4\all\land j=\min\{\,j\,|\,\al=(i_{A(0)})_{k+j}\,\} \end{cases}
\end{equation*}%
we have $\sig_0=\trn{M_0}\hrp(\all,i_{A(0)})$.

Otherwise, $m_0=j_A\,\lor m_0=(i_B,\ixi)$, a store-H. If there exists an $\al\in\NA{C}$ such that $\sig_0$ answers to $[i_{A(0)}\,*\,\ixi\,m_0\,\al]$
then, by Decomposition Lemma,
\[ \sig_0 = \ang{\Delta,\sig_\al}\1\tau\1T(\id\ten\phi\1\tau)\1\mu\1T\sig'\1\mu \]
with $|\trunc(\sig_\al)|\,,|\trunc(\sig')|<|\trunc(\sig_0)|$\,. By IH, there exist $\Gseq{\all}{M_\al:C}$ and $\Gseq{\all}{M':B}$ such that
$\sig_\al=\trn{M_\al}$ and $\sig'=\trn{M'}$. Taking
\begin{equation*}
M_0\defn\begin{cases} (\al:= M_\al);M'   &\text{ , if }\al\in\supp(\all)\\
                      (z_j':= M_\al);M'  &\text{ , if }\al\4\all\land j=\min\{\,j\,|\,\al=(i_{A(0)})_{k+j}\,\} \end{cases}
\end{equation*}%
we obtain $\sig_0=\trn{M_0}$\,. Note here that $\sig_\al$ blocks initial moves $[\all,i_A]\neq[\all,i_{A(0)}]$ and hence we do not need the
restriction.

We are left with the case of $m_0$ being as above and $\sig_0$ not answering to any store-Q, which corresponds to the case of Player not updating any
names before playing $m_0$.  
\begin{aDescription}{}
\item[If $m_0=(i_B,\ixi)$] then we need to derive a value term $\ang{V_1,...,V_m}$ (as $B=B_1\times\cdots\times B_m$).
    For each $p$, if $B_p$ is a base or reference type then we can choose a $V_p$ canonically so that its denotation be $i_{B_p}$
    (the only interesting such case is this of $i_{B_p}$ being a name $\al\4\all$, where we take $V_p$ to be $z_j'$\,,
    for $j=\min\{\,j\,|\,\al=(i_{A(0)})_{k+j}\,\}$).
    Otherwise, $B_p=B_p'\tote B_p''$ and from $\sig_0$ we obtain the (tidy) viewfunction $f:\prn(\trn{A}\ten\trn{B_{p}'})\paei T\trn{B_{p}''}$ \ by:
    \[ f\defn\{\,[\,(\all,i_{A(0)},i_{B_{p}'})\,*\,\ixi\, s\,]\,|\,
        [\,(\all,i_{A(0)})\,*\,\ixi\,(i_B,\ixi)\,(i_{B_{p}'},\ixi)\, s\,]\in\viewf(\sig_0)\,\}. \]
    Note that, for any $[(\all,i_A)\,*\,\ixi\,(i_B,\ixi)\,(i_{B_{p}'},\ixi)\, s]\in\viewf(\sig_0)$, $s$ cannot contain store-Q's justified
    by $\ixi\,$, as these would break (TD2).
    Hence, $f$ fully describes $\sig_0$ after $(i_{B_{p}'},\ixi)\,$. By IH, there exists $\Gseq[y\o B_{p}']{\all}{N:B_{p}''}$ such that
    $\trn{N}=\strat(f)$\,; take then $V_p\defn\la y.N$. Hence, taking
    \begin{equation*} M_0\defn \ang{V_1,...,V_m} \end{equation*}%
    we obtain $\sig_0=\trn{M_0}\hrp (\all,i_{A(0)})$.
\item[If $m_0=j_A$,] played in some $A_{k+k'+i}=A_i'\tote A_i''$, then $m_0=(i_{A_i'},\ixi)\,$.
    Assume that $A_i'=A_{i,1}'\times\cdots\times A_{i,n_i}'$ with $A_{i,p}'$'s being non-products.
    Now, O can either ask some name $\al$ (which would lead to a store-CC), or answer at $A_i''$, or play at some $A_{i,p}'$ of arrow type,
    say $A_{i,p}'=C_{i,p}\tote C'_{i,p}$\,. Hence,
    \[ \viewf(\sig_0)=f_A\cup\bigcup\nolimits_{p=1}^{n_i}f_p  \text{ \ where:} \]
    \begin{leftalign}
    f_A&\defn f_0\cup \{\,[\,(\all,i_{A(0)})\,*\,\ixi\,(i_{A'_i},\ixi)\,(i_{A''_i},\ixi)\, s\,]\in\viewf(\sig_0)\,\} \\
    f_p&\defn f_0\cup \{\,[\,(\all,i_{A(0)})\,*\,\ixi\,(i_{A'_i},\ixi)\,(i_{C_{i,p}},\ixi)\, s\,]\in\viewf(\sig_0)\,\} \\
    f_0&\defn\{\,[\,(\all,i_{A(0)})\,*\,\ixi\,(i_{A'_i},\ixi)\, s]\,|\,[\ixi\,\ixi\, s\,]\in\viewf(\id_\xi)\,\}
    \end{leftalign}%
    and where we assume $f_p\defn f_0$ if $A'_{i,p}$ is not an arrow type.
    It is not difficult to see that $f_A,f_p$ are viewfunctions. Now, from $f_A$ we obtain:
\begin{align*}
  f'_A:\prn(\trn{A}\ten\trn{A_i''})\paei T\trn{B}
 &\defn\{\,[\,(\all,i_{A(0)},i_{A_i''})\,*\,\ixi\, s\,]\,|\,\\
 &\phantom{{}\defn\{\,}[(\all,i_{A(0)})\, *\,\ixi\, (i_{A_i'},\ixi)\,
        (i_{A_i''},\ixi)\, s\,]\in f_A\,\}\,. 
\end{align*}
    It is not difficult to see that $f'_A$ is indeed a viewfunction (note that P cannot play a store-Q under $\ixi$ on the RHS
    once $(i_{A_i''},\ixi)$ is played, by tidiness). By IH, there exists some $\Gseq[y\o A_i'']{\all}{M_A:B}$ such that
    $\trn{M_A}=\strat(f'_A)$. 

    From each $f_p\neq f_0$ we obtain a viewfunction $f'_p:\prn(\trn{A}\ten\trn{C_{i,p}})\paei T\trn{C'_{i,p}}$ \ by:
    \[ f'_p\defn\{\,[\,(\all,i_{A(0)},i_{C_{i,p}})\,*\,\ixi\, s\,]\,|\,
        [\,(\all,i_{A(0)})\,*\,\ixi\,(i_{A_i'},\ixi)\,(i_{C_{i,p}},\ixi)\, s\,]\in f_p\,\}\,. \]
    By IH, there exists some $\Gseq[y'\o C_{i,p}]{\all}{M_p:C'_{i,p}}$ such that $\trn{M_p}=\strat(f'_p)$\,, so take $V_p\defn\la y'.M_p$.
    For each $A_{i,p}'$ of non-arrow type, the behaviour of $\sig_0$ at $A_{i,p}'$ is fully described by $(i_{A_i'})_p$\,, so we choose $V_p$
    canonically as previously. $\ang{V_1,...,V_{n_i}}$ is now of type
    $A_i'$ and describes $\sig_0$'s behaviour in $A_i'$.

    Now, taking
    \begin{equation*}
        M_0\defn (\la y.M_A)(z_i''\ang{V_1,...,V_{n_i}})
    \end{equation*}%
    we obtain $\sig_0=\trn{M_0}\hrp (\all,i_{A(0)})$. \qed
\end{aDescription}%
Finally, using the definability result and proposition~\ref{NR:p:trunc} we can now show the following.

\becor\label{c:defn}%
$\T=\MODEL[]{\T,T,\prn[],\obs[]}$ satisfies ip-definability.
\encor%
\proof For each $\all,A,B$, define $\dfny{A,B}\defn\{\,f:\prn\trn{A}\paei T\trn{B}\,|\,f\text{ is finitary}\,\}$\,. By definability, every
$f\in\dfny{A,B}$ is definable. We need also show:
\[ (\forall\rho\in \dfny{A\tote B,\N}\,.\ \Laa(f)\1\rho\in\obs\implies\Laa(g)\1\rho\in \obs)\implies f\ypoc[\all] g\,. \]
Assume the LHS assertion holds and let $\Laa(f)\1\rho\in \obs$, some \hbox{$\rho:\prn(\trn{A}\him T\trn{B})\paei T\GN$.} Then, let
$[s\1t]=[(\all,*)\,*\,\ixi\,(0,\ixi)^{\bee}]\in\Laa(f)\1\rho$\,, $[s]\in\Laa(f)$ and $[t]\in\rho$. By proposition~\ref{NR:p:trunc},
$[t]\in\rho\trnat{t}$\,, so $\Laa(f)\1\rho\trnat{t}\in \obs$. Moreover, $\rho\trnat{t}\in \dfny{A\tote B,\N}$\,, so $\Laa(g)\1\rho\trnat{t}\in \obs$,
by hypothesis. Finally, $\rho\trnat{t}\ypoo\rho$ implies $\Laa(g)\1\rho\trnat{t}\ypoo\Laa(g)\1\rho$\,, hence the latter observable, so $f\ypoc g$.%
\qed\noindent%
Hence, we have shown full abstraction.
\bethm\label{t:p-laNR}
$\T=\MODEL[]{\T,T,\prn[],\obs[]}$ is a fully abstract model of $\nurho$. \qed \enthm %

\subsection{An equivalence established semantically}\label{NR:s:equiv}

In this last section we prove that the following terms $M$ and $N$ are equivalent. The particular equivalence exemplifies the fact that exceptional behaviour cannot be simulated in general by use of references, even of higher-order.
\[  M\defn \la f.\,\sstop:(\ena\tote\ena)\tote\ena\,,\qd
  N\defn \la f.\,f\,\sskip\1\sstop:(\ena\tote\ena)\tote\ena\,. \]
By full-abstraction, it suffices to show $\trn{M}\Eypoc\trn{N}$, where the latter are
given as follows.
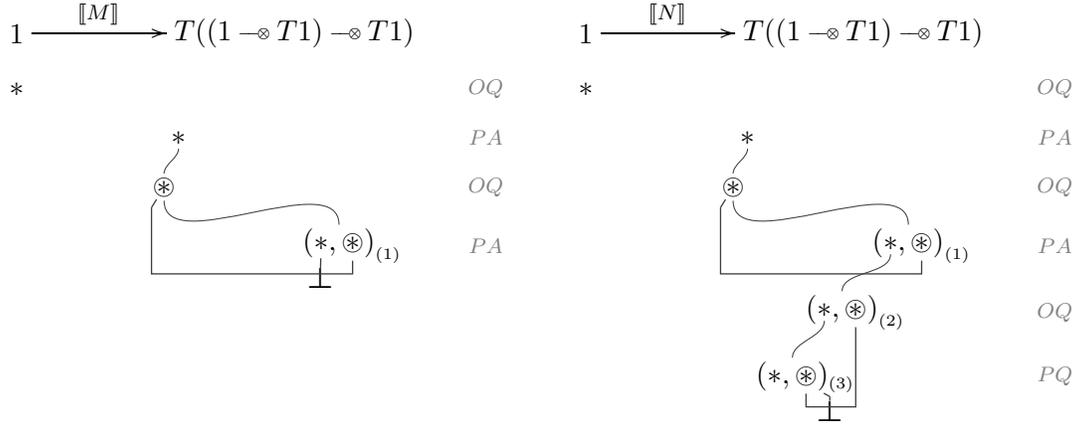
\[
\xymatrix@C=2mm@R=2mm{1\ar[rrrr]^-{\trn{M}} &&&& T((1\him T1)\him T1) \\
{*} &&&&& \GOQ\\
&&&& \rnode{A}{*}\qd[8] & \GPA\\
&&&& \xinode{B}\qd[9]   & \GOQ\\
&&&& \qd[4]\rnode{C}{(\rput[b](.085,-.5){\rnode{C1P}{\boldsymbol\bot}}\rnode{C1}{*},\xinode{C2})}\indx{1} & \GPA\\
\nccdu{B}{A}\nccdu{C}{B}\xiccl{B}{C2}}%
\ncline[nodesepA=3pt,linewidth=0.5pt,linecolor=darkgray]{C1}{C1P}%
\qd[2]%
\xymatrix@C=2mm@R=2mm{1\ar[rrrr]^-{\trn{N}} &&&& T((1\him T1)\him T1) \\
{*} &&&&& \GOQ\\
&&&& \rnode{A}{*}\qd[8] & \GPA\\
&&&& \xinode{B}\qd[9]   & \GOQ\\
&&&& \qd[4]\rnode{C}{(\rnode{C1}{*},\xinode{C2})}\indx{1} & \GPA\\
&&&& \,\;\rnode{D}{(\rnode{D1}{*},\xinode{D2})}\indx{2}\qd[1] & \GOQ\\
&&&& \xinode[(\rnode{E1}{*},\xinode{E2})]{E}\rput[b](.014,-.5){\rnode{E2P}{\boldsymbol\bot}}\indx{3}\qd[4] & \GPQ\\
\nccdu{B}{A}\nccdu{C}{B}\nccdu{D}{C1}\nccdu{E}{D1}\xiccl{B}{C2}\cc{E2}{D2}}%
\ncline[linewidth=0.5pt,linecolor=darkgray,nodesepA=1pt]{E}{E**}\ncline[nodesep=0pt,linewidth=0.5pt,linecolor=darkgray]{E**}{E2P}%
\]
Bottom links stand for deadlocks: if Opponent plays a move $(*,\ixi)\indx{2}$ under the last ${*}$ in $\trn{M}$ (thus providing the function $f$) then Player must play $\trn{\sstop}$, i.e.~remain idle. Similarly for $\trn{N}$: if Opponent gives an answer to $(*,\ixi)\indx{3}$
(providing thus the outcome of $f\sskip$) then Player deadlocks the play.

We have that $\trn{M}\ypoo\trn{N}$ and therefore, by~\eqref{e:ypooypoc}, $\trn{M}\ypoc[]\trn{N}$\,. Conversely, let $\rho:T((1\him T1)\him T1)\paei T\GN$ be a tl4 tidy strategy such that
$[*\,*\,\ixi\,(0,\ixi)^\all]\in\trn{N}\1\rho$ for some $\all$. Then, because of the form of $\trn{N}$, $\rho$ can only play initial moves up to $(*,\ixi)\indx{1}$, then possibly ask some names to $(*,\ixi)\indx{1}$, and finally play $(0,\ixi)^\all$. Crucially, $\rho$ cannot play $(*,\ixi)\indx{2}$ under $*$: this would introduce a question that could never be answered by $\trn{N}$, and therefore $\rho$ would not be able
to play $(0,\ixi)^\all$ without breaking well-bracketing. Hence, $\trn{M}$ and $\rho$ can simulate the whole interaction and therefore $[*\,*\,\ixi\,(0,\ixi)^\all]\in\trn{M}\1\rho$.

\section{Conclusion}\noindent
%
%
Until recently, names used to be bypassed in Denotational Semantics: most approaches focussed on the effect achieved by use of names rather than names themselves. Characteristic of this attitude was the `object-oriented'  modelling of references~\cite{Abramsky_McCusker:IA97,Abramsky+:GamesReferences} and exceptions~\cite{Laird:Exceptions} as products of their effect-related \emph{methods} (in the spirit of~\cite{Reynolds:IA}). These approaches were  unsatisfactory to some extent, due to the need for `bad' syntactic constructors in the examined languages. Moreover, they could not apply to the simplest nominal language, the $\nu$-calculus~\cite{Pitts_Stark}, since there the achieved effect could not be given an extensional, name-free description. These issues revealed the need that names be treated as a proper computational effect~\cite{Stark:PhD}, and led to the advent of nominal games~\cite{AGMOS,Laird:fossacs04}.

In this paper we have taken some further steps in the semantics of nominal computation by examining the effect of (nominal) general references. We have shown that nominal games provide a framework expressive enough that, by use of appropriate monadic (and comonadic) constructions, one can model general references without moving too far from the model of the $\nu$-calculus~\cite{AGMOS}. This approach can be extended to other nominal effects too; e.g.~in~\cite{Tze:apal08} it is applied to exceptions (with and without references). Moreover, we have examined abstract categorical models for nominal computation, and references in particular (in the spirit of~\cite{Stark:CatMod,Stark:PhD}).

There are many threads in the semantics of nominal computation which need to be pursued further. Firstly, there are many nominal games models to build yet: research in this direction has already been undertaken in~\cite{Laird:Names_Pointers,Laird:06:GamesConcur,Tze:apal08,Tze_Murawski:RML}. By constructing models for more nominal languages we better understand the essential features of nominal computation (e.g.~\emph{name-availability}~\cite{Tze_Murawski:RML}) and build stronger intuitions on nominal games. Another direction for further research is that of characterising
the nominal effect\HY i.e.~the computational effect that rises from the use of names\HY in abstract categorical terms. Here we have pursued this task to some extent by introducing the monadic-comonadic description of nominal computation, but it is evident that the description needs further investigation. We see that there are more monad-comonad connections to be revealed, which will simplify and further substantiate the presentation. The work of Sch{\"o}pp which examines categories with names~\cite{Schoepp_PhD}
seems to be particularly helpful in this direction.

A direction which has not been pursued here is that of decidability of observational equivalence in nominal languages.
The use of denotational methods, and game semantics in particular, for attacking the problem has been extremely successful in the `non-nominal' case, having characterised decidability of (fragments of) Idealized Algol~\cite{Ghica_McCusker:IA:00,Ong:lics02,Murawski:lics03}.
It would therefore be useful to `nominalise' that body of work and apply it to nominal calculi. Already from~\cite{Murawski:lics03} we can deduce that nominal languages with ground store are undecidable, and from~\cite{Pitts_Stark} we know that equivalence is decidable for programs of first-order type in the $\nu$-calculus, but otherwise the problem remains open.

\paragraph{\it Acknowledgements} I would like to thank Samson Abramsky for his constant encouragement, support and guidance. I would also like to thank Andy Pitts, Andrzej Murawski, Dan Ghica, Ian Stark, Luke Ong, Guy McCusker, Jim Laird, Paul Levy, Sam Sanjabi and the anonymous reviewers for fruitful discussions, suggestions and criticisms.

%

\appendix
\section{Deferred proofs}
\noindent\textbf{I. Proof of closure of tidiness under composition.}

\belem\label{l:tidy} %
Let $\sig:A\paei B$ and $\tau:B\paei C$ be tidy strategies, and let $[s\1t]\in\st$, $[s]\in\sig$ and $[t]\in\tau$, with $\pv{s\comp t}=s\comp t$ ending in a generalised O-move in $AB$ and $x$, an O-move, being the last store-H in $\pv{s}$. Let $x$ appear in $s\comp t$ as $\tilde x$. Then, $\tilde x$ is the last store-H in $s\comp t$ and if $x$ is in $A$ then all moves after $\tilde x$ in $s\comp t$ are in $A$.
Similarly for $BC$ and $t$.
\enlem %
\proof We show the $(AB,s)$ case, the other case being entirely dual. Let $s=s_1xs_2$ and let $x$ appear in $s\comp t$ as some $\tilde x$. If $x$
is in $A$ then we claim that $s_2$ is in $A$. Suppose otherwise, so $s=s_1xs_{21}ys_{22}$ with $s_{21}$ in $A$ and $y$ a P-move in $B$. Since $x$
appears in $\pv{s}$, the whole of $s_{21}y$ appears in it, as it is in P-view mode already. Since $x$ is last store-H in $\pv{s}$, $s_{21}y$ is
store-H-less. If $y$ a store-Q then it should be justified by last O-store-H in $\pv{s_{<y}}$, that is $x$, which is not possible as $x$ is in $A$.
Thus, $y$ must be a store-A, say to some O-store-Q $q$ in $B$. Now, since $q$ wasn't immediately answered by P, tidiness dictates that $\pv{s}$ be a
copycat from move $q$ and on. But then the move following $x$ in $s$ must be a copy of $x$ in $B$, \contra. Hence, $s_2$ is in $A$ and therefore it
appears in $\pv{s}$, which implies that it is store-H-less. Thus, $\tilde x$ is last store-H in $s\comp t$.

If $x$ is in $B$ then we do induction on $|s\comp t|$. The base case is encompassed in the case of $s_2$ being empty, which is trivial.
So let $s_2=s_{21}ys_{22}z$ with $y$ justifying $z$ (since $x$ appears in $\pv{s}$, $z$ has to be justified in $s_2$). $z$ is not a store-H and neither is it a store-Q, as then $y$ would be a store-H after $x$ in
$\pv{s}$. Thus $z$ a store-A and $y$ a store-Q, the latter justified by last O-store-H in $\pv{s_{<y}}=\pv{s}_{<y}$\,, that is $x$, so $y,z$ in $B$.
Now, $s=s_1xs_{21}ys_{22}z$ and $t=t_1x't_{21}y't_{22}z'$\,; we claim that $s_{21}$ and $t_{21}$ are store-H-less. Indeed, $s_{<y}\comp t_{<y'}$
ends in a generalised O-move in $AB$ and $x$ is still the last store-H in $\pv{s_{<y}}$\,, from which we have, by IH, that $\tilde x$
is the last store-H in $s_{<y}\comp t_{<y'}$.

Thus, $s\comp t=(s_1\comp t_1)\tilde{x}v\tilde{y}u\tilde{z}$ with $v$ store-H-less. It suffices to show that $u$ is also store-H-less.
In fact,
$u=\underbrace{\tilde{y}\dots\tilde{y}}_n\,\underbrace{\mathstrut\tilde{z}\dots\tilde{z}}_n$
for some $n\geq0$.  Indeed, by tidiness of $\tau$,
$(t_{22}z').1$ is either an answer to $y'$, whence $t_{22}=u=\ee$, or a copy of it under the last O-store-H  in $\pv{t_{\leq y'}}$.
If the latter is in $B$ then $\sig$ reacts analogously, and so on, so there is initially a sequence $\tilde{y}\dots\tilde{y}$ in $u$, played in $B$.
As $u$ finite, at some point $\sig$ (or $\tau$) either answers $y$ ($y'$) or copycats it in $A$ (in $C$). In the latter case,
O immediately answers, as $s$ ($t$) is in P-view mode in $A$ (in $C$). Hence, in either cases there is an answer that is copycatted to all open
$\tilde y$ in $u$, yielding thus the required pattern. Therefore, $u$ is store-H-less. \qed
\belem\label{l:tidy2}%
Let $\sig:A\paei B$ and $\tau:B\paei C$ be tidy strategies, and let $[s\1t]\in\st$, $[s]\in\sig$ and $[t]\in\tau$, with $\pv{s\comp t}=s\comp t$ ending in a generalised O-move. If there exists $i\geq1$ and store-Q's $\tilde{q}_1,...,\tilde{q}_i$ with $\tilde{q}=\tilde{q}_j$, all $1\leq j\leq i$, and
$\tilde{q}_1,...,\tilde{q}_{i-1}$ in $B$ and $\tilde{q}_i$ in $AC$ and $[(s\comp t)\tilde{q}_1...\tilde{q}_i]\in\sig\comp\tau$, then $\tilde{q}_i$ is
justified by the last O-store-H in $s\1t$.%
\enlem%
\proof By induction on $|s\comp t|$. The base case is encompassed in the case of $s\1t$ containing at most one O-store-H, which is trivial. Now let without loss of generality $(s\comp t)\tilde{q}_1...\tilde{q}_i=(sq_1...q_i)\comp(tq'_1...q'_{i-1})$ with $[sq_1...q_i]\in\sig$ and $[tq'_1...q'_{i-1}]\in\tau$, and let each $q_j$ be justified by $x_j$ and each $q'_j$ by $x'_j$\,. Moreover, by hypothesis, $\ul{x_j}=\ul{x'_j}$, for $1\leq j\leq i-1$, and therefore each such pair $x_j,x'_j$ appears in $s\comp t$ as some $\tilde{x}_j$, the latter justifying $\tilde{q}_j$ in $s\comp t$.

Now, assume without loss of generality that $s\comp t$ ends in $AB$. Then, by tidiness of $\sig$ and $\tau$ we have that, for each $j\geq1$,
\[ q_{2j+1}=q_{2j} \qd,\qd q'_{2j}=q'_{2j-1} \qd,\qd \ul{q_j}=\ul{q'_j} \]
For each $j\geq1$, $q_{2j+1}$ is a P-move of $\sig$ justified by some store-H, say $x_{2j+1}$. By tidiness of $\sig$, $x_{2j+1}$ is the last O-store-H
in $\pv{s_{<q_{2j+1}}}=\pv{s_{\leq q_{2j}}}$, and therefore $x_{2j+1}$ is the last store-H in $\pv{s_{<x_{2j}}}$. Then, by previous lemma,
$\tilde{x}_{2j+1}$ is the last store-H in $s_{<x_{2j}}\comp t_{<x'_{2j}}=(s\comp t)_{<\tilde{x}_{2j}}$. Similarly, $\tilde{x}_{2j}$ is the last store-H in $(s\comp t)_{<\tilde{x}_{2j-1}}$. Hence, the store-H subsequence of $(s\comp t)_{\leq\tilde{x}_1}$ ends in $\tilde{x}_i...\tilde{x}_1$.

Now, by tidiness of $\sig$, $x_1$ is the last O-store-H in $\pv{s}$. If $x_1$ is also the last store-H in $\pv{s}$ then, by previous lemma, $\td[x]_1$
is the last store-H in $s\comp t$, hence $\td[x]_i$ is the last store-H in $s\1t$. Otherwise, by corollary~\ref{c:TidyDiscipline}, $q_1$ is a copy
of $s.\arnt1=q_0$\,. If $q_0$ is in $A$ then its justifier is $s.\arnt2=x_0$ and, because of CC-mode, the store-H subsequence of $s\comp t$ ends in
$\td[x]_i...\td[x]_1\td[x]_0$\,, so $\td[x]_i$ is the last O-store-H in $s\1t$.
If $q_0$ is in $B$ then we can use the IH on $s^-\comp t^-$ and $\tilde{q}_0,\tilde{q}_1,...,\tilde{q}_{i}$, and obtain that
$\tilde{x_i}$ is the last O-store-H in $s^-\1t^-=s\1t$. \qed
\beprop\label{p:tidys}%
If $\sig:A\paei B$ and $\tau:B\paei C$ are tidy strategies then so is $\st$.%
\enprop
\proof Take odd-length $[s\1t]\in\st$ with not both $s$ and $t$ ending
in $B$, $\pv{s\comp t}=s\comp t$ and $|s\1t|$ odd. We need to show
that $s\1t$ satisfies (TD1-3). As (TD2) is a direct consequence of the
previous lemma, we need only show the other two conditions. Assume
without loss of generality that $s\1t$ ends in $A$.

For (TD1), assume $s\1t$ ends in a store-Q $\tilde q$. Then $s$ ends in some $q$, which is justified by the P-store-H $s.\arnt2=x$ (also in $A$). $q$
is either answered or copied by $\sig$\,; in particular, if $\tilde{q}=\al^\all$ with $\al\4\pv{s\1t}^-=s^-\1t$ then $\al\4s^-,t$\,, so $\sig$ copies
$q$. If $\sig$ answers $q$ with $z$ then $z$ doesn't introduce new names, so $[(s\1t)\tilde z]\in\st$ with $\nlist(\tilde z)=\nlist(\tilde q)$ and
$\ul{\tilde z}=\ul{z}$\,, as required.

Otherwise, let $\sig$ copy $q$ as $q_1$\,, say, under last O-store-H in $\pv{s}$, say $x_1$. If $x_1$ is in $B$ then
$sq_1\asymp tq'_1$, with $q_1,q'_1$ in $B$ and $q'_1$ being $\underline{q_1}$ with name-list that of its justifier, say $x'_1$, where
$\ul{x_1}=\ul{x'_1}$\,. Now $[tq'_1]\in\tau$ and it ends in a store-Q, so $\tau$ either answers it or copies it under last O-store-H in $\pv{tq'_1}$.
In particular, if $q=\al^\all$ with $\al\4\pv{s\1t}$ then, as above, $\al\4t$ and $\tau$ copies $q'_1$. This same reasoning can be applied
consecutively, with copycats attaching store-Q's to store-H's appearing each time earlier in $s$ and $t$. As the latter are finite and initial
store-H's are third moves in $s$ and $t$, at some point either $\sig$ plays $q_i$ in $A$ or answers it in $B$, or $\tau$ plays $q'_i$ in $C$ or answers
it in $B$. If an answer occurs then it doesn't introduce new names (by tidiness), so it is copycatted back to $q$ closing all open $q_j$'s and
$q'_j$'s. Otherwise, we need only show that, for each $j$, $\tilde{q}_j=\tilde{q}$, which we do by induction on $j$: $\tilde{q}_1=\ul{q}^{s\mix t,\ee}$
and $\tilde{q}_{j+1}=\ul{q}^{(s_{\leq q_j})\mix (t_{\leq q'_j}),\ee}=\tilde{q}_j\overset{IH}{=}\tilde{q}$. This proves (TD1).

For (TD3), assume $s\1t=u\tilde{q}\indx{O}\tilde{q}\indx{P}v\tilde{y}$ with $\tilde{q}\indx{O}\tilde{q}\indx{P}v$ a copycat. Then, either both
$\tilde{q}\indx{O},\tilde{q}\indx{P}$ are in $A$, or one is in $A$ and the other in $C$. Let's assume $\tilde{q}\indx{O}$ in $A$ and
$\tilde{q}\indx{P}$ in $C$\HY the other cases are shown similarly. Then, $\tilde{q}\indx{O}$ her(editarily)-justifies $\tilde{y}$, and let $s.\arnt 1=y$ be justified by some $x$ in $s$. Now, as above, $\tilde{q}\indx{O}\tilde{q}\indx{P}$ is witnessed by some $\td\indx{O}\td_1\dots\td_i\td\indx{P}$ in $s\comp t$, with odd $i\geq1$ and all $\td_j$'s in $B$. We show by induction on $1\leq k\leq i$ that there exist
$x_1,...,x_k,x'_1,...,x'_k,y_1,...,y_k,y'_1,...,y'_k$ in $B$ such that $(sy_1\dots y_k\comp ty'_1\dots y'_k)\in\sig\comp\tau$ and, for each relevant
$j\geq1$,
\[ \ul{y_j}=\ul{y'_j}=\ul{y}\qd,\qd y_1=y\qd,\qd y_{2j}=y_{2j+1}\qd,\qd y'_{2j-1}=y'_{2j}\qd,\qd \ul{x_j}=\ul{x'_j} \]
with $q_j$ her-justifying $x_j$ in $s$ and $x_j$ justifying $y_j$ (and $q'_j$ her-justifying $x'_j$ in $t$ and $x'_j$ justifying $y'_j$), and
$\td[x]_{j+1},\td[x]_{j}$ consecutive in $s\comp t$, and $\td[x]_1,\td[x]$ also consecutive.

For $k=1$, let $s=s_1q\indx{O}q_1s_2y$. Now, $\td\indx{O}$ her-justifying $\td[y]$ implies that $q\indx{O}$ her-justifies $y$, hence it appears in
$\pv{s}$. Thus $\pv{s}=s_1'q\indx{O}q_1s_2'y$, so, by (original definition of) tidiness, $[sy_1]\in\sig$ with $y_1=y$ justified by
$x_1=\pv{s}.\arnt3=s.\arnt3$. Then, $[ty'_1]\in\tau$ with $\ul{y'_1}=\ul{y_1}$. By proposition~\ref{p:tidy_CC},
$q\indx{O}q_1s_2'$ is a copycat, so $q_1$ her-justifies $x_1$ and therefore $x_1,y_1$ in $B$. Finally, $x=\pv{s}.\arnt2=s.\arnt2$ is a P-move so
$\td[x]_1,\td[x]$ are consecutive in $s\comp t$.

For even $k>1$ we have, by IH, that $(sy_1\dots y_{k-\!1}\comp ty'_1\dots y'_{k-\!1})\in\sig\comp\tau$ with $y'_{k-\!1}$ an O-move her-justified by
$q'_{k-\!1}$, an O-move. Then, $q'_{k-\!1}$ appears in $\pv{ty'_1...y'_{k-\!1}}$, so $\pv{ty'_1...y'_{k-\!1}}=t_1q'_{k-\!1}q'_{k}t_2y'_{k-\!1}$, thus
(by tidiness) $[ty'_1...y'_{k-\!1}y'_{k}]\in\tau$ with $y'_{k}=y'_{k-\!1}$ justified by $x'_{k}=\pv{ty'_1...y'_{k-\!1}}.\arnt3$. Now,
$q'_{k-\!1}q'_{k}t_2$ is a copycat so $q'_{k}$ her-justifies $x'_{k}$. Moreover, $x'_{k},x'_{k-\!1}$ are consecutive in $\pv{t}$, so, as $x'_{k-\!1}$ a P-move, they are consecutive in $t$, and therefore $\td[x]_k,\td[x]_{k-\!1}$ consecutive in $s\comp t$. Finally,
$[sy_1\dots y_{k-\!1}y_{k}]\in\sig$ with $\ul{y_{k}}=\ul{y'_{k}}$. The case of $k$ odd is entirely dual.

Now, just as above, we can show that there exist $x'_{i+1},y'_{i+1}$ in $C$ such that $[ty'_1...y'_{i}y'_{i+1}]\in\tau$ and $y'_{i+1}$ justified by
$x'_{i+1}$, $x'_{i+1}$ her.~justified by $q\indx{P}$, etc. Then $[(s\1t)\td[y]_{i+1}]\in\st$ with $\td[x]_{i+1},\td[x]_i,...,\td[x]_1,\td[x]$ consecutive in $s\comp t$, so
$\td[x]_{i+1}=(s\1t).\arnt3$. Finally, as above, $\td[y]_{i+1}=\td[y]_j=\td[y]$, all $j$, as required. \myqed[2]

\noindent\textbf{II.}
\proof[ of Decomposition Lemma~\ref{l:Decomp}] 1 is straightforward: we just partition $\sigma$ into $\sigma_0$ and $\sigma'$ and recover it by use of $[x\overset{\all}{=}i_{A(0)}]$ and $\cnd$. For 2, we
just use the definition of name-abstraction for strategies and the condition on $\sig$.

For 3, it is clear that $m_0$ is either a store-Q $\al$ under $\ixi$, or a store-H $j_A$, or a store-H $(i_B,\ixi)$.

In case $m_0=\al$ with $\al\in\NA{C}$, we define $\sig_\al:\prn(\trn{A}\ten\trn{C})\paei T\trn{B}\defn\strat(f_\al)$\,, where
\begin{align*}
f_\al &\defn\{\,[\,(\all,i_{A(0)},i_C)\,*\,\ixi\, s\,]\,|\,[\,(\all,i_{A(0)})\,*\,\ixi\,\al\, i_C\,s\,]\in\viewf(\sig)\,\}\,.
\end{align*}
To see that $f_\al$ is a viewfunction it suffices to show that its elements are plays, and for that it suffices to show that they are legal. Now, for
any $[(\all,i_{A(0)},i_C)*\,\ixi\, s]\in f_\al$ with $[(\all,i_{A(0)})*\,\ixi\,\al\,i_C\,s]\in\viewf(\sig)$, $(\all,i_{A(0)},i_C)*\,\ixi\, s$ is a
justified sequence and satisfies well-bracketing, as its open Q's outside $s$ are the same as those in $(\all,i_{A(0)})*\,\ixi\,\al\, i_C\,s$\,, i.e.
$\ixi$. Moreover, visibility is obvious.
Hence, $f_\al$ is a viewfunction, and it inherits tidiness from $\sig$. Moreover, we have the following diagram.
{\small
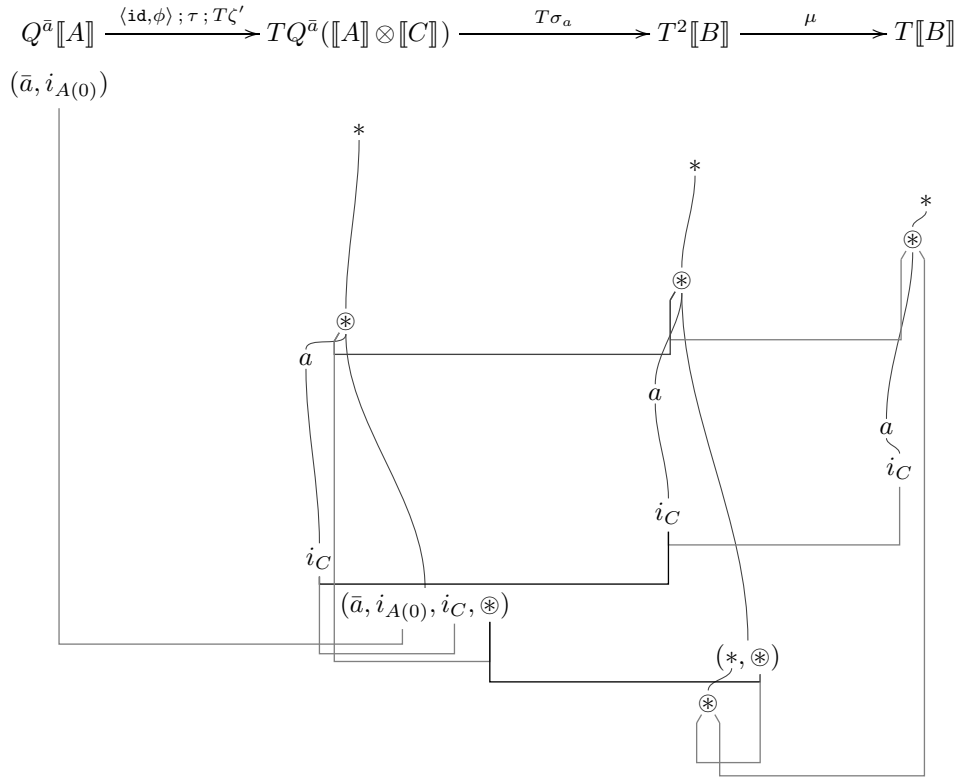
\[ \xymatrix@C=11mm@R=1mm{
\prn\trn{A}\ar[r]^-{{\ang{\id,\phi}\1\tau\1T\zet'}} & T\prn(\trn{A}\ten\trn{C})\ar[r]^-{{T\sig_{\al}}} & \TT\trn{B}\ar[r]^{{\mu}} & T\trn{B}\\
\rnode{A1}{(\all,i_{A(0)})} \\
&   \rnode{B2}{*} \\
&                   &   \rnode{C3}{*} \\
&                   &                   &   \rnode{D4}{*} \\
&                   &                   &   \xinode{D5}\qd[1] \\
&                   &   \xinode{C6}\qd[1] \\
&   \xinode{B7}\qd[1] \\
&   \rnode{B8}{\al}\qd[4] \\
&                   &   \rnode{C9}{\al}\qd[3] \\
&                   &                   &   \rnode{D10}{\al}\qd[3] \\
&                   &                   &   \rnode{D11}{i_C}\qd[2] \\
&                   &   \rnode{C12}{i_C}\qd[2] \\
&   \rnode{B13}{i_C}\qd[3] \\
&   \qd[5]\rnode{B14}{(\all,\rnode{B14a}{i_{A(0)}},\rnode{B14b}{i_C},\xinode{B14c})} \\
&                   &   \qd[4]\rnode{C15}{(\rnode{C15a}{*},\xinode{C15b})} \\
&                   &   \qd[1]\xinode{C16} \\
\nccdu{B7}{B2}\nccdu{C6}{C3}\nccdu{D5}{D4}\nccdu{B8}{B7}\nccdu{C9}{C6}\nccdu{D10}{D5}\nccdu{D11}{D10}\nccdu{C12}{C9}\nccdu{B13}{B8}
\nccdu{B14}{B7}\nccdu{C15}{C6}\nccdu{C16}{C15a}
\ccA[2mm]{A1}{B14a}\ccA{D11}{C12}\ccB[1mm]{C12}{B13}\ccA[4mm]{B13}{B14b}\xicc[-15pt]{D5}{C6}\xicc{C6}{B7}\xiccl[-15pt]{B7}{B14c}\ccB[1mm]{B14c}{C15b}
\xiccl[-15pt]{C16}{C15b}\xxicc[-20pt]{C16}{D5} }
\]}%
Because of the copycat links, we see that
\begin{multline*}
\viewf(\ang{\id,\phi}\1\tau\1T\zet'\1T\sig_\al\1\mu)\hrp (\all,i_{A(0)})\\
= \{[(\all,i_{A(0)})*\ixi\,\al\, i_C\,s]\,|\,
    [(\all,i_{A(0)},i_C)*\ixi\, s]\in\viewf(\sig_\al)\}
    = \viewf(\sig)\,,
\end{multline*}
as required. Note that the restriction to initial moves $[\all,i_{A(0)}]$ taken above is necessary in case $\phi$ contains a projection (in which case
it may also answer other initial moves).

In case $m_0=j_A$ (so $m_0$ a store-H) and $[(\all,i_{A(0)})\,*\,\ixi\, m_0\,\al\,i_C]\in\sig$, we have that
\[ \sig=\strat(f_\al\cup(f'\plhn f'_\al))\,, \]
where $f_\al,f'$ are viewfunctions of type $\prn\trn{A}\paei T\trn{B}$, so that $f_\al$ determines $\sig$'s behaviour if O plays $\al$ at the given
point, and $f'\plhn f'_\al$ determines $\sig$'s behaviour if O plays something else. That is,
\begin{align*}
f_\al  &\defn \{\,[\,(\all,i_{A(0)})\,*\,\ixi\, j_A\,\al\, i_C\,s\,]\in\viewf(\sig)\,\}\\
f'_\al &\defn\{\,[\,(\all,i_{A(0)})\,*\,\ixi\, j_A\,\al\, s\,]\,|\,[\ixi\,\ixi\,\al\, s]\in\viewf(\id_\xi)\,\} \\
f'     &\defn f'_\al\cup\{\,[\,(\all,i_{A(0)})\,*\,\ixi\, j_A\, y\,s\,]\in\viewf(\sig)\,|\,y\neq\al\,\}\,.
\end{align*}
$f'$ differs from $\viewf(\sig)$ solely in the fact that it doesn't answer $\al$ but copycats it instead; it is a version of $\viewf(\sig)$ which has
forgotten the name-update of $\al$. On the other hand, $f_\al$ contains exactly the information for this update. It is not difficult to see that
$f',f_\al$ are indeed viewfunctions. We now define
\begin{align*}
  f''_\al:\prn\trn{A}\paei T\trn{C}
 &\defn\{\,[\,(\all,i_{A(0)})*\ixi(i_C,\ixi)\,s\,]\,|\,\\
 &\phantom{{}\defn\{\,}[\,(\all,i_{A(0)})*\ixi j_A\,\al\, i_C\,s\,] \in f_\al \lor [\ixi\ixi s]\in\viewf(\id_\xi)\,\} \\
  \sig_\al:\prn\trn{A}\paei T\trn{C}
 &\defn\strat(f''_\al) \\
  \sig':\prn\trn{A}\paei T\trn{B}
 &\defn\strat(f') \\
  \sig'':\prn\trn{A}\paei T\trn{B}
 &\defn\ang{\De,\sig_\al}\1\tau\1T(\id\tenn\phi\1\tau)\1\mu\1\isom\1T\sig'\1\mu\,.
\end{align*}
We can see that $\sig'$ is a tidy strategy. For $\sig_\al$, it suffices to show that $f''_\al$ is a viewfunction, since tidiness is straightforward.
For that, we note that even-prefix closure and single-valuedness are clear, so it suffices to show that the elements of $f''_\al$ are plays.

So let $[(\all,i_{A(0)})\,*\,\ixi\,(i_C,\ixi)\, s]\in f''_\al$ with $[(\all,i_{A(0)})*\,\ixi\, j_A\,\al\, i_C\,s]\in\viewf(\sig)$. We have that
$(\all,i_{A(0)})\,*\,\ixi\,(i_C,\ixi)\,s$ is a justified sequence, because $s$ does not contain any moves justified by $j_A$ or $\al$. In the former
case this holds because we have a P-view, and in the latter because $\al$ is a closed (answered) Q. Note also that there is no move in $s$ justified by $\ixi$: such a move $(i_B,\ixi)$ would be an A ruining well-bracketing as $j_A$ is an open Q, while a store-Q under $\ixi$ is disallowed by tidiness as $s.1$ is an O-store-H. Finally, well-bracketing, visibility and NC's are straightforward.

We now proceed to show that $\sig=\sig''$. By the previous analysis on $f''_\al$ we have that $\sig_\al=\sig'_\al\1\eta$ (modulo totality) where
$\sig'_\al$ is the possibly non-total strategy
\[ \sig'_\al:\prn\trn{A}\paei\trn{C}\defn\strat \{\,[\,(\all,i_{A(0)})\,i_C\,s\,]\,|\,[\,(\all,i_{A(0)})\,*\,\ixi\, j_A\,\al\, i_C\,] \in f_\al\,\}\,, \]
and hence $\sig''\hrp(\all,i_{A(0)})=\ang{\De,\sig'_\al}\1\id\tenn\phi\1\tau\1\isom\1T\sig'\1\mu$\,.
Analysing the behaviour of the latter composite strategy and observing that the response of $\sigma''$ to inputs different than $[\all,i_{A(0)}]$ is merely the initial answer $*$ imposed by totality, we obtain:
\begin{align*}
\viewf(\sig'') &= \{\,[\,(\all,i_{A(0)})\,*\,\ixi\, j_A\,\al\, s\,],[(\all,i_{A(0)})\,*\,\ixi\, j_A\,y\,s]\in\viewf(\sig'')\,|\,y\neq\al\,\} \\
    &= \{\,[\,(\all,i_{A(0)})\,*\,\ixi\, j_A\,\al\, i_C\,s\,]\,|\,[\,(\all,i_{A(0)})\, *\, \ixi\,(i_C,\ixi)\, s]\in f''_\al\Land s.1\in J_{\trn{C}}\,\} \\
    &\qd[2] \cup\{\,[\,(\all,i_{A(0)})\,*\,\ixi\, j_A\,y\,s\,]\in f'\,|\,y\neq\al\,\} \\
    &= f_\al\cup(f'\plhn f'_\al)=\viewf(\sig)
\end{align*}
as required.

In case $x=(i_B,\ixi)$ we work similarly as above.\qed

\bibliographystyle{acm}
\bibliography{bgrafia}
\end{document}